\documentclass[
 reprint,
superscriptaddress,
 amsmath,amssymb,
 aps,
prb,
]{revtex4-2}

\usepackage{graphicx}
\usepackage{dcolumn}
\usepackage{bm}
\usepackage{color}
\usepackage[colorlinks,allcolors=blue]{hyperref}

\usepackage{physics}
\usepackage{subfigure}

\begin{document}

\title{Antiparticle of exciton in semimetals}
\author{Lingxian Kong}
\affiliation{International Center for Quantum Materials, School of Physics, Peking University, Beijing 100871, China}

\author{Ryuichi Shindou}
\email{Corresponding author. rshindou@pku.edu.cn}
\affiliation{International Center for Quantum Materials, School of Physics, Peking University, Beijing 100871, China}%

\author{Yeyang Zhang}%
\email{Corresponding author. yeyzhang@pku.edu.cn}
\affiliation{International Center for Quantum Materials, School of Physics, Peking University, Beijing 100871, China}%

\date{\today}
\begin{abstract}
An emergent quantized field enriches quantum many-body systems. 
We propose an antiparticle analog of the exciton 
in semimetals as an emergent collective mode in interacting electron 
systems. 
We show that inter-band excitations in semimetals  are generally comprised of 
both excitons and antiparticles of excitons.
These two stand for two distinct 
inter-band collective modes in semimetals, having different energies  
and opposite conserved charges. The conserved charge here is a 
quantity conjugate to a joint U(1) symmetry of two electron's 
bands associated with the inter-band excitations. 
The opposite charges foster fertile scattering processes 
among the inter-band collective modes. In spin-polarized systems, they also suggest possible experimental 
detections of the antiparticles.
We clarify  that the effective theory of the inter-band excitations
is given by a generalized Klein-Gordon theory. 
Our theory provides a comprehensive understanding of 
excitonic spectra in generic semimetals, 
bringing a new insight into electronic collective phenomena in solids.
\end{abstract}

\maketitle


\section{\label{sec1}Introduction}
The existence of antiparticles is one of the most striking predictions by the quantum field theory~\cite{srednicki_2007}. It renews the basic concept of particles by showing that particle numbers 
are no longer conserved quantities. In elementary particle physics, every particle has a corresponding 
antiparticle in some forms. 
Some particles are their own antiparticles, such as photons and Higgs bosons; while others are not, 
such as electrons and quarks~\cite{griffiths2020introduction}. A pair of a particle and an antiparticle can 
be simultaneously created and/or annihilated. A typical example is an annihilation of an electron-positron pair producing two photons. 

Electromagnetic and optical properties of solid-state materials 
are related to emergent collective modes in quantum many-particle systems of electrons and cations. 
Energy scales of the collective modes 
are typically much lower than 
those in high-energy experiments, while 
effective field theories of the emergent degrees of freedom often share rich mathematical structures 
with elementary particles in high-energy physics. Thus, 
it is an important 
quest to find collective modes in condensed matter systems 
as analogs of antiparticles. 

\begin{figure}[b]
    \centering
    \includegraphics[height=5.5cm]{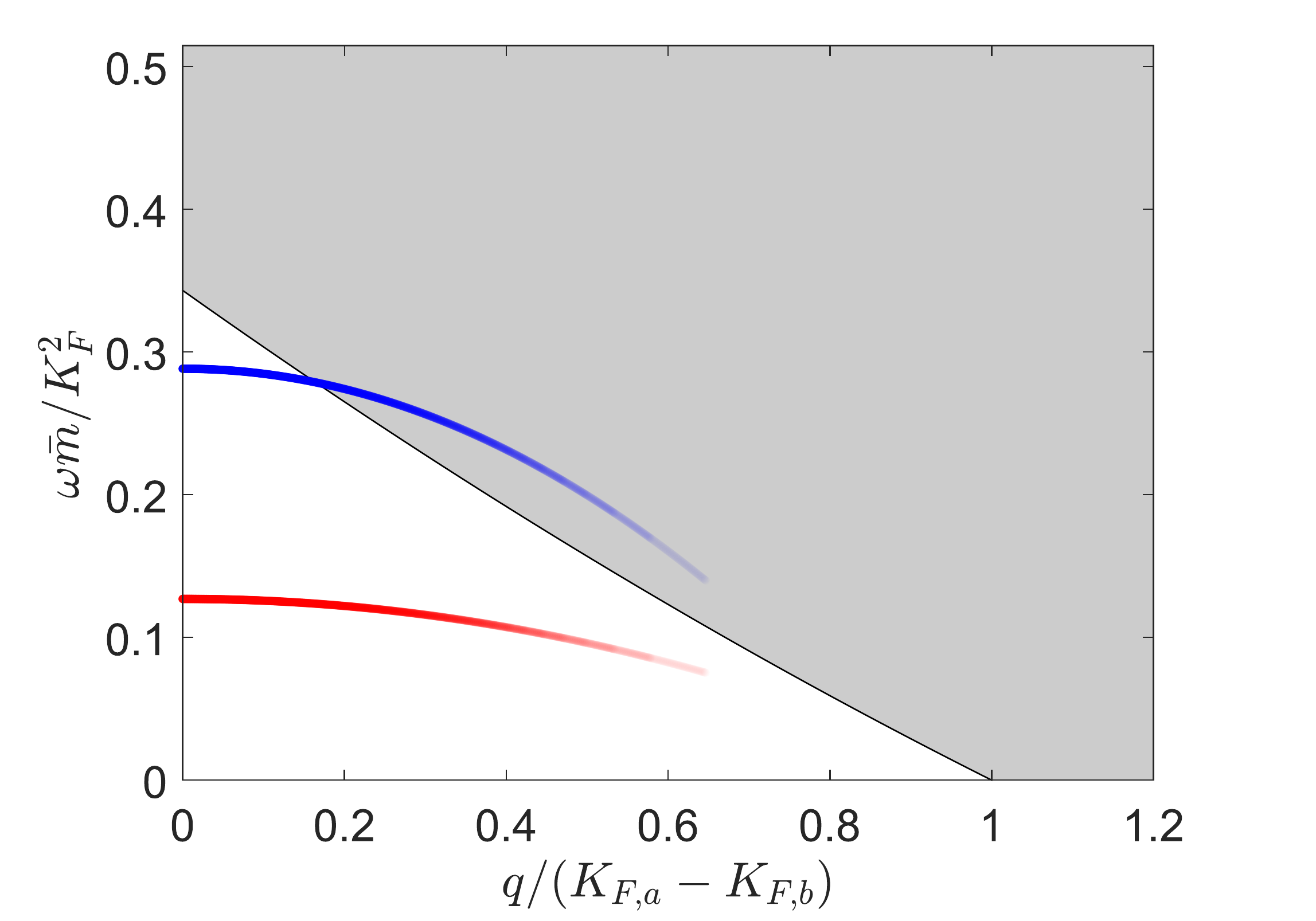}%
    \caption{ Energy-momentum ($\omega$-$q$) dispersions of the $s$-wave exciton (the red line) and antiexciton (the blue line) bands together with an energy-momentum region of inter-band individual excitations (the shaded region) in a two-bands model in 3D given by Eqs.~(\ref{band1}-\ref{Hint}). In the case of ${\bm k}_c \ne 0$, the momentum ${\bm Q}$ of the inter-band excitations is given by ${\bm Q} = {\bm q}+{\bm k}_c$. The parameters set in this Figure are the same as in Fig.~\ref{data}(a). In this paper, the energy dispersions are always calculated perturbatively in small $q$ around the $\Gamma$ point. To emphasize this point, we use more transparent colors for the dispersions with larger $q$ in this Figure and other Figures below. 
    }
    \label{dispersion}
\end{figure}

In this paper, we demonstrate  universal presence of  
the antiparticle analogs of excitons that are 
distinguishable from their counterpart excitons and that coexist with the counterpart excitons (see Fig.~\ref{dispersion}).
In semiconductors, an exciton is a non-relativistic boson that 
describes a bound state of a conduction-band electron 
and a valance-band hole. The 
bound state is an eigenstate of a two-body Hamiltonian 
of the electron and hole that interact through the long-range Coulomb 
interaction~\cite{PhysRevB.62.4927,martin_reining_ceperley_2016,strinati1988application}. 
Excitons play crucial roles in optoelectronic properties in semiconductors~\cite{koch2006semiconductor,kaindl2003ultrafast,PhysRevLett.81.3263}. 
In semimetals or narrow-gap semiconductors where the conduction and 
valance bands overlap or nearly overlap, excitons may undergo Bose-Einstein condensation~\cite{Halperin1968TheES,RevModPhys.40.755,kozlov1965metal,HANAMURA1977209, doi:10.1126/science.aam6432}. The condensates are described by a Ginzburg-Landau  
theory as in superconductivity~\cite{PhysRev.158.462,PhysRev.126.1691,Littlewood_2004}. 
There is no antiparticle analog in such conventional theories of excitons.  

 There were attempts to define an antiparticle of the exciton~\cite{PhysRevB.44.1368,PhysRevB.44.3638,arseev1998,lerner1981two,PhysRevB.59.5627}. Excitonic spectra in the condensates~\cite{lerner1981two} or under external pumps~\cite{PhysRevB.44.1368,PhysRevB.44.3638} which induce a hopping between the conduction and valence band have been studied previously. 
Thereby, the two-body Hamiltonian has pairs of degenerate positive- and 
negative-energy eigenstates~\cite{PhysRevB.62.4927,martin_reining_ceperley_2016,PhysRevB.44.1368,PhysRevB.44.3638}, 
which are related to each other by a generic particle-hole symmetry of the Hamiltonian. 
They are sometimes dubbed as pairs of ``exciton'' and ``antiexciton'' respectively in 
literature~\cite{PhysRevB.44.1368,PhysRevB.44.3638}, while such pairs of ``two'' states actually 
characterize identical physical excitations (see Appendix 
\ref{appendix:b}). 
Besides, inter-band excitations in 
semiconductors with positive and ``negative" band gaps are characterized
as excitons and antiexcitons in other literature~\cite{lerner1981two,PhysRevB.59.5627}, while coexistence of 
these two kinds of inter-band collective modes in the same bulk has not been explored (Appendix \ref{appendix:b}). 
%

\begin{figure}[t]
    \centering
    \subfigure[ ]{\includegraphics[height= 1.2in]{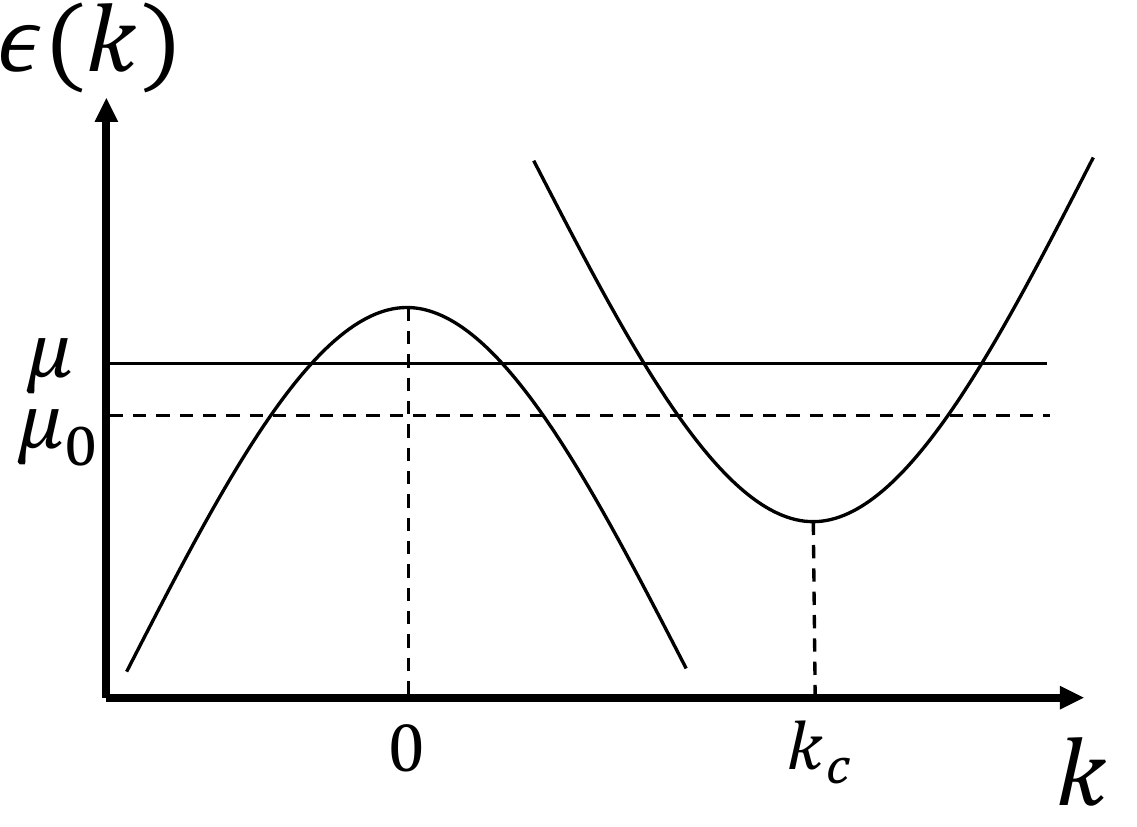}}
    \subfigure[ ]{\includegraphics[height= 1.2in]{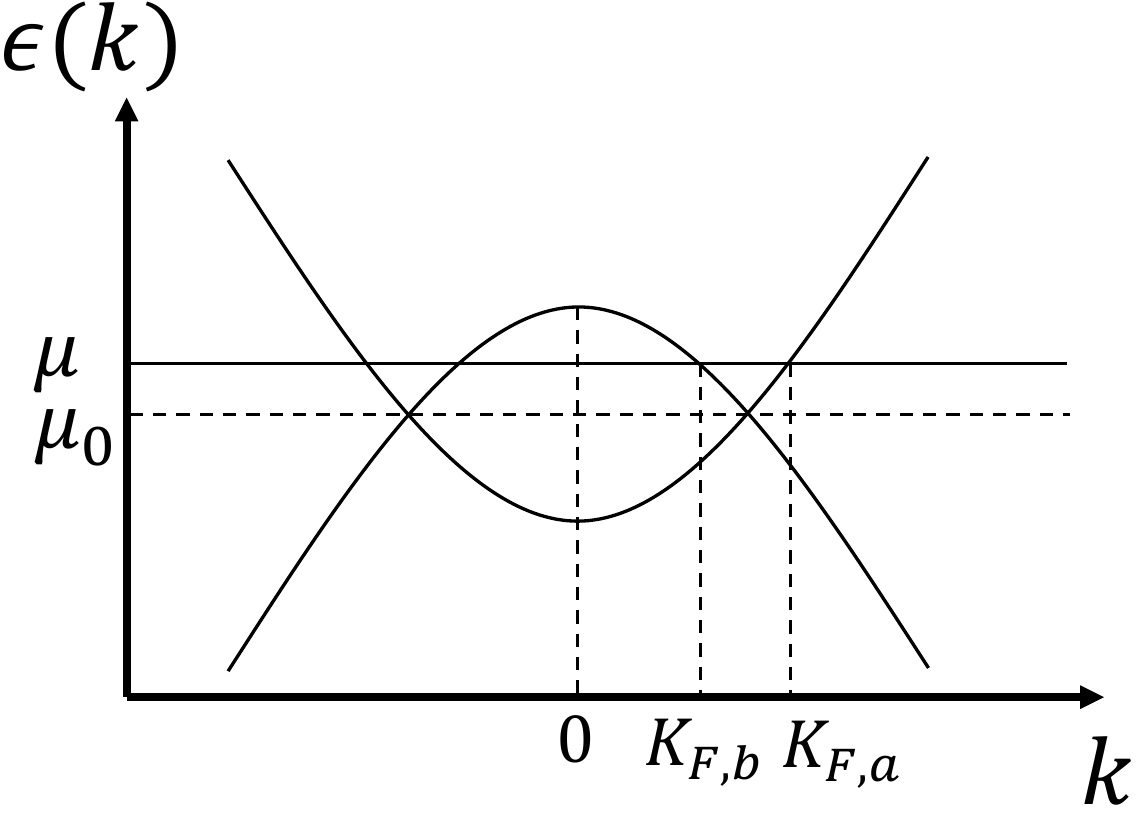}}
    
    \subfigure[ ]{\includegraphics[height= 1.in]{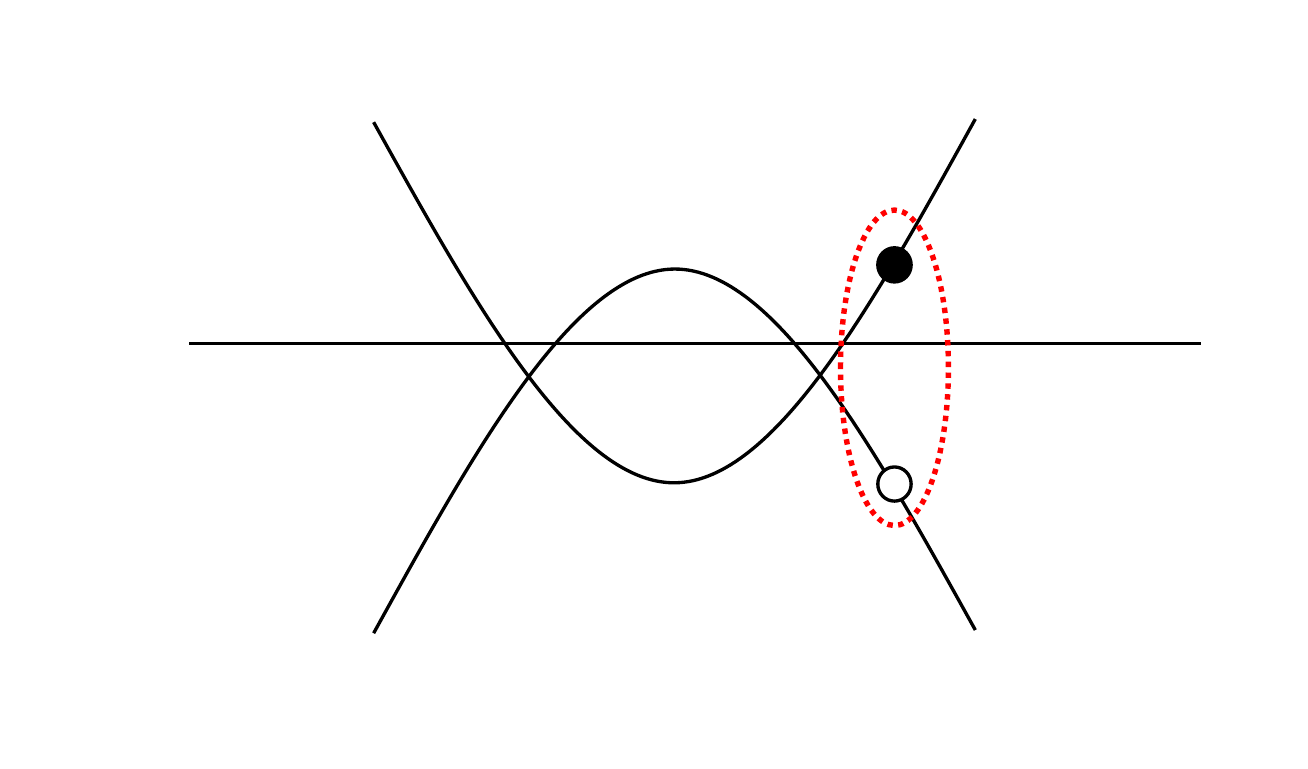}} 
    \subfigure[ ]{\includegraphics[height= 1.in]{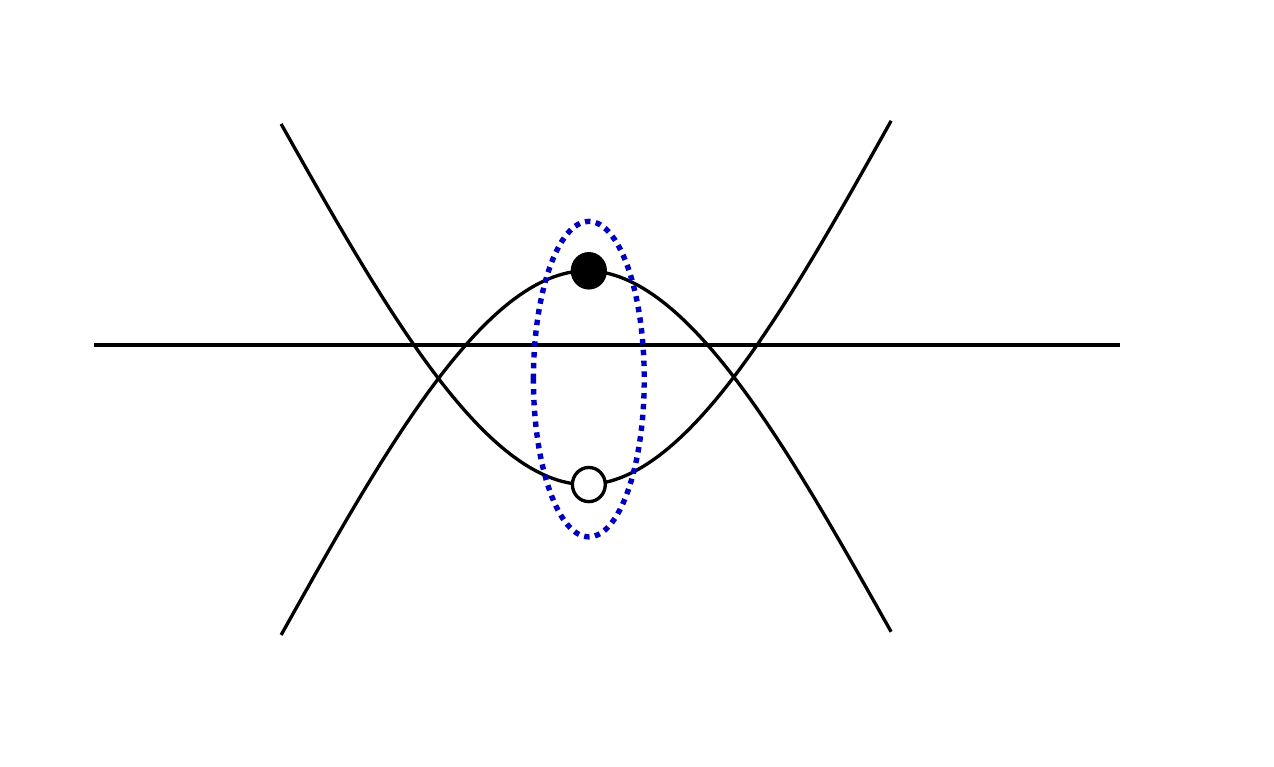}}
    \caption{ 
(a) A two-band semimetal model with $k_c \ne 0$. (b) A semimetal with $k_c=0$. In a ground state, the conduction band ($a$ band) and valence band ($b$ band) has $N_a$ and $N_b$ electrons, respectively. $K_{F,a}$ and $K_{F,b}$ denote the Fermi momentum of a circular electron pocket in the $a$ band and the Fermi momentum of a circular hole pocket in the $b$ band. (c,d) Two types of inter-band collective modes in the semimetal with the U(1) $\times$ U(1) symmetry; (c) Exciton, an inter-band collective mode in the space of $|N_a+1,N_b-1\rangle$. A wave
function of an exciton state has a weight mainly in a region of $|k|>K_{F,a}$. (d) Antiexiciton, an
inter-band collective mode in the space of $|N_a-1,N_b+1\rangle$. A wave
function of an exciton state has a weight mainly in a region of $|k|<K_{F,b}$. } 
    \label{band}
\end{figure}

To reveal the coexistence of the exciton and the antiparticle of the exciton as two distinguishable collective modes in semimetals, we solve the Bethe-Salpeter (BS) equation for a two-band model 
with screened Coulomb interaction in two and three dimensions~(Fig.~\ref{band}). 
An inter-band two-particle Green's function obtained from the BS 
equation has a pair of {\it two} poles for the $1s$ exciton levels 
($1s$ hydrogen-atom-type orbitals for the relative coordinates between an
electron and a hole). 
The inverse of the two-particle Green's function plays a role of the Lagrangian of 
free excitons~\cite{martin_reining_ceperley_2016}. The Lagrangian takes 
a form of a generalized Klein-Gordon field theory without the Lorentz symmetry. By quantizing the effective 
field theory~\cite{srednicki_2007}, we show that the pair of the poles 
can be interpreted as an exciton and an antiparticle of the exciton (\textit{antiexciton}), respectively. 
A calculation of conserved charge of the effective field theory 
shows that the exciton and antiexciton carry opposite charges, 
where the conserved charge is conjugate to a joint U(1) phase of the two bands. The opposite charge enables 
pair annihilation of an exciton and an antiexciton 
which produces a pair of density waves in conduction 
and valence bands. When the two bands have opposite 
spin polarization, e.g. one band with up-spin polarization $(s_z=1/2)$ 
and the other band with down-spin polarization ($s_z=-1/2$), 
the exciton and antiexciton carries $S_z=+1$ and $S_z=-1$. 
The opposite spin polarization can be utilized for 
distinguishing antiexcitons from excitons experimentally in the spin-polarized case.

The organization of this paper is as follows. In the next section, we introduce model Hamiltonians for semimetals studied in this paper. In Sec. \ref{sec3}, we introduce the concept of the antiexciton in semimetal in terms of an inter-band two-particle Green's function. In Sec. \ref{sec4}, we explain how we calculate the Green's function in a limit of dilute carrier densities. In Sec. \ref{sec5}, we demonstrate universal coexistence of exciton and antiexciton states based on calculations of energy-momentum dispersions of the inter-band collective modes. In Sec. \ref{sec6}, we show that the $CP$-violated Klein-Gordon theory describes a pair of an exciton and an antiexciton. In Sec. \ref{sec7}, we discuss possible optical spectroscopy experiment for detecting the antiexciton states in semimetals. A brief summary and relevant experimental materials are listed 
in Sec. \ref{sec8}. In Appendix \ref{appendix:a}, we provide technical details for calculations of the energy-momentum dispersion of exciton and antiexciton states in the semimetals. In Appendix \ref{appendix:b}, we clarify the physical difference between our concept of the antiexciton and those in the literature~\cite{PhysRevB.44.1368,PhysRevB.44.3638,arseev1998,lerner1981two,PhysRevB.59.5627}.    

\section{\label{sec2}Models of semimetals}
We study a two-band semimetal Hamiltonian with a valence band maximum at $\bm{k}=0$ and a conduction band minimum at $\bm{k}=\bm{k_c}$ (Figs.~\ref{band} (a),(b)
)~\cite{PhysRev.158.462,CLOIZEAUX1965259}. The kinetic energy part of the electronic Hamiltonian is given by
\begin{align}
    \hat{K_0} = \sum_{\bm{k}} \Big[\Big(\epsilon_a(\bm{k}) - \mu \Big)a^{\dagger}_{\bm{k}}a_{\bm{k}} + \Big(\epsilon_b(\bm{k})-\mu\Big)b^{\dagger}_{\bm{k}}b_{\bm{k}}\Big] \label{band1}
\end{align}
with 
\begin{align}
\epsilon_{a}\left(\bm{k}\right)= \frac{k^2}{2m_a} +\frac{E_g}{2} , \   \
\epsilon_{b}\left(\bm{k}\right)= -\frac{k^2}{2m_b} - \frac{E_g}{2}. \label{band2}
\end{align}
Here $m_a$ and $m_b$ are effective masses of $a$ and $b$ bands, respectively. 
The reduced Planck constant $\hbar$ is set to 1.
$E_g$ is an energy difference between the $a$-band energy minimum and the $b$-band energy maximum. $E_g$ is negative for the semimetal case. We define $a^{\dagger}_{\bm{k}}$ and $a_{\bm{k}}$ as creation and annihilation operators for electrons of the conduction band with wave vector $\bm{k}+\bm{k_c}$, while $b^{\dagger}_{\bm{k}}$ and $b_{\bm{k}}$ for electrons of the valence band with wave vector $\bm{k}$. 
A charge neutrality (the electron density equals  
the hole density) can be realized by a chemical potential 
of $\mu_0 = \frac{E_g(m_a - m_b)}{2(m_a + m_b)}$. To study the inter-band bound states 
with their crystal momenta ${\bm Q}$ around $\bm{k_c}$, we put the chemical potential 
away from the charge neutrality point, $\mu \ne \mu_0$, where inter-band 
electron-hole individual excitations have a finite gap at $\bm{Q}=\bm{k_c}$ 
(see Fig.~\ref{band} with ${\bm Q}={\bm q}+{\bm k}_c$). The radii of the Fermi 
surfaces of $a$ and $b$ bands are denoted as $K_{F,a}$ and 
$K_{F,b}$ ($K_{F,a}\ne K_{F,b}$). 
For clarity of presentation, 
we put $\bm{k_c}=\bm{0}$ (${\bm Q}={\bm q}$), and choose $\mu>\mu_0$ ($K_{F,a}>K_{F,b}$) henceforth. The following argument can be 
directly applied to the case of $\bm{k_c} \ne \bm{0}$~\cite{PhysRev.158.462}.

Electrons in the two bands interact through the long-range 
Coulomb interaction. The interaction takes the following form in the momentum 
representation, 
\begin{align}
    \hat{V} = \frac{1}{2 \Omega}\sum_{\bm{q}} v(\bm{q})\hat{\rho}(\bm{q})\hat{\rho}(-\bm{q}),  \label{Hint}
\end{align}
with a total volume of the system $\Omega$. Here $\hat\rho({\bm q})$ stands for
the density operator with momentum ${\bm q}$~\cite{PhysRev.158.462,CLOIZEAUX1965259},
\begin{align}
    \hat{\rho}(\bm{q})=\sum_{\bm{k}} \Big( 
    a_{\bm{k}+\bm{q}}^{\dagger} a_{\bm{k}}+b_{\bm{k}+\bm{q}}^{\dagger} b_{\bm{k}} \Big), 
\end{align}
and $v(\bm{q})$ is the Fourier transform of the bare Coulomb 
potential,
\begin{equation}
v({\bm q}) = 
    \begin{cases}
    \frac{4\pi}{q^2},  & \text{in three dimensions (3D),} \\
    \frac{2\pi}{q}, & \text{in two dimensions (2D).}
\end{cases}
\end{equation} 
The elementary charge $e$ and the Coulomb constant $(4\pi \epsilon_0)^{-1}$  are set to 1. 
Since the Hamiltonian, $\hat{K}_0+\hat{V}$, has no 
single-particle inter-band hopping terms, the interacting 
model has a U(1) $\times$ U(1) symmetry. 

\section{\label{sec3}exciton and antiexciton in semimetals}

Suppose that a many-body ground state $|0\rangle$ of the interacting 
electron system does not break the U(1) $\times$ U(1) symmetry. Therefore, it is in the eigenspace of total particle-number 
operators of $a$- and $b$-band electrons, 
$|N_a, N_b\rangle$, where $N_a$ and $N_b$ denote the 
electron numbers in $a$ and $b$ bands, respectively. 
Then, the inter-band excited 
eigenstates in the semimetals can be either in 
$|N_a + 1, N_b - 1 \rangle$ or in $|N_a-1,N_b+1\rangle$.   
Excitons and antiexcitons are nothing but 
bound states living in the former and later eigenspace, respectively. 
These excitations can be characterized by 
a time-ordered inter-band two-particles Green's function 
in the zero-temperature ($T=0$) field theory: 
\begin{align}
&G^{ex}(\bm{x} - \bm{x^{\prime}}, t-t^{\prime})_{\bm{y} \bm{y^{\prime}}} \nonumber \\
& =  -(-i)^2\langle 0 | \mathcal{T} \{  a_{\bm{x}}(t) b^{\dagger}_{\bm{x}+\bm{y}}(t) b_{\bm{x^{\prime}} + \bm{y^{\prime}}}(t^{\prime})a^{\dagger}_{\bm{x^{\prime}}}(t^{
\prime}) \}|0 \rangle.  \label{two-particle-Green}
\end{align}
Here the many-body ground state $|0\rangle$ is in the $|N_a,N_b\rangle$ Hilbert space.  ${\cal T}$ denotes the time-ordered product~\cite{fetter2012quantum}. $a_{\bm x}$ and $b_{\bm x}$ are annihilation 
operators in the $a$ and $b$ bands. $a_{\bm k}$ and $b_{\bm k}$ in Eq.~(\ref{band1}) are Fourier transforms of $a_{\bm x}$ and 
$b_{\bm x}$, 
\begin{align}
    a_{\bm x} = \frac{1}{\sqrt{\Omega}} \sum_{\bm k} e^{i{\bm k}{\bm x}} a_{\bm k}, 
    \!\ \!\ b_{\bm x} = \frac{1}{\sqrt{\Omega}} \sum_{\bm k} e^{i{\bm k}{\bm x}} b_{\bm k},
\end{align}
with a total volume of the system $\Omega$. 
${\bm y}$ and ${\bm y}^{\prime}$ in Eq.~(\ref{two-particle-Green}) 
are relative distances between the particle and hole that form a bound state. 
$\bm{x}+\frac{m_b}{m_a+m_b}\bm{y}$, $\bm{x^{\prime}}+\frac{m_b}{m_a+m_b}\bm{y^{\prime}}$ can be regarded as the center-of-mass coordinates of the particle and hole.  
The Fourier-transform of the two-particles Green's function is defined by 
\begin{align}
&G^{ex}\left(\boldsymbol{q}, \omega\right)_{ \boldsymbol{k} \boldsymbol{k}^{\prime}} =  \int d(t-t^{\prime}) \int 
d^d(\bm{x-x^{\prime}}) \int d^d \bm{y} \int d^d \bm{y^{\prime}} 
\nonumber \\
& e^{ i\omega (t-t^{\prime}) - i\bm{q}\cdot (\bm{x-x^{\prime}}) + i\bm{k}\cdot\bm{y} - i\bm{k^{\prime}}\cdot\bm{y^{\prime}}} G^{ex}(\bm{x} - \bm{x^{\prime}}, t-t^{\prime})_{\bm{y} \bm{y^{\prime}}}. \label{ft} 
\end{align}

In the semimetals, a spectral representation of the zero-temperature time-ordered Green's function for the inter-band excitations 
can be decomposed not only by the excited eigenstates 
in $|N_a + 1, N_b - 1 \rangle$ but also by those 
in $|N_a-1,N_b+1\rangle$ as
\begin{align}
    G^{ex}(\bm{q},\omega)_{\bm{kk^{\prime}}} = & \sum_n \frac{i \langle 0 | b_{\bm{k}}^{\dagger}a_{\bm{q}+\bm{k}} |n\rangle \langle n| a^{\dagger}_{\bm{q}+\bm{k^{\prime}}} b_{\bm{k^{\prime}}} |0\rangle }
    { \omega - (E_n - E_0) + i0^+ } \nonumber\\
    - & \sum_{n^{\prime}}  \frac{i \langle 0 | a^{\dagger}_{\bm{q}+\bm{k^{\prime}}} b_{\bm{k^{\prime}}} |n^{\prime}\rangle \langle n^{\prime}|b_{\bm{k}}^{\dagger}a_{\bm{q}+\bm{k}}  |0\rangle }
    { \omega + (E_{n^{\prime}} - E_0) - i0^+ }.\label{eqn_Lehmann}
\end{align}
Here $E_0$ is a ground-state energy in the $|N_a,N_b\rangle$ Hilbert space. $\bm{q}$ and $\omega$ correspond to the total 
momentum and frequency of the inter-band bound states, and $\bm{k},\bm{k^{\prime}}$ are relative momenta. 
$|n\rangle$ and $|n^{\prime}\rangle$ are the 
excited eigenstates with the momentum ${\bm q}$ and $-{\bm q}$ and with the energy $E_{n}$ and $E_{n^{\prime}}$ in the eigenspaces of 
$|N_a + 1, N_b - 1 \rangle$ and $|N_a-1,N_b+1\rangle$,  
respectively. Excitons and antiexcitons are  
bound states comprised of inter-band excitations in 
$|N_a + 1, N_b - 1 \rangle$ and $|N_a-1,N_b+1\rangle$, respectively. In the spectral representation, they can be detected as poles in the fourth 
and second quadrant in the complex-$\omega$ plane, respectively.

In a semiconductor with $E_g>0$, $N_a=0$ and $N_b=N$, the inter-band excited eigenstates are only in the space of $|1, N - 1 \rangle$, where there is no antiparticle-type exciton. Nevertheless, Lerner and Lozovik previously studied inter-band collective modes in two-dimensional (2D) electron-hole gas (EHG) 
under magnetic field. They described
the collective modes in two different field regimes as excitons and 
antiexcitons, respectively~\cite{lerner1981two}. The 2D EHG under the 
field can approximately realize its ground states either in the space 
of $|0,N\rangle$ (positive-band-gap semiconductor regime) or in the space of 
$|N,0\rangle$ (negative-band-gap semiconductor regime). 
Interband collective modes in these two semiconductor 
regimes live in $|1,N-1\rangle$ and in $|N-1,1\rangle$, 
which can be also regarded as excitons and antiexcitons, 
respectively~\cite{lerner1981two} (see Appendix \ref{appendix:b}). 


\section{\label{sec4}Semimetals with dilute carrier densities}
\begin{figure}[t]
    \centering
    \includegraphics[height=1.6cm]{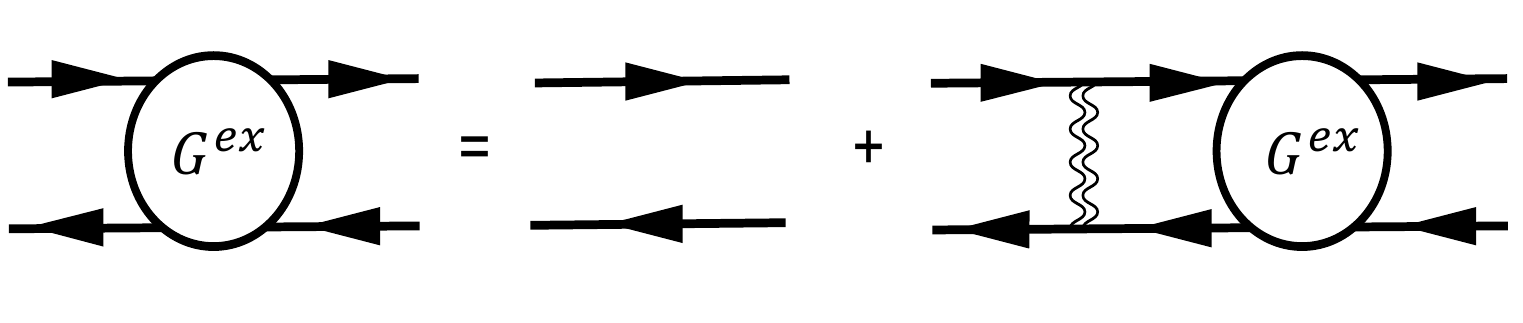}%
    \caption{The Feynman diagram of the screened-ladder approximation for $G^{ex}$. The upper and lower solid lines with rightward and leftward arrows are electron propagators in the $a$ band and $b$ band, respectively. The double-wavy line is the screened Coulomb interaction $w({\bm y})$ or $w({\bm k}-{\bm k}^{\prime})$ represented by the Feynman diagram Fig.~\ref{RPA}.}
    \label{feyn}
\end{figure}

In the dilute limit of the carrier densities, the two-particle 
Green's function can be evaluated in terms of the ladder 
approximation represented by the Feynman diagram 
Fig.~\ref{feyn} ~\cite{fetter2012quantum},
\begin{align}
& G^{ex}(\bm{x-x^{\prime}}, t-t^{\prime})_{\bm{y}\bm{y^{\prime}}} =  G^{ex}_0(\bm{x-x^{\prime}}, t-t^{\prime})_{\bm{y}\bm{y^{\prime}}} \nonumber \\
& + i\int d^d \overline{\bm{x}} \int d^d \overline{\bm{y}}\int d \overline{t}\  \ G^{ex}_0(\bm{x-\overline{\bm{x}}}, t-\overline{t})_{\bm{y}\overline{\bm{y}}}\nonumber \\
&  \hspace{2cm} \times w(\overline{\bm{y}}) G^{ex}(\overline{\bm{x}}-\bm{x^{\prime}}, \overline{t}-t^{\prime})_{\overline{\bm{y}}\bm{y^{\prime}}}, \label{bs1}
\end{align}
where $G^{ex}_0(\bm{x-x^{\prime}}, t-t^{\prime})_{\bm{y}\bm{y^{\prime}}}$ denotes 
the two-particle Green's function in a free theory ($V=0$), 
\begin{align}
&G^{ex}_0(\bm{x-x^{\prime}}, t-t^{\prime})_{\bm{y}\bm{y^{\prime}}} 
= G^a_0(\bm{x-x^{\prime}},t-t^{\prime}) \nonumber \\
& \hspace{1.8cm} \times G^b_0(\bm{x^{\prime}-x} + \bm{y^{\prime}-y}, t^{\prime}-t). 
\end{align}
Here $G_0^a$ and $G_0^b$ are single-particle Green's functions of 
the $a$ and $b$ bands in the free theory, respectively,  
\begin{align}
     & iG_0^{a}(\bm{x-x^{\prime}},t-t^{\prime}) = \langle 0 |\mathcal{T}\{a_{\bm{x}}(t) a_{\bm{x^{\prime}}}^{\dagger}( t^{\prime})\} |0 \rangle_{|V=0} \nonumber \\
     & = \Omega^{-1} 
     \sum_{\bm k} e^{i{\bm k}({\bm x}-{\bm x}^{\prime})-i \epsilon_a({\bm k})(t-t^{\prime})}\nonumber \\ 
&\big[\theta(t-t^{\prime})\theta(|{\bm k}|-K_{F,a})
 - \theta(t^{\prime}-t) \theta(K_{F,a}-|{\bm k}|)\big], \\
    & iG_0^{b}(\bm{x-x^{\prime}},t-t^{\prime}) = \langle 0 |\mathcal{T}\{b_{\bm{x}}(t) b_{\bm{x^{\prime}}}^{\dagger}( t^{\prime})\} 
    | 0\rangle_{|V=0}.\nonumber \\
    & = \Omega^{-1}  
     \sum_{\bm k} e^{i{\bm k}({\bm x}-{\bm x}^{\prime})-i \epsilon_b({\bm k})(t-t^{\prime})} \nonumber \\
 & \big[\theta(t-t^{\prime})\theta(K_{F,b}-|{\bm k}|) 
- \theta(t^{\prime}-t) \theta(|{\bm k}|-K_{F,b})\big],  
\end{align}
with Fermi momenta $K_{F,a}$, $K_{F,b}$ of the conduction-band and 
valence-band Fermi surface (Fig.~\ref{band}(b)). 
Their Fourier transforms are as follows,
\begin{align}
\widetilde{G}^{ex}_0\left(\boldsymbol{q}, \omega\right)_{ \boldsymbol{k} \boldsymbol{k}^{\prime}} 
&= -i \delta_{{\bm k}{\bm k}^{\prime}} \int \frac{d\omega_1}{2\pi} 
G^a_{0}({\bm k}+{\bm q},\omega_1+\omega) G^b_0({\bm k},\omega_1) \nonumber \\
&=  \delta_{{\bm k}{\bm k}^{\prime}}  \bigg\{\frac{\theta(|{\bm k}+{\bm q}|-K_{F,a})
\theta(|{\bm k}|-K_{F,b})}{\omega - [\epsilon_{a}({\bm k}+{\bm q})-\epsilon_{b}({\bm k})]+i0^{+}} \nonumber \\ 
& - \frac{\theta(K_{F,a}-|{\bm k}+{\bm q}|)
\theta(K_{F,b}-|{\bm k}|)}{\omega - [\epsilon_{a}({\bm k}+{\bm q})-\epsilon_{b}({\bm k})]-i0^{+}} \bigg\}. \label{free-2pG} 
\end{align}
with  
\begin{align}
    G^{a}_{0}({\bm k},\omega) &= \frac{\theta(|{\bm k}|-K_{F,a})}{\omega - 
    \epsilon_a({\bm k})+i0^+} + \frac{\theta(K_{F,a}-|{\bm k}|)}{\omega - 
    \epsilon_a({\bm k})- i0^+}, \nonumber \\ 
    G^{b}_{0}({\bm k},\omega) &= \frac{\theta(K_{F,b}-|{\bm k}|)}{\omega - 
    \epsilon_b({\bm k})+i0^+} + \frac{\theta(|\bm k|-K_{F,b})}{\omega - 
    \epsilon_b({\bm k})- i0^+}. 
\end{align}

\begin{figure}[t]
    \centering
    \includegraphics[height=1.6cm]{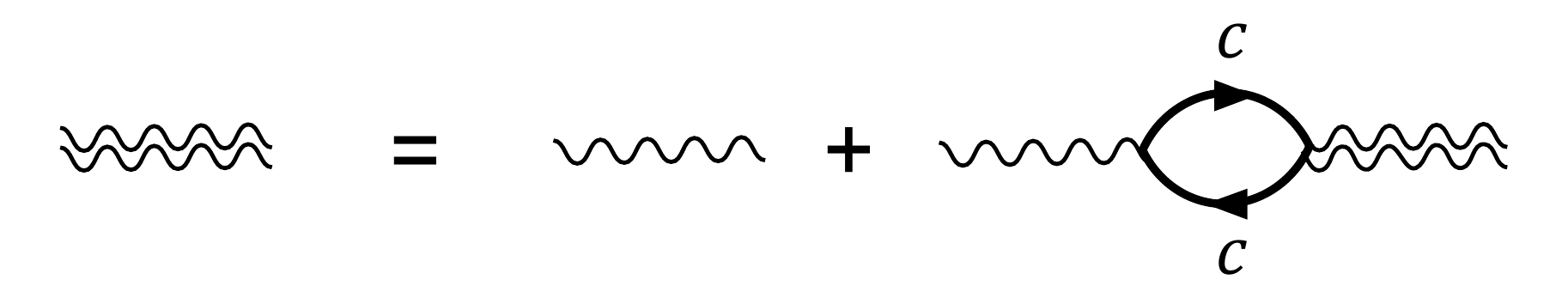}%
    \caption{Random phase approximation (RPA) for screened Coulomb interaction. The single- and double-wavy lines are bare (long-ranged) and screened Coulomb interactions. Solid lines of bubbles correspond to free single-particle Green's functions. A bubble with label $c$ contains a summation of bubbles consisting of free single-particle Green's functions of the $a$ and $b$ band ($c=a,b$).}
    \label{RPA}
\end{figure}

$w({\bm y})$ in Eq.~(\ref{bs1}) stands for an effective interaction between the
electron and hole. In a semiconductor regime ($E_g>0$), it is the long-ranged Coulomb interaction. In a semimetal regime ($E_g<0$), the Coulomb interaction is screened by carrier densities. The screened Coulomb interaction can be evaluated by the random phase approximation (Fig.~\ref{RPA}). 
In the approximation, the Fourier transform of $w({\bm y})$ is given by a static limit of the bare polarization function $\Pi_0(\bm{q},\omega)$, 
\begin{equation}
    w(\bm{q}) = \frac{v(\bm{q})}{1-v(\bm{q})\Pi_0({\bm 0},0)} = 
    \begin{cases}
    \frac{4\pi}{q^2 - 4\pi\Pi_0(\bm{0},0)},            & \text{in 3D,} \\
    \frac{2\pi}{q - 2\pi\Pi_0(\bm{0},0)},              & \text{in 2D.}
\end{cases}
\end{equation}
The bare polarization function in our two-band model is given by 
\begin{align}
    \Pi_0(\boldsymbol{q}, \omega)&= \sum_{c=a,b}\Pi_0^c(\boldsymbol{q}, \omega),  \nonumber \\
    \Pi_0^c(\boldsymbol{q}, \omega) &=  -i
    \int  \int \frac{d\omega_1 d^{d} \boldsymbol{k}}{(2 \pi)^{d+1}} G_{0}^{c}\left(\boldsymbol{k}+\boldsymbol{q}, \omega_{1}+\omega\right) G_{0}^{c}\left(\boldsymbol{k}, \omega_{1}\right).
    \label{bare_polar}
\end{align}
The static limit of the polarization function is given by 
\begin{align}
    \Pi_0(\bm{0},0) \equiv 
\begin{cases}
    -\frac{1}{4\pi}\big(\frac{2K_{F,a}m_a}{\pi} + \frac{2K_{F,b}m_b}{\pi}\big),            & \text{in 3D,} \\
    -\frac{1}{2\pi} \big( m_a + m_b \big).              & \text{in 2D.}
\end{cases}
\end{align}
The screened Coulomb potential is given by the Thomas-Fermi (TF) wavelength $k_{\rm TF}$ as $w(\bm{k}) = 4\pi/(k^2 + k^2_{\rm TF})$ in 3D and $w(\bm{k}) = 2\pi/(k + k_{\rm TF})$ in 2D. 
The TF wavevector is calculated as 
$k_{\rm TF}^2 = 2 (K_{F, a} m_{a}+ K_{F, b} m_{b})/\pi$ in 3D and $k_{\rm TF}=m_{a}+m_{b}$ in 2D. 

In the momentum-frequency space, Eq.~(\ref{bs1}) takes a form of  
\begin{align}
   & \widetilde{G}^{ex}\left(\boldsymbol{q}, \omega\right)_{ \boldsymbol{k} \boldsymbol{k}^{\prime}} = \widetilde{G}^{ex}_0 \left(\boldsymbol{q}, \omega\right)_{ \boldsymbol{k} \boldsymbol{k}^{\prime}} \nonumber \\
& - \frac{1}{\Omega} \sum_{\bm{k_1}\bm{k_2}} \widetilde{G}^{ex}_0 \left(\boldsymbol{q}, \omega\right)_{ \boldsymbol{k} \boldsymbol{k_1}} w(\bm{k_1}-\bm{k_2}) \widetilde{G}^{ex} \left(\boldsymbol{q}, \omega\right)_{ \boldsymbol{k_2} \boldsymbol{k}^{\prime}}, \label{bs2}
\end{align}
with $i\Omega \widetilde{G}^{ex}\left(\boldsymbol{q}, \omega\right)_{ \boldsymbol{k} \boldsymbol{k}^{\prime}} \equiv G^{ex}\left(\boldsymbol{q}, \omega\right)_{ \boldsymbol{k} \boldsymbol{k}^{\prime}}$. Eq.~(\ref{bs2}) can be written into a following matrix form~\cite{martin_reining_ceperley_2016}
\begin{align}
    \widetilde{G}^{ex}\left(\boldsymbol{q}, \omega\right)^{-1} = \widetilde{G}^{ex}_0\left(\boldsymbol{q}, \omega\right)^{-1} + W, \label{bs3}
\end{align}
with $W_{{\bm k}{\bm k}^{\prime}} \equiv \Omega^{-1} w({\bm k}-{\bm k}^{\prime})$.  
Suppose that $|\phi_j({\bm q},\omega)\rangle$ and $\xi_j({\bm q},\omega)$ are 
eigenvectors and eigenvalues of Eq.~(\ref{bs3}),
\begin{align}
    \widetilde{G}^{ex}({\bm q},\omega)^{-1} |\phi_j({\bm q},\omega)\rangle 
    = \xi_{j}({\bm q},\omega)|\phi_j({\bm q},\omega)\rangle. \label{bs4}
\end{align}
Then, the inter-band two-particle Green's function is 
given by the eigenvectors and eigenvalues, 
\begin{align}
    \widetilde{G}^{ex}({\bm q},\omega) = \sum_j 
    |\phi_j({\bm q},\omega)\rangle \xi_j({\bm q},\omega)^{-1} \langle 
    \phi_j({\bm q},\omega)|. \label{solution}
\end{align}

 In the next section, we solve Eq.~(\ref{bs4}) at ${\bm q}=0$, where we can 
employ the spherical and circular symmetry in 3D and 2D, respectively, 
and use the irreducible representations of Eq.~(\ref{bs4}). 
From $\xi_j({\bm 0},\omega)$ thus calculated, we determine 
exciton and antiexciton levels at the $\Gamma$ point. 
Then we treat small ${\bm q}$ around the 
$\Gamma$ point as a perturbation and calculate band curvatures 
of the exciton and antiexciton bands, using the irreducible representations 
at ${\bm q}=0$. 

Before closing this section, let us use the Feynman-Hellman theorem and 
derive a useful relation between $\partial_{\omega} \xi_{j}$ 
and wave functions of exciton and antiexciton excitation 
at ${\bm q}=0$.  The $\omega$-derivative of
Eq.~(\ref{bs4}) at the $\Gamma$ point leads to 
\begin{align}
    \frac{d\xi_j(\bm{0},\omega)}{d\omega} & = \langle \phi_j(\bm{0},\omega) | \Big[\frac{d\widetilde{G}^{ex}\left(\boldsymbol{0}, \omega\right)^{-1}}{d\omega}\Big] |\phi_j(\bm{0},\omega)\rangle ,
    \label{FH1}
\end{align}
where $\widetilde{G}^{ex}(\bm{0},\omega)^{-1}$ and its $\omega$-derivate 
are given by  
\begin{align}
    & [\widetilde{G}^{ex}(\bm{0},\omega)^{-1}]_{\bm{k}\bm{k^{\prime}}} = \delta_{\bm{k}\bm{k^{\prime}}} \Big\{ \theta(|\bm{k}|-K_{\rm{out}})\Big( \omega- (\epsilon_a(\bm{k}) - \epsilon_b(\bm{k}))\Big) \nonumber \\
& \!\ \!\  - \theta(K_{\rm{in}}-|\bm{k}|)\Big( \omega- (\epsilon_a(\bm{k}) - \epsilon_b(\bm{k}))\Big) \Big\} + \frac{w(\bm{k}-\bm{k^{\prime}})}{ \Omega}, \nonumber \\
   & \Big[\partial_{\omega} \widetilde{G}^{ex}\left(\boldsymbol{0}, \omega\right)^{-1} \Big]_{\bm{k}\bm{k^{\prime}}} = \delta_{\bm{k}\bm{k^{\prime}}} \Big\{ \theta(|\bm{k}|-K_{\rm out}) - \theta(K_{\rm in}-|\bm{k}|) \Big\}.
    \label{FH2}
\end{align}
Here $K_{\rm{out}} \equiv  \mathrm{max}(K_{F,a},K_{F,b})=K_{F,a}$, and 
$K_{\rm{in}} \equiv  \mathrm{min}(K_{F,a},K_{F,b}) = K_{F,b}$. 
From this expression, we can relate 
the derivative with the momentum-space wavefunctions,  
\begin{align}
    \partial_{\omega} \xi_j =  \sum_{|\bm{k}|>K_{\rm out}}
    |\langle {\bm k}|\phi_j\rangle |^2  - \sum_{|\bm{k}|<K_{\rm in}}  
    |\langle {\bm k}|\phi_j\rangle |^2 , \label{FH3}
\end{align}
with $K_{\rm out} = K_{F,a}  > K_{\rm in} = K_{F,b}$. 

As shown in the next section, $\xi_j({\bm 0},\omega)$ 
as a function of $\omega$ crosses zero at both a positive $\omega$ 
($\omega=\omega_{+}$) with $\partial_{\omega} \xi_j({\bm 0},\omega)_{|\omega=\omega_+}>0$ and a negative $\omega$ 
($\omega=-\omega_{-}$) with $\partial_{\omega} \xi_j({\bm 0},\omega)_{|\omega=-\omega_-}<0$. The Lehmann representation 
dictates that the positive zero corresponds to a bound state in 
$|N_a+1,N_b-1\rangle$ (exciton) and the negative zero corresponds to 
a bound state in $|N_a-1,N_b+1\rangle$ (antiexciton). In fact, 
Eq.~(\ref{FH3}) shows that  
the eigenvectors of the positive-$\omega$ 
(negative-$\omega$) bound states has 
larger spectral weight in $|\bm{k}|>K_{\rm out}$ 
($|\bm{k}|<K_{\rm in}$), suggesting that the former and latter 
bound states are of $a^{\dagger}_{\bm k}b_{\bm k}|0\rangle$ 
type and of $b^{\dagger}_{\bm k}a_{\bm k}|0\rangle$ type, 
respectively (Fig.~\ref{band}(c,d)).

\begin{figure}[t]
    \centering
    \subfigure[ ]{\includegraphics[width= 2.7in]{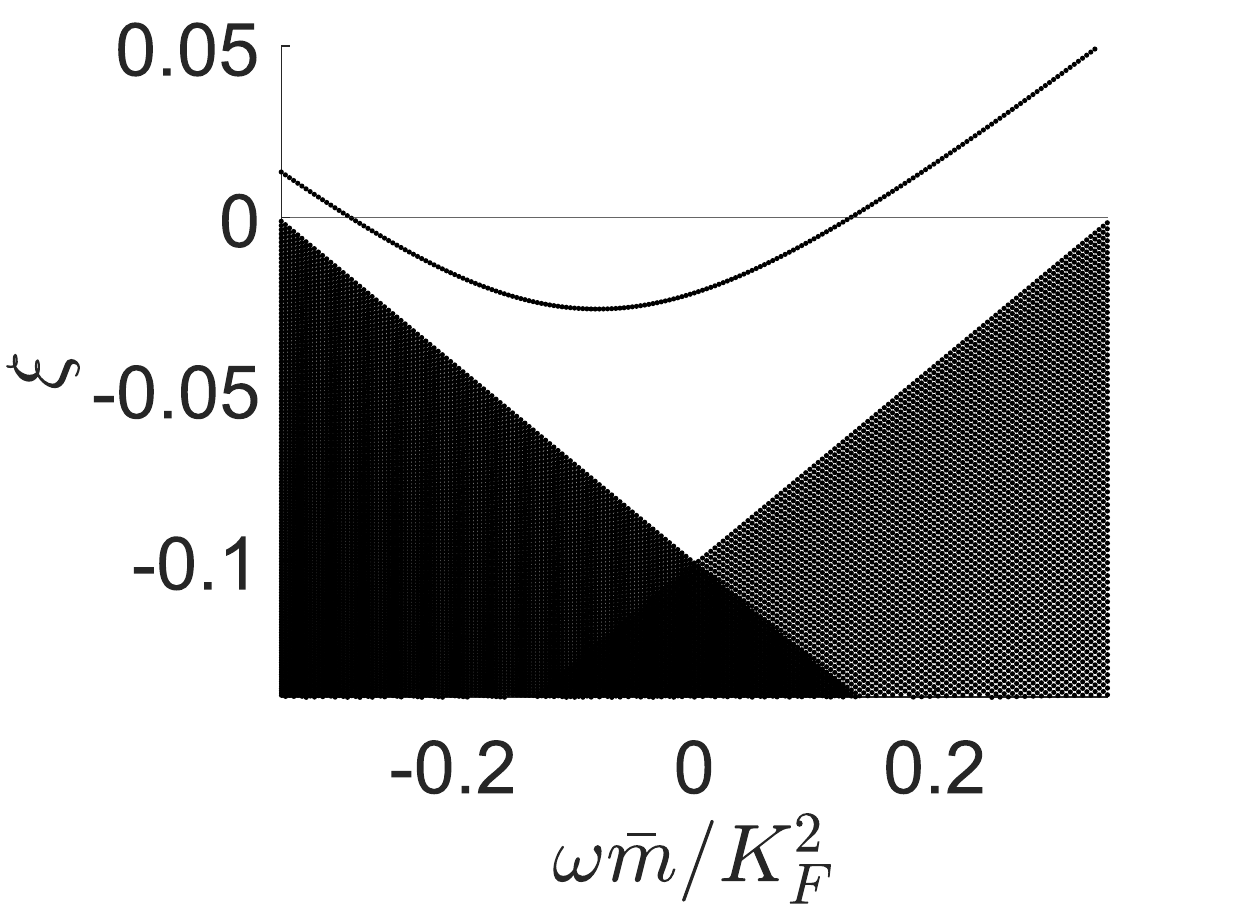}}
    
    \subfigure[ ]{\includegraphics[width= 2.7in]{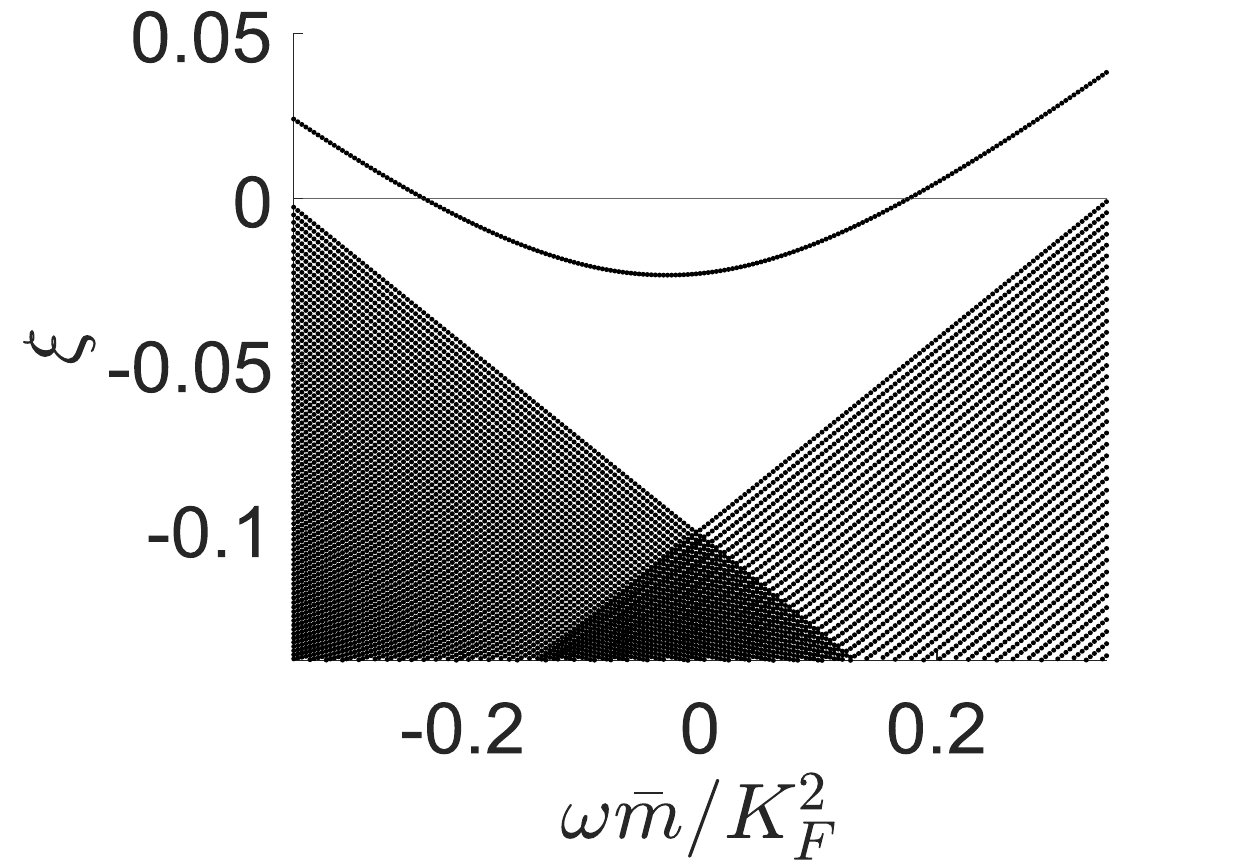}}
    \caption{$\xi_j(\bm{q=0},\omega)$ as a function of $\omega$, with $m_a = m_b \equiv \bar{m}$, $K_F \equiv (K_{F,a}+K_{F,b})/2$. 
    Locations of the zeros ($\xi_j(\bm{q=0},\omega)=0$) are determined by three dimensionless quantities, $K_{F}\alpha_B$, $\tilde{E}_g \equiv E_g \bar{m}/K^2_F$ and $\tilde{\mu} \equiv \mu \bar{m}/K^2_{F}$, where the effective Bohr radius $\alpha_B = 1/\bar{m}$. The electron rest mass $m_e$ is set to 1. (a) 3D case: eigenvalues with $l=0$, $m=0$ are plotted for $K_{F}\alpha_B=K_{F}/3.5=0.289$, $\tilde{E}_g = -0.3\bar{m}/K_{F}^2= -1.03$, and $\tilde{\mu} = 0.05\bar{m}/K_{F}^2 = 0.172$. (b) 2D case: eigenvalues with $m=0$ are plotted for $K_{F}\alpha_B=K_{F}= 0.540$, $\tilde{E}_g =-0.3\bar{m}/K_{F}^2=-1.03$, and $\tilde{\mu}=0.05\bar{m}/K_{F}^2 = 0.172$.}
    \label{data}
\end{figure}

\section{\label{sec5}Energies of inter-band excitations in semimetals}

Solving Eq.~(\ref{bs4}) in 3D and 2D  is computationally expensive for general ${\bm q}$. 
Thus, we first focus on the solutions at ${\bm q}=0$. $G^{ex}\left(\bm{0}, \omega\right)$ 
has the spatially rotational symmetry;  
\begin{align}
G^{ex}\left(\bm{0}, \omega\right)_{{\bm k}{\bm k}^{\prime}} 
= G^{ex}\left(\bm{0}, \omega\right)_{\widetilde{\bm k} \widetilde{\bm k}^{\prime}},
\end{align}  
where $\widetilde{\bm k}$ and $\widetilde{\bm k}^{\prime}$ are transformed into  
${\bm k}$ and ${\bm k}^{\prime}$ respectively by the same rotation.  
The eigenvalue problem at the $\Gamma$ point is decomposed by 
the irreducible representations of the rotational symmetry group (Appendix \ref{appendix:a}). 
The Green's function is expanded by spherical harmonics in the 3D case, 
\begin{align}
   -iG^{ex}\left(\bm{0}, \omega\right)_{\bm{k}\bm{k^{\prime}}} = \sum_{nlm} \frac{Y_{l m}(\theta, \varphi) f_{nl}(\omega;k) f_{nl}(\omega;k^{\prime}) Y_{lm}^{*}\left(\theta^{\prime}, \varphi^{\prime}\right)}{\xi_{nl}(\omega)} \label{3d-q=0}
\end{align}
and by trigonometric functions in the 2D case,  
\begin{align}
    -iG^{ex}\left(\bm{0}, \omega\right)_{\bm{k}\bm{k^{\prime}}} = \sum_{nm} \frac{f_{nm}(\omega;k) f_{nm}(\omega;k^{\prime}) e^{im(\varphi-\varphi^{\prime})}}{\xi_{nm}(\omega)}.  
\label{2d-q=0}
\end{align}
Here $\bm{k} = k(\sin \theta \cos \varphi, \sin \theta \sin \varphi, \cos \theta)$ in the 3D case and $\bm{k} = k(\cos \varphi,\sin \varphi)$ in the 2D case. $f_{nl}(\omega;k)$ and $f_{nm}(\omega;k)$ stand for radial wavefunctions and $Y_{lm}(\theta,\varphi)$ are the spherical harmonics. In the 3D and 2D case, $\langle{\bm k}|\phi_j(\bm{0},\omega)\rangle =\frac{1}{\sqrt{\Omega}} f_{nl}(\omega;k)Y_{lm}(\theta,\varphi)$ and $\langle{\bm k}|\phi_j(\bm{0},\omega)\rangle = \frac{1}{\sqrt{\Omega}} f_{nm}(\omega;k)e^{im\varphi}$, respectively. Here $j$ is the combination of principal 
quantum number $n$, azimuthal quantum number $l$ in the 3D case, and magnetic quantum number $m$ in the 2D case.

Fig.~\ref{data} plots $\xi_j(\bm{0},\omega)$ as a function of 
$\omega$. Individual inter-band excitations with ${\bm q}=0$ 
form continuum spectra in  certain ranges of $\omega$, which correspond 
to the shaded regions in the figure. 
Apart from the continuum spectra, a branch of $s$-wave ($l=0$) bound states in 3D 
and a branch of $m=0$ bound states in 2D form a parabolic curve of $\omega$ outside the shaded region. 
The branch crosses zero at both a positive $\omega$ ($\omega=\omega_+$) 
and a negative $\omega$ ($\omega=-\omega_-$).  The spectral representation dictates that the positive zero corresponds 
to a bound state in $|N_a+1,N_b-1\rangle$ (exciton) and 
the negative zero corresponds to a bound state in 
$|N_a-1,N_b+1\rangle$ (antiexciton). 

\begin{figure}[t]
\centering
\subfigure[ ]{\includegraphics[width= 3.5in]{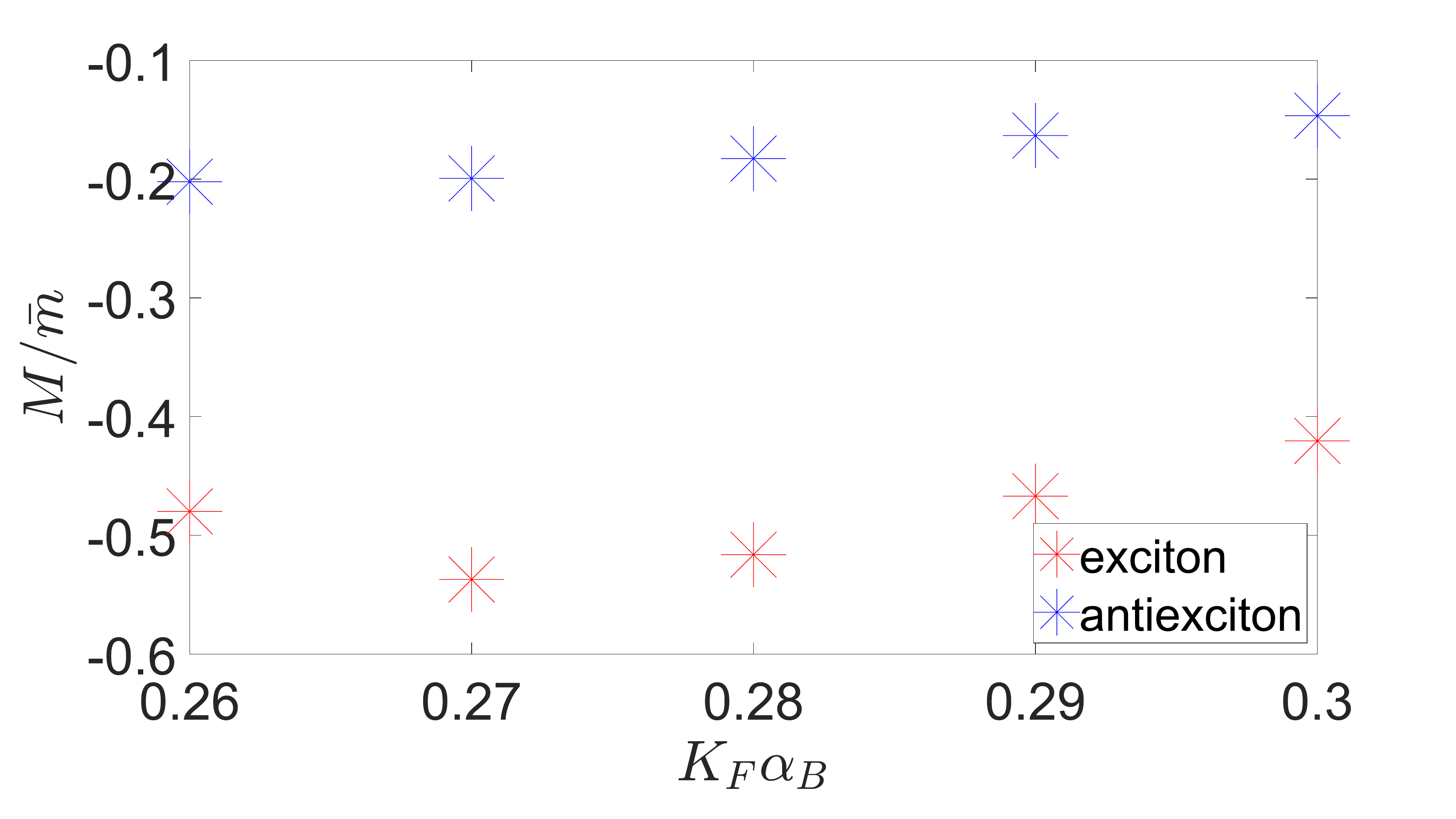}}

\subfigure[ ]{\includegraphics[width= 3.5in]{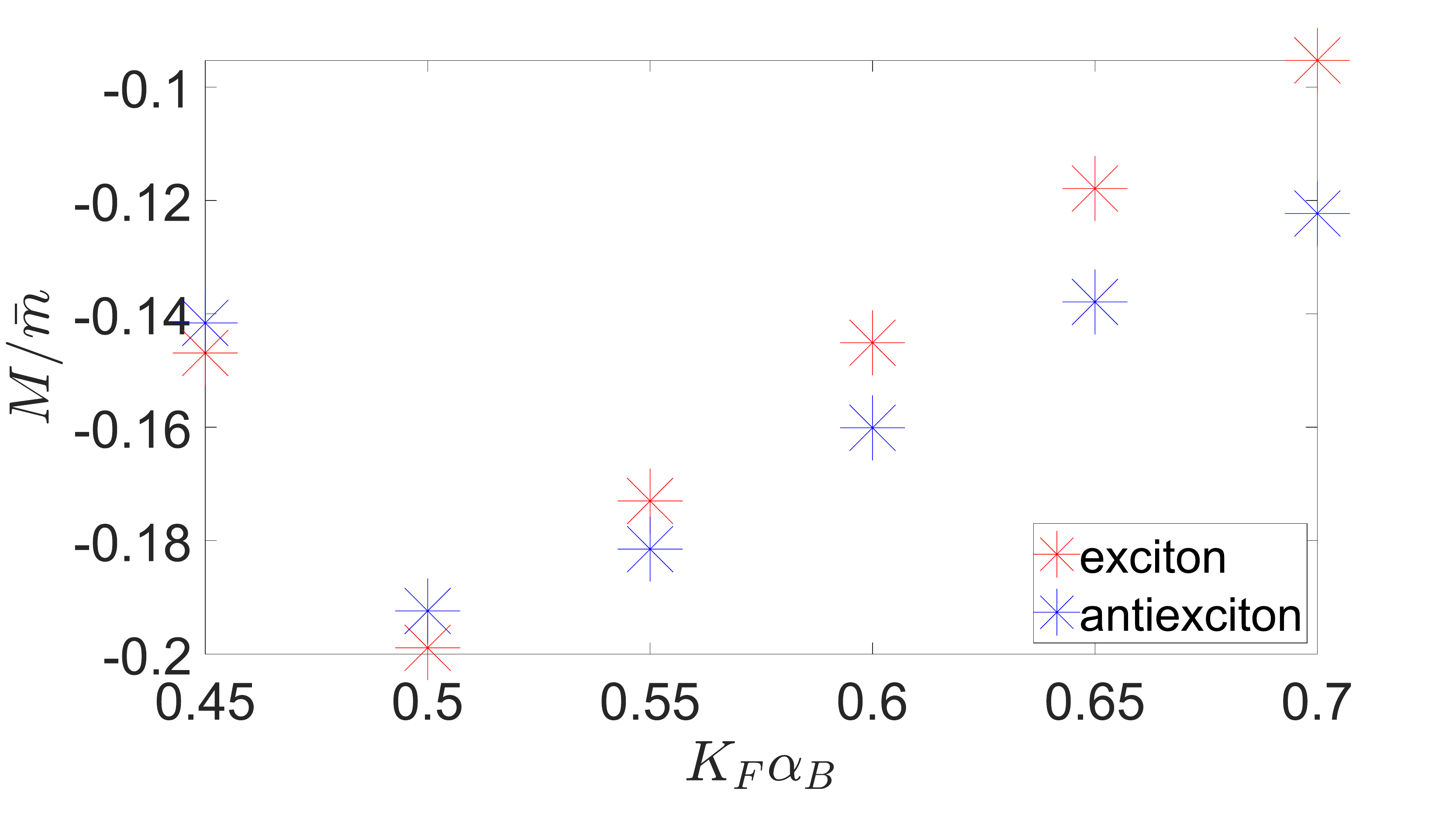}}
\caption{Ratios between curvatures of the $s$-wave exciton and antiexciton bands and the mass of the electron and hole bands $\bar{m}\equiv m_a=m_b$. The ratios are plotted as a function of a dimensionless quantity $K_{F}\alpha_B$, where $K_{F} \equiv (K_{F,a} + K_{F,b})/2$ and $\alpha_{B} \equiv 1/\bar{m}$ (the effective Bohr radius of the electron system).
The exciton's and antiexciton's band curvatures are defined in the expansion $\omega_{\pm}(\bm{q})$ around the $\Gamma$ point, $\omega_{\pm}({\bm q}) \equiv \omega_{\pm} + q^2/(2M)$. 
(a) The 3D case. (b) The 2D case. In these two plots, we choose $\tilde{E_g}  = -1.03$ and $\widetilde{\mu} =  0.172$.}
\label{curv}
\end{figure}
Energies of excitons and antiexcitons at finite $\bm{q}$ 
can be expanded as $\omega_{\pm}(\bm{q}) = \omega_{\pm} + q^2/(2M) $ to the lowest order in $q$. The band curvature 
$1/(2M)$ can be calculated from the irreducible representations of Eq.~(\ref{bs3}) (see Appendix \ref{appendix:a}). 
In parameter ranges studied in this paper, 
the band curvatures always take negative values both for the exciton and antiexciton bands (Fig.~\ref{dispersion} and Fig.~\ref{curv}). When energies of these inter-band bound states touch 
zero at finite ${\bm q}$ outside the energy-momentum 
region of inter-band individual excitations, the system can 
undergo Bose-Einstein condensation. When the exciton 
or antiexciton condensates, the two bound-state modes 
become a Goldstone mode and a Higgs mode~\cite{annurev-conmatphys-031214-014350,PhysRevLett.47.811,PhysRevB.26.4883,varma2002higgs}.

\section{\label{sec6}Effective field interpretation}
$\xi_j(\bm{0},\omega)$ is an eigenvalue of the 
inverse of the inter-band two-particle Green's function $\widetilde{G}^{ex}\left(\boldsymbol{0}, \omega\right)^{-1}$. 
To the quadratic level in $\omega$, it can be regarded as an effective  
Lagrangian for inter-band collective modes at ${\bm q}=0$. Unlike in the semiconductor case, the $\omega$-dependent part of the $\widetilde{G}^{ex}\left(\boldsymbol{0}, \omega\right)^{-1}$ matrix in the semimetal case is not simply proportional to an identity matrix. Namely, 
both of the two terms in the right hand side of Eq.~(\ref{eqn_Lehmann}) do not vanish in the semimetal case. Thus, $\xi_j(\bm{0},\omega)$ becomes 
a nonlinear function of $\omega$. The simplest effective 
Lagrangian for the inter-band collective excitations in semimetals contains an $\omega^2$ term 
in addition to an $\omega$ term, 
\begin{align}
\xi_j ({\bm 0},\omega)=\gamma \omega^2 + \alpha \omega -\beta 
\end{align}
with $\gamma>0$ and $\beta>0$ (see Fig.~\ref{data}). 
Thus, the effective Lagrangian contains a second-order 
time derivative of a complex scalar field 
$\varphi(t)$ for the ${\bm q}=0$ inter-band collective modes,   
\begin{align}
    \int dt \!\ \mathcal{L}&=
    \int dt \!\ 
    \varphi^{\dagger}(t) \big( -\gamma\partial_t^2 + i\alpha \partial_t - \beta \big) \varphi(t),   
    \label{Lagrangian}
\end{align}
with the field defined by 
\begin{align}
\varphi(t)\equiv\sum_{\bm{k}}\int^{+\infty}_{-\infty} dt' 
\langle {\bm k}|\phi_j(\bm{0},t-t')\rangle b^{\dagger}_{\bm k}(t')a_{\bm k}(t'). 
\end{align}
Here we omit $j$-indices in $\varphi(t)$ and in 
the effective Lagrangian. The complex field is 
decomposed into two real fields, $\varphi_1$ and $\varphi_2$, as 
$\varphi \equiv \varphi_1 + i\varphi_2$. 
Two conjugate momenta are introduced as 
$ \pi_{1} \equiv \frac{\partial \mathcal{L}}{\partial\left(\partial_{t} \varphi_{1}\right)} $ and 
$\pi_{2} \equiv \frac{\partial \mathcal{L}}{\partial\left(\partial_{t} \varphi_{2}\right)}$. This leads to an effective Hamiltonian for the ${\bm q}=0$ inter-band collective modes in semimetals as:
\begin{align}
\mathcal{H} &= \pi_{1} \partial_{t} \varphi_{1}+\pi_{2} \partial_{t} \varphi_{2}-\mathcal{L} \nonumber\\
            &= \frac{1}{2 \lambda}\left(\pi_{1}^{2}+\pi_{2}^{2}\right)+\frac{1}{2} \lambda \eta^{2}\left(\varphi_{1}^{2}+\varphi_{2}^{2}\right)+\frac{\alpha}{2\gamma}\left(\pi_{2} \varphi_{1}-\pi_{1} \varphi_{2}\right).
\end{align}
Here $\lambda = 2 \gamma$, and $\eta = \sqrt{\frac{\alpha^{2}}{4 \gamma^2}+\frac{\beta}{\gamma}}$. The Hamiltonian takes a form of two coupled harmonic oscillators and is bosonized by two boson fields:
\begin{align}
    \mathcal{H} = \nu_{+} a_{+}^{\dagger} a_{+}+\nu_{-} a_{-}^{\dagger} a_{-},
\end{align}
with
\begin{align}
    a_{1,2} \equiv \sqrt{\frac{\lambda \eta}{2}}\left(\varphi_{1,2}+\frac{i}{\lambda \eta} \pi_{1,2}\right), \quad a_{\pm} \equiv \frac{1}{\sqrt{2}}\left(a_{1}\pm i a_{2}\right), \label{a-def}
\end{align}
and 
\begin{align}
\nu_{\pm} =\sqrt{\frac{\alpha^{2}}{4 \gamma^2}+\frac{\beta}{\gamma}}\mp\frac{\alpha}{2 \gamma}.
\end{align}
Note that within the quadratic expansion of $\xi_j({\bm 0},\omega)$ in $\omega$, 
the two zeros of $\xi_j(\bm{0},\omega)$ correspond to the 
quantized energies of the two bosons, 
$\nu_{\pm} = \omega_{\pm} (>0)$. This concludes that 
$a_{+}$ and $a_{-}$ boson operators represent the exciton and antiexciton annihilation operators, respectively. When $\alpha \ne 0$, the Hamiltonian describes two non-degenerate harmonic oscillators, whereas the Klein-Gordon theory (the $\alpha=0$ case) describes two degenerate harmonic oscillators~\cite{srednicki_2007}. Thus, we can view the inter-band collective modes in semimetals 
as a ``$CP$-violated" Klein-Gordon field without the Lorentz symmetry.

\subsection{Conserved charge}
The Lagrangian in Eq.~(\ref{Lagrangian}) is invariant under a U(1) transformation $\varphi \rightarrow \varphi e^{i\theta}$. By Noether's theorem, it has a conserved charge density \cite{srednicki_2007,altland_simons_2010}:
\begin{align}
    j^0 &=-i \varphi \frac{\partial \mathcal{L}}{\partial\left(\partial_{t} \varphi\right)}+i \varphi^{\dagger} \frac{\partial \mathcal{L}}{\partial\left(\partial_{t} \varphi^{\dagger}\right)} \nonumber \\
& = i \gamma\left[\varphi^{\dagger}\left(\partial_{t} \varphi\right)-\left(\partial_{t} \varphi^{\dagger}\right) \varphi\right]+\alpha \varphi^{\dagger} \varphi. \label{charge-a}
\end{align}
From Eq.~(\ref{a-def}), $\varphi$ is given by a linear 
combination of the annihilation of the exciton ($a_{+}$) 
and the creation of the antiexciton ($a^{\dagger}_{-}$); 
\begin{align}
    \varphi=\frac{1}{\sqrt{\lambda \eta}}\left(a_{+}+a_{-}^{\dagger}\right). \label{varphi-a}
\end{align}
$a_{+}$ and $a_{-}$ have their dynamical evolutions in the interaction picture for a 
quantum-mechanical problem; $a_{+}(t)=a_{+} e^{-i \nu_{+} t}$, $a_{-}(t)=a_{-} e^{-i \nu_{-} t}$. 
From Eqs.~(\ref{charge-a},\ref{varphi-a}) together with the time evolutions, one can readily see that 
the density is time-independent and is given by the difference between the exciton and antiexciton density; 
\begin{align}
    j^0(t)&=\frac{\gamma}{\lambda \eta}\Big[\left(a_{+}^{\dagger} e^{i \nu_{+} t}+a_{-} e^{-i \nu_{-} t}\right)\nonumber\\
&\ \ \left(\nu_{+} a_{+} e^{-i \nu_{+} t}-\nu_{-} a_{-}^{\dagger} e^{i \nu_{-} t}\right)+ {\rm h.c.} \Big]  \nonumber \\
 &\ \ +\frac{\alpha}{\lambda \eta}\left(a_{+}^{\dagger} e^{i \nu_{+} t}+a_{-} e^{-i \nu_{-} t}\right)\left(a_{+} e^{-i \nu_{+} t}+a_{-}^{\dagger} e^{i \nu_{-} t}\right) \nonumber \\
    & =\frac{\alpha+2 \gamma \nu_{+}}{\lambda \eta} a_{+}^{\dagger} a_{+}+\frac{\alpha-2 \gamma \nu_{-}}{\lambda \eta} a_{-}^{\dagger} a_{-} \nonumber \\
  &  \ \ +\frac{\gamma\left(\nu_{+}-\nu_{-}\right)+\alpha}{\lambda \eta}\left[a_{+}^{\dagger} a_{-}^{\dagger} e^{i\left(\nu_{+}+\nu_{-}\right) t}+ {\rm h.c.}\right] \nonumber \\
     &= a_{+}^{\dagger} a_{+}-a_{-}^{\dagger} a_{-}.
\end{align}
Here $\nu_{\pm} = \eta \mp \alpha/(2\gamma)$ and $\lambda=2\gamma$. Thus, the particle $a_{+}$ carries charge $+1$, 
while the antiparticle $a_-$ carries charge $-1$. 

When the $a$-band and $b$-band electrons have an opposite physical property such as spin, 
the conserved charge carries the physical property. This is 
because a joint U(1) transformation,  
$a_{\bm{k}} \rightarrow a_{\bm{k}}e^{\frac{i\theta}{2}}$ and $b_{\bm{k}} \rightarrow b_{\bm{k}}e^{-\frac{i\theta}{2}}$, leads to $\varphi\rightarrow \varphi e^{i\theta}$ with $\varphi \propto b^{\dagger}a$. 
Suppose that the $a$-band electrons are with spin-up polarization 
along $z$-direction, and the $b$-band electrons are with spin-down polarization. Then exciton states carry $S_z = 1$ and antiexciton states carry $S_z = -1$. This can be also seen from the spectral representation of the Green's function, Eq.~(\ref{eqn_Lehmann}). In this case, spin-polarized excitation spectroscopy could distinguish exciton states from antiexciton states experimentally~\cite{MAIALLE1994256,PhysRevLett.115.257403}.

\begin{figure}[t]
    \centering
    \subfigure[ ]{\includegraphics[width= 3.5in]{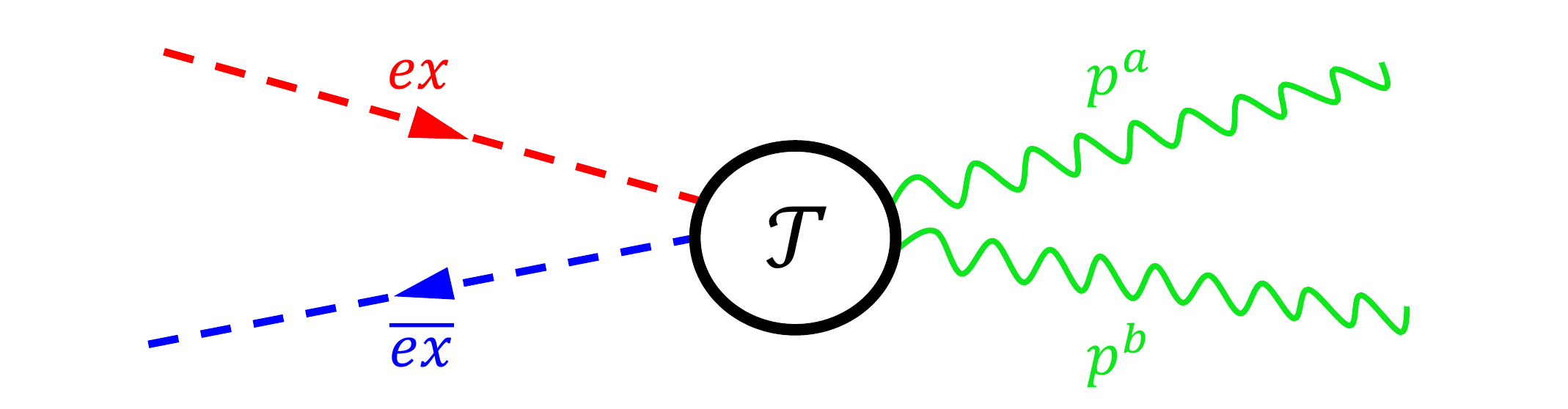}}
    
   \subfigure[ ]{\includegraphics[width= 3.5in]{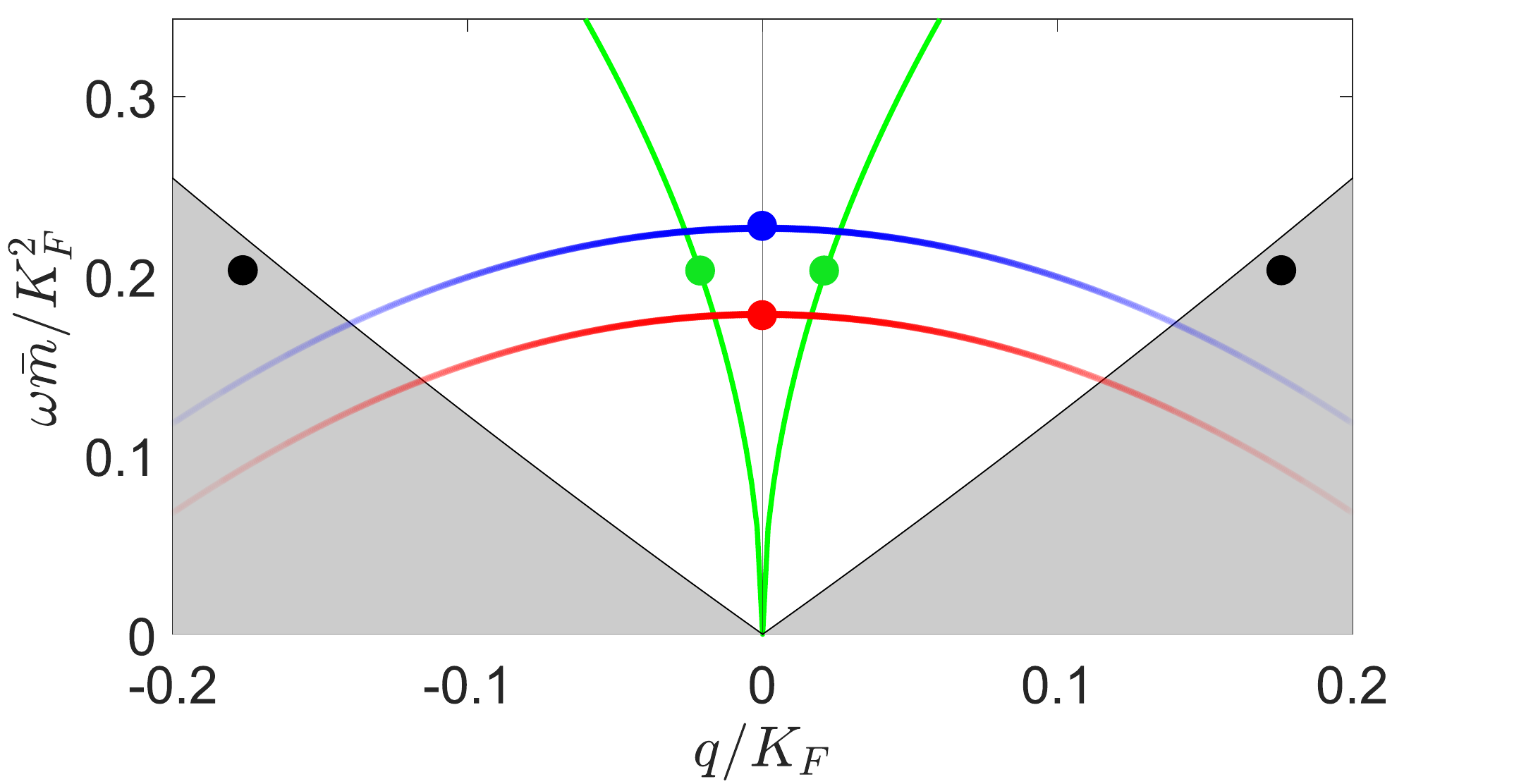}}
    \caption{(a) An annihilation process of an exciton-antiexciton pair. 
    $ex$ and $\overline{ex}$ stand for an exciton and an antiexicton, while $p^a$ and $p^b$ 
    represent an intra-band particle-hole excitation in the $a$ and $b$ band, respectively. 
    (b) Dispersions of the $s$-wave exciton (red) and antiexction (blue) bands, plasmon band (green) and intra-band individual excitations (the grey shaded area) around the $\Gamma$ point in the 2D case for
    the same parameters as in Fig.~\ref{data}(b). The pair annihilation of 
    ${\bm q}=0$ exciton and antiexciton (the red and blue points) 
    leads to two intra-band particle-hole excitations, which can be either individual excitations (the black points) or plasmons (the green points).}
    \label{anni}
\end{figure}

\section{\label{sec7}Physical consequences}
The antiexciton proposed in this paper represents 
a distinct inter-band collective excitation from its
counterpart exciton having different energies ($\omega_+ \ne \omega_-$) and opposite physical charges. 
They manifest themselves as distict peaks in optical spectroscopy experiments. 

When a single inter-band excitation process by a photon 
is allowed by the symmetry and the charge dual to the joint U(1) phase, 
the optical spectroscopy experiment distinguishes the exciton and antiexciton from 
each other in a form of two distinct absorption peaks. 
When the direct excitation process is prohibited by the symmetry, 
an exciton-antiexciton pair can be excited by higher-order scattering processes. 
For example, in the eigenspace of
$|N_a,N_b\rangle$, the energy-momentum conservation allows the pair to 
decay into two {\it intra}-band collective or individual  
excitations~\cite{fetter2012quantum,giuliani_vignale_2005} (Fig.~\ref{anni}), 
being analogous to the electron-positron pair annihilation 
that releases the two photons. The reverse process of the pair 
annihilation enables conversion from two photon-excited 
intra-band excitations into an exciton-antiexciton pair. 

\subsection{Possible annihilation processes of an exciton-antiexciton pair}
The conversion between an exciton-antiexction 
pair and multiple intra-band collective excitations or 
individual excitations is constrained by the momentum and energy 
conservation and momentum-energy dispersions 
of the inter-band/intra-band collective/individual excitations. 
In this subsection, we discuss the conversion process based on calculations of 
the momentum-energy dispersions of intra-band and 
inter-band collective excitations around the $\Gamma$ point for a 
specific set of parameters with 
$m_a = m_b = \bar{m}$, $\mu>0$, and $K_{F,a}>K_{F,b}$. 

In the presence of finite $K_{F,a}$ and $K_{F,b}$, individual excitations 
form continuum spectra on the $\omega-q$ plane. Borders of the continuum spectra 
for the intra-band individual excitations for the $a$ and $b$ bands  
are given by $\omega = (q^2 \pm 2K_{F,a}q)/(2m)$ and 
$\omega = (q^2 \pm 2K_{F,b}q)/(2m)$, respectively~\cite{giuliani_vignale_2005}. 
Intra-band density-wave modes form momentum-energy dispersions outside 
the continuum spectra. The 
dispersion of the intra-band density modes can be calculated from 
the polarization functions of the $a$-band and $b$-band densities, 
\begin{align}
&\Pi_c(\bm{x} - \bm{x^{\prime}}, t-t^{\prime}) \nonumber \\
& \ = -i\langle 0 |\mathcal{T} \{ [\hat{\rho}^c(\bm{x},t) - \rho_0^c][\hat{\rho}^c(\bm{x^{\prime}},t^{\prime})- \rho_0^c]\} | 0 \rangle.
\end{align}
Here $\hat{\rho}^c$ and $\rho_0^c$ represent the $c$-band density operator 
and its ground-state average, respectively ($c=a,b$). Let $\Pi_c(\bm{q},\omega)$  be the Fourier transform of 
$\Pi_c(\bm{x} - \bm{x^{\prime}}, t-t^{\prime})$.  
\begin{figure}[t]
    \centering
\includegraphics[height=1.7cm]{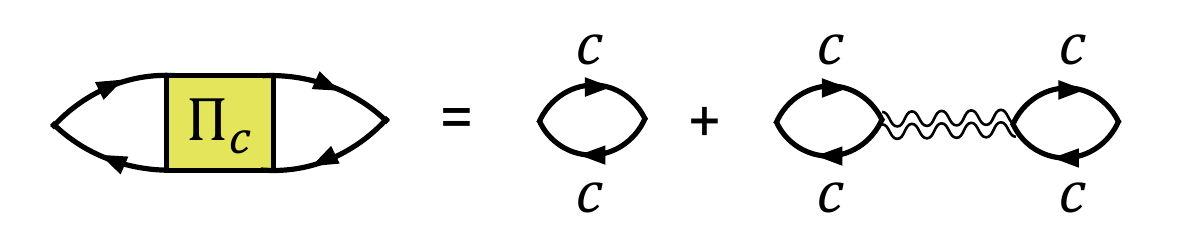}
     \caption{Feynman diagrams of the RPA contributions to the polarization function $\Pi^c(\bm{q},\omega)$ ($c=a$ or $b$). The 
 double-wavy lines represent screened Coulomb interactions given in Fig.~\ref{RPA}. Solid lines with labels $c$ stand for the free single-particle Green's functions of the $c$ band ($c= a,b$).}
    \label{ring}
\end{figure}
In terms of random phase approximation (RPA), $\Pi^c(\bm{q},\omega)$ 
is calculated as follows (Fig.~\ref{ring}):
\begin{align}
    \Pi^c(\bm{q},\omega) &= \Pi_0^c(\bm{q},\omega) + \Pi_0^c(\bm{q},\omega) w(\bm{q},\omega)\Pi_0^c(\bm{q},\omega) \nonumber \\
    w({\bm q},\omega)&=\frac{v(\bm{q})}{1-\Pi_0(\bm{q},\omega)v(\bm{q})},
\end{align}
where $\Pi_0^c(\bm{q},\omega)$ and $\Pi_0(\bm{q},\omega)$ are given by  
Eq.~(\ref{bare_polar}). Thus, the intra-band density-wave modes for 
the two bands share the same denominator and the zeros of the denominator 
determine the momentum-energy dispersions of the plasmon mode,
\begin{align}
    1-\Pi_0(\bm{q},\omega)v(\bm{q}) = 0.
    \label{plasmon_eq}
\end{align}
Using the Lindhard function at the limit of
$q\rightarrow0$ for 3D and $q\rightarrow0$, $\omega\rightarrow0$ for 2D ~\cite{fetter2012quantum,mihaila2011lindhard,giuliani_vignale_2005}, 
\begin{equation}
    \Pi^c_0(q,\omega) = 
    \begin{cases}
    -\frac{m_c K_{F,c}}{4\pi^2}\Big( 2- x_c \ln(\frac{x_c +1}{x_c -1}) \Big)            & \text{in 3D,} \\
       -\frac{m_c}{2\pi}(1-\frac{|x_c|}{\sqrt{x_c^2 - 1}})    & \text{in 2D,}
    \end{cases}
\end{equation}
with $x_c \equiv \frac{m_c \omega}{K_{F,c}q}$ ($c=a,b$), we further take $\omega\gg q$ and expand 
$\Pi_0(\bm{q},\omega) \equiv \Pi^a_0(q,\omega) + \Pi^b_0(q,\omega)$ up to the fourth order of 
$\frac{1}{x_a}$ and $\frac{1}{x_b}$. Then we solve Eq.~(\ref{plasmon_eq}) for $\omega$ up to the subleading order in small $q$. This gives out 
\begin{equation}
    \omega = 
    \begin{cases}
    \sqrt{A_3}(1+\frac{B_3}{2A_3^2}q^2)          & \text{in 3D,} \\
        \sqrt{A_2 q}(1 + \frac{B_2}{2A_2^2}q)  & \text{in 2D,}
    \end{cases}
\end{equation}
with 
\begin{equation}
\begin{cases} 
A_3 = \sum\limits_{c=a,b} \frac{2K_{F,c}^3}{3\pi m_c},  
& A_2 = \sum\limits_{c=a,b} \frac{K_{F,c}^2}{2m_c},\\ 
B_3 = \sum\limits_{c=a,b} \frac{2K_{F,c}^5}{5\pi m_c^3}, &  
B_2 = \sum\limits_{c=a,b} \frac{3K_{F,c}^4}{8m_c^3}. \\
\end{cases}
\end{equation}

\begin{figure}[t]
    \centering
    \subfigure[ ]{\includegraphics[width= 3.5in]{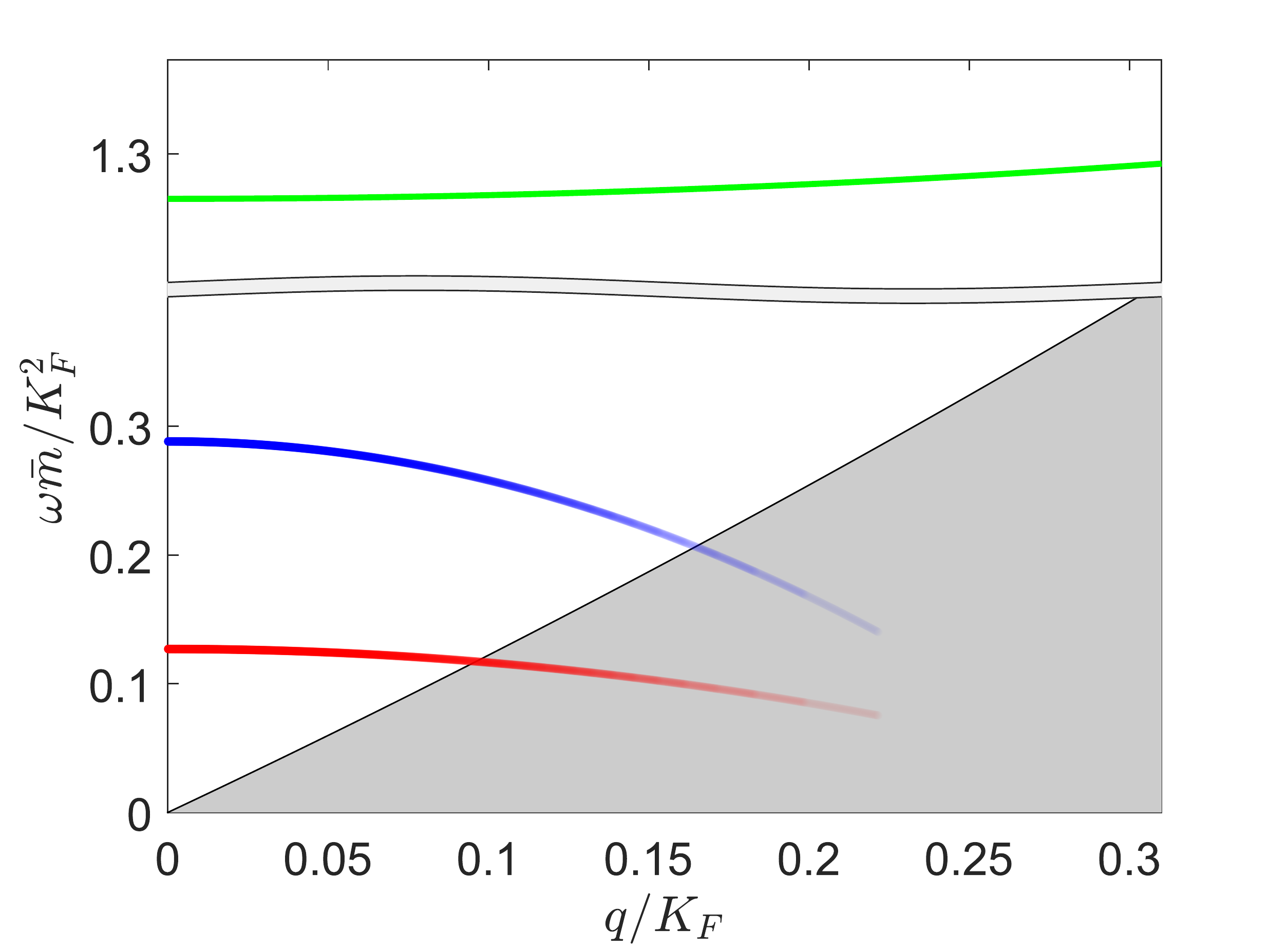}}
    
    \subfigure[ ]{\includegraphics[width= 3.5in]{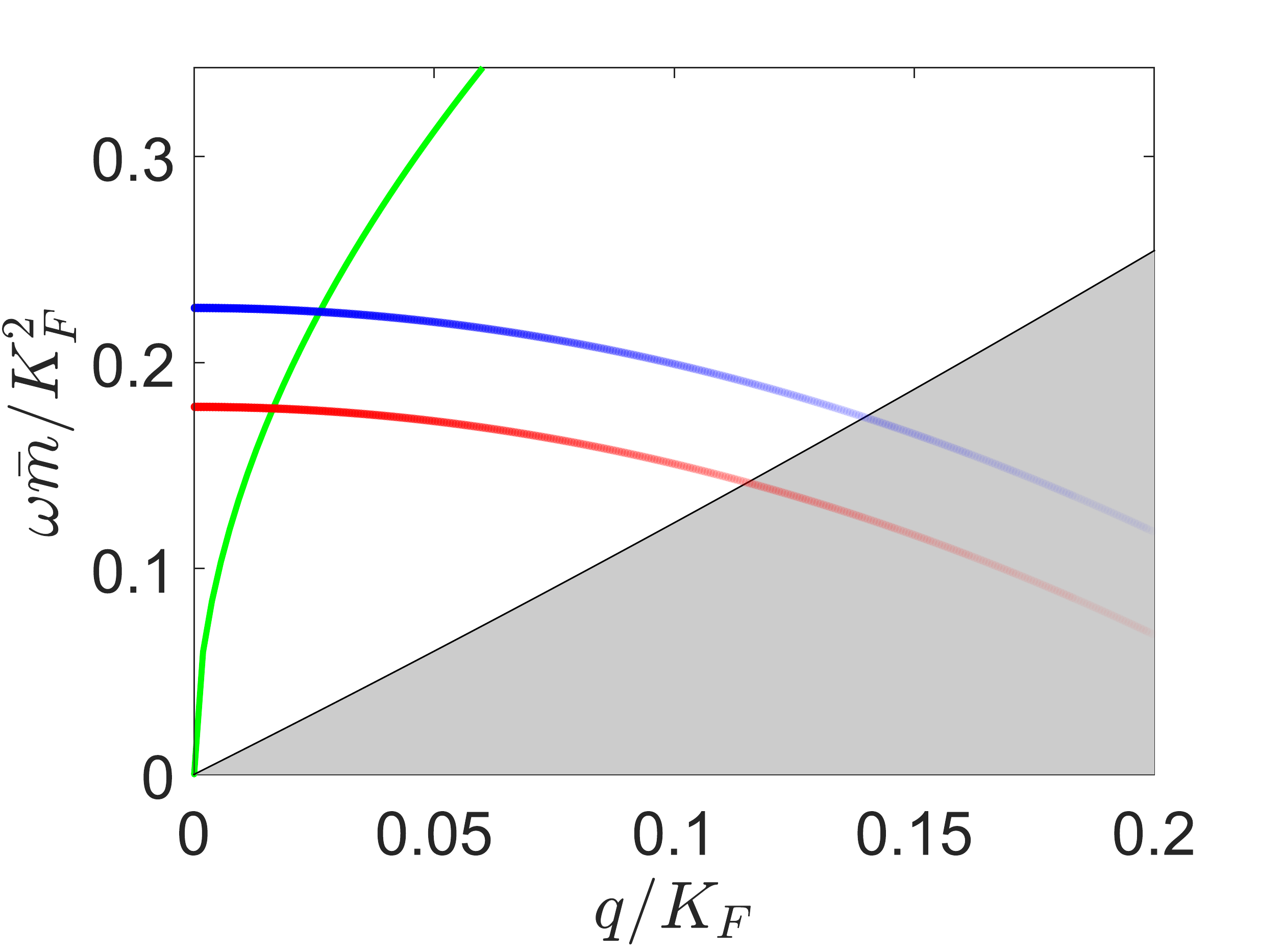}}
    \caption{Energy-momentum dispersions of the $s$-wave exciton and antiexciton bands (the red and blue lines), plasmon bands (the green lines) and continuum spectra of the intra-band individual excitations (the grey shaded regions). The dispersions for $\bar{m}=m_a=m_b$ are determined by three dimensionless quantities, $K_{F}\alpha_B$, $\tilde{E}_g \equiv E_g \bar{m}/K^2_F$ and $\tilde{\mu} \equiv \mu \bar{m}/K^2_{F}$, where $K_{F} \equiv (K_{F,a}+K_{F,b})/2$ and the effective Bohr radius $\alpha_B \equiv  1/\bar{m}$.  (a) The 3D case with $K_{F}\alpha_B= 0.289, \tilde{E}_g =-1.03$, and $\tilde{\mu}= 0.172$. (b) The 2D case with $K_{F}\alpha_B= 0.540$, $\tilde{E}_g =-1.03$, and $\tilde{\mu}= 0.172$. The quantum numbers of the $s$-wave exciton and antiexciton are $(nlm) = (100)$ in the 3D case and $(nm) = (10)$ in the 2D case (see the text).}
    \label{plasmon_fig}
\end{figure}

Fig.~\ref{plasmon_fig} shows the region of the continuum spectra of the intra-band individual excitations, the dispersions of the plasmon modes, and momentum-energy dispersions of the $s$-wave exciton and antiexciton bands. The band dispersions for the exciton and antiexciton bands are calculataed only around the $\Gamma$ point (see Eqs.~(\ref{mass2}) and Appendix \ref{appendix:a}). 
The figure shows that the 
plasmon oscillation in the 3D case appears at much higher energies than 
the exciton and antiexciton bands. Thereby, it is likely that the exciton-antiexciton pair in the 3D case only decays into intra-band individual excitations. In the 2D case, the plasmon dispersion is gapless, where an exciton-antiexciton pair decays either into intra-band individual excitations or into density-wave modes.

\section{\label{sec8}Summary} 
In this paper, we demonstrate the universal coexistence of the exciton and the
antiparticle analog of the exciton (antiexciton) in semimetals 
as two distinguishable collective modes. 
The concept of the antiparticle in semimetals is introduced for the first time by a spectral representation 
of the inter-band two-particle Green's function and physical differences between our 
concept of the antiexciton and 
those in literature~\cite{PhysRevB.44.1368,PhysRevB.44.3638,lerner1981two,PhysRevB.59.5627} are clarified. 
Evaluations of the Green's function in the dilute carrier-density 
limit show that the exciton and the antiexciton coexist in the inter-band 
excitation spectra of doped semimetals in 2D and 3D. 
The effective Lagrangian 
of the exciton and antiexciton is given by the $CP$-violated Klein-Gordon theory. 
The physical consequence of the coexistence of exciton and antiexciton is  
discussed in the optical spectroscopy experiment. Our theory is relevant to 
inter-band excitation spectra in semimetal materials, such 
as As, Sb, and HgTe, and electron-hole double-layer systems such as 
semiconductor heterostructures~\cite{Jain2013,PhysRevB.99.085307,PhysRevB.95.045116,du2017evidence,
li2017excitonic,PhysRevB.100.035130,zhang2021dissipationless,kroemer20046} and bilayer graphene under an external 
perpendicular electric field~\cite{Perali2013,Conti2017,Rios2018}. 
Optical absorption and photoluminescence can be experimental 
probes of the antiparticle analog of the exciton in these materials~\cite{Liang_1970,PhysRevLett.82.3544,doi:10.1063/1.109053,PhysRevLett.103.026402}. 

\section*{Acknowledgement}
We thank Zhenyu Xiao, Weiliang Qiao, and Qingzheng Qiu for helpful discussions. The work was supported by the National Basic Research Programs of China (No.~2019YFA0308401) and by National Natural Science Foundation of China (No.~11674011 and No.~12074008). 

\appendix 
\section{Energy bands of the exciton and antiexciton around the $\Gamma$ point}
\label{appendix:a}
In this appendix, we explain how we diagonalize Eq.~(\ref{bs3}) around ${\bm q}=0$ and 
obtain energy bands of the inter-band bound states around the $\Gamma$ point. For clarity of presentation, 
we consider the case with $K_{F,a}>K_{F,b}$ ($\mu>\mu_0$). 
A generalization to the other case is straightforward. 

For $|{\bm q}|$ smaller than $K_{F,a}-K_{F,b}$ 
(including $\bm{q}=\bm{0}$), the diagonal matrix $[\widetilde{G}^{ex}_0(\bm{q},\omega)^{-1}]_{{\bm k}{\bm k}^{\prime}}$ has no finite matrix element for $K_{F,a}>|{\bm k}+{\bm q}|$ and $|{\bm k}|>K_{F,b}$; 
\begin{align}
& \widetilde{G}^{ex}_0(\bm{q},\omega)_{\bm{k}\bm{k^{\prime}}}^{-1} = \nonumber \\
& 
\delta_{\bm{k}\bm{k^{\prime}}} 
     \begin{cases}
    \omega - [\epsilon_a(\bm{k}+\bm{q}) - \epsilon_b(\bm{k})],            & |\bm{k} + \bm{q}| > K_{F,a} \\
    -\{\omega - [\epsilon_a(\bm{k}+\bm{q}) - \epsilon_b(\bm{k})]\},                      & |\bm{k}| < K_{F,b}
     \end{cases}. 
\end{align}
Therefore, $[\widetilde{G}^{ex}(\bm{q},\omega)^{-1}]_{{\bm k}{\bm k}^{\prime}}$ has finite matrix elements only within a domain of (i) 
$|{\bm k}+{\bm q}|>K_{F,a}$ or $K_{F,b}>|{\bm k}|$ and (ii) 
$|{\bm k}^{\prime}+{\bm q}|>K_{F,a}$ or 
$K_{F,b}>|{\bm k}^{\prime}|$;
\begin{align}
          \widetilde{G}^{ex}\left(\boldsymbol{q}, \omega\right)^{-1} = \widetilde{G}^{ex}_0\left(\boldsymbol{q}, \omega\right)^{-1} + 
          \eta \!\ W \!\ \eta. \label{etaWeta}
\end{align}
Here $\eta$ is a diagonal matrix,
\begin{align}
    [\eta]_{{\bm k}{\bm k}^{\prime}} \equiv 
    \delta_{{\bm k}{\bm k}^{\prime}} \Big[1-\theta(K_{F,a}-|{\bm k}+{\bm q}|)\theta(|{\bm k}|-K_{F,b})\Big]. \label{eta}
\end{align}
One can readily see this from a Taylor expansion of Eq.~(\ref{bs3}) 
in $W$ for $\widetilde{G}^{ex}$; 
\begin{align}
\widetilde{G}^{ex} = 
\widetilde{G}^{ex}_0 - \widetilde{G}^{ex}_0 \!\ W \!\ 
\widetilde{G}^{ex}_0 + \widetilde{G}^{ex}_0 \!\ W \!\  
\widetilde{G}^{ex}_0 \!\ W \!\  
\widetilde{G}^{ex}_0 + \cdots,  
\end{align}
together with $\widetilde{G}^{ex}_0=\eta \widetilde{G}^{ex}_0 \eta$. 
In the next two sections, we diagonalize $\widetilde{G}^{ex,-1}$ 
within the domain specified by Eq.~(\ref{eta}). 

\subsection{Inter-band excitation energies at the $\Gamma$ point}
When ${\bm q}=0$, $G^{ex}\left(\boldsymbol{0},\omega\right)^{-1}$ becomes 
real symmetric and it has continuous spatial rotation symmetries. 
Eq.~(\ref{bs4}) can be block-diagonalized in terms of spherical 
harmonics in 3D and trigonometric functions in 2D. This leads to 
Eq.~(\ref{3d-q=0}) and Eq.~(\ref{2d-q=0}).  In this section, we will 
describe this deduction and how the radial functions 
in Eq.~(\ref{3d-q=0}) and Eq.~(\ref{2d-q=0}) should be calculated. 
With $\overline{G^{ex}}({\bm q},\omega;{\bm k},{\bm k}^{\prime})\equiv 
\Omega \widetilde{G}^{ex}({\bm q},\omega)_{{\bm k}{\bm k}^{\prime}}$, 
the BS equation at ${\bm q}=0$ is given by 
\begin{widetext}
\begin{align}
    \int \frac{d^d \bm{k^{\prime\prime}}} {(2\pi)^d} \Big( D(\omega;\bm{k})(2\pi)^d \delta(\bm{k}-\bm{k^{\prime\prime}})  + w(\bm{k}-\bm{k^{\prime\prime}}) \Big) \overline{G^{ex}} \left(\boldsymbol{0}, \omega; \bm{k^{\prime\prime}} ,\bm{k^{\prime}}\right) = (2\pi)^d \delta(\bm{k}-\bm{k^{\prime}}),
    \label{BS_16}
\end{align}
and  
\begin{align}
    D(\omega;\bm{k}) =  \theta(|\bm{k}|-K_{\rm out})\Big( \omega- (\epsilon_a(\bm{k}) - \epsilon_b(\bm{k}))\Big) - \theta(K_{\rm in}-|\bm{k}|)\Big( \omega- (\epsilon_a(\bm{k}) - \epsilon_b(\bm{k}))\Big).
    \label{Dk}
\end{align}
\end{widetext}
Note that $\bm{k}$, ${\bm k}^{\prime}$ and ${\bm k}^{\prime\prime}$ in the equations are 
in a range $|\bm{k}|>K_{F,a} \equiv K_{\mathrm{out}}$ or $|\bm{k}|< K_{F,b} \equiv K_{\mathrm{in}}$. $\widetilde{G}^{ex}(\bm{q},\omega)_{\bm{k}\bm{k}'}=0$ 
if $\bm{k}$ or $\bm{k}'$ is outside the range.

\subsubsection{3D case}
In the 3D case, the delta function in the right hand side of 
Eq.~(\ref{BS_16}) as well as the screened Coulomb interaction 
can be decomposed in terms of the spherical harmonics, 
\begin{align}
    \delta\left(\boldsymbol{k}-\boldsymbol{k}^{\prime}\right)&=\frac{1}{k^2} \delta\left(k-k^{\prime}\right) \delta\left(\cos \theta-\cos \theta^{\prime}\right) \delta\left(\varphi-\varphi^{\prime}\right) \nonumber \\
     &= \frac{1}{k^2} \delta\left(k-k^{\prime}\right) \sum_{lm} Y_{lm}(\theta,\varphi)Y^{*}_{lm}(\theta^{\prime}, \varphi^{\prime}) 
     \label{delta_3d}, 
\end{align}
\begin{align} 
    w\left(\boldsymbol{k}-\boldsymbol{k}^{\prime}\right)&=\frac{4 \pi }{k^{2}+{k^{\prime}}^{2}-2 k k^{\prime} \cos \gamma+k_{TF}^{2}} \nonumber \\ 
& = \sum_{lm} a_l(k,k^{\prime}) \frac{4\pi}{2l+1}  Y_{lm}(\theta,\varphi)Y^{*}_{lm}(\theta^{\prime}, \varphi^{\prime}).
    \label{wk_3d}
\end{align}
Here $\bm{k} \equiv (k\sin \theta \cos \varphi, k\sin \theta \sin \varphi, k\cos \theta)$.  $\gamma$ is an angle between $\bm{k}$ and $\bm{k^{\prime}}$. $P_l(\cos \gamma)$ is the Legendre polynomial ($l=0,1,\cdots$). $Y_{lm}(\theta,\varphi)$ is the spherical harmonics ($m=-l,-l+1,\cdots,l$). The spherical harmonics are defined with normalization and completeness relations, 
\begin{align}
    &\int^1_{-1} d(\cos\theta) \int^{2\pi}_{0}d\varphi  \!\
    Y^{*}_{lm}(\theta,\varphi) Y_{l^{\prime}m^{\prime}}(\theta,\varphi) = \delta_{ll^{\prime}}\delta_{mm^{\prime}}, \nonumber \\ 
    &\sum_{lm}Y_{lm}(\theta,\varphi) Y^{*}_{lm}(\theta^{\prime},\varphi^{\prime}) = \delta(\cos\theta-\cos\theta^{\prime}) \delta(\varphi-\varphi^{\prime}). \label{nc-sp}
\end{align}
$a_l(k,k^{\prime})$ is the coefficient of the Legendre expansion. The lowest and second lowest order coefficients are 
calculated as 
\begin{align}
    a_0(k,k^{\prime}) &= \frac{\pi}{kk^{\prime}} \Big\{ \ln\Big( (k+k^{\prime})^2 + k_{TF}^2 \Big)  \nonumber \\
    &\ \ \ - \ln\Big( (k-k^{\prime})^2 + k_{TF}^2 \Big) \Big\}, \\
    a_1(k,k^{\prime}) &= \frac{3\pi}{kk^{\prime}} \Big\{ \frac{k^2+k^{\prime 2} + k_{TF}^2}{2kk^{\prime}} \Big[ \ln\Big( (k+k^{\prime})^2+k_{TF}^2 \Big) \nonumber \\
&  \ \ \ - \ln\Big(  (k-k^{\prime})^2+k_{TF}^2 \Big) \Big] -2 \Big\}.
\end{align}
In terms of the harmonics, the solution of 
Eq.~(\ref{BS_16}) is given by 
\begin{align}
   & \overline{G^{ex}} \left(\boldsymbol{0}, \omega; \bm{k^{\prime\prime}} ,\bm{k^{\prime}}\right) = \nonumber \\
& \ \ \sum_{nlm}  \frac{Y_{l m}(\theta^{\prime\prime}, \varphi^{\prime\prime}) f_{nl}(\omega;k^{\prime\prime}) f_{nl}(\omega;k^{\prime}) Y_{lm}^{*}\left(\theta^{\prime}, \varphi^{\prime}\right)}{\xi_{nl}(\omega)}.
\end{align}
Here $f_{nl}(\omega;k)$ and $\xi_{nl}(\omega)$ are 
eigenvector and eigenvalue of a one-dimensional integral equation, 
\begin{align}
\int^{\infty}_{0} \frac{{k^{\prime}}^2 dk^{\prime}}{(2\pi)^3} 
h_{l}(\omega;k,k^{\prime}) f_{nl}(\omega;k^{\prime}) 
= \xi_{nl}(\omega) f_{nl}(\omega;k),   
\end{align}
with 
\begin{align}
    h_l(\omega;k,k^{\prime}) \equiv \frac{D(\omega;k)}{k^2} (2\pi)^3 \delta(k-k^{\prime}) + \frac{4\pi}{2l+1}a_l(k,k^{\prime}),  
    \label{BS_16b}
\end{align}
and normalization and completeness relations, 
\begin{align}
    & \sum_{n} f_{nl}(\omega;k) f_{nl}(\omega;k^{\prime}) 
    = \frac{(2\pi)^3}{k^2} \delta(k-k^{\prime}), \nonumber \\ 
    & \int^{\infty}_{0} \frac{k^2 dk}{(2\pi)^3} f_{nl}(\omega;k) 
    f_{n^{\prime}l}(\omega;k) = \delta_{nn^{\prime}}. \label{nc-f}
\end{align}
To solve the one-dimensional integral equation numerically, 
$k$ is discretized by $2\pi/L$ with large $L$; 
\begin{align}
    \int^{\infty}_{0} dk  = \frac{2\pi}{L} \sum_{k}, \!\ \!\ \!\ \delta(k-k^{\prime}) = \frac{L}{2\pi} \delta_{kk^{\prime}}.  \label{discrete}
\end{align}
With the discretization, the integral equation takes a form of 
\begin{align}
    \sum_{k^{\prime}}  H_{l,kk^{\prime}}^{\omega} V_{nl,k^{\prime}}^{\omega}= \xi_{nl}(\omega)V_{nl,k}^{\omega}, \label{3d-diag}
\end{align}
and  
\begin{align}
    &H_{l,kk^{\prime}}^{\omega} \equiv  D(\omega;k)\delta_{kk^{\prime}} + \frac{1}{L}\frac{kk^{\prime}}{\pi}\frac{a_l(k,k^{\prime})}{2l+1}, \nonumber \\
  & V_{nl,k}^{\omega} \equiv  \frac{k}{2\pi\sqrt{L}} f_{nl}(\omega;k), \label{HandV}
\end{align}
where $\sum\limits_{k} V_{nl,k}^{\omega}V_{n^{\prime}l,k}^{\omega} = \delta_{nn^{\prime}}$ and $\sum\limits_{n} V_{nl,k}^{\omega}V_{nl,k^{\prime}}^{\omega} = \delta_{kk^{\prime}}$.

\subsubsection{2D case}
In the 2D case, the delta function and 
the screened Coulomb interaction are expanded in terms of the trigonometric functions, 
\begin{align}
    \delta\left(\boldsymbol{k}-\boldsymbol{k}^{\prime}\right) &=\frac{\delta\left(k-k^{\prime}\right)}{k} \delta\left(\varphi-\varphi^{\prime}\right) \nonumber \\
& =\frac{\delta\left(k-k^{\prime}\right)}{k}\frac{1}{2 \pi} \sum_{m} e^{i m(\varphi-\varphi^{\prime})} \label{delta_2d}, \\ 
   w\left(\boldsymbol{k}-\boldsymbol{k}^{\prime}\right)&=\frac{2\pi}{\sqrt{k^{2}+k^{\prime 2}-2 k k^{\prime} \cos \left(\varphi-\varphi^{\prime}\right)}+k_{T F}} \nonumber \\
 &=\sum_{m} F_{m}\left(k, k^{\prime}\right) e^{im(\varphi-\varphi^{\prime})},
   \label{wk_2d}
\end{align}
where $\bm{k} \equiv (k\cos\varphi, k\sin\varphi)$, and
\begin{align}
    F_{m}\left(k, k^{\prime}\right) 
    &= \int_{0}^{\pi} d \phi \frac{2 \cos (m \phi)}{\sqrt{k^{2}+k^{\prime 2}-2 k k^{\prime} \cos \phi}+k_{T F}}.
    \label{Fm}
\end{align}
In terms of the expansion, the solution of Eq.~(\ref{BS_16}) is given by 
\begin{align}
  & \overline{G^{ex}} \left(\boldsymbol{0}, \omega; \bm{k^{\prime\prime}} ,\bm{k^{\prime}}\right) = \nonumber \\
 & \sum_{nm} \frac{ f_{nm}(\omega;k^{\prime\prime}) f_{nm}(\omega;k^{\prime}) e^{im(\varphi^{\prime\prime}-\varphi^{\prime})}}{\xi_{nm}(\omega)}.
\end{align}
Here $f_{nm}(\omega;k)$ and $\xi_{nm}(\omega)$ are the eigenvector and eigenvalue of a one-dimensional integral equation, 
\begin{align}
    \int_0^{+\infty} \frac{k^{\prime}dk^{\prime}}{2\pi}  h_m(\omega;k,k^{\prime})f_{nm}(\omega;k^{\prime}) 
    = \xi_{nm}(\omega) f_{nm}(\omega;k), 
\end{align}
with 
\begin{align}
    h_m(\omega;k,k^{\prime}) = \frac{D(\omega;k)}{k} (2\pi) \delta(k-k^{\prime}) +  F_m(k,k^{\prime}),
\end{align}
and normalization and completeness relation
\begin{align}
    & \sum_{n} f_{nm}(\omega;k) f_{nm}(\omega;k^{\prime}) 
 =  \frac{2\pi}{k} \delta\left(k-k^{\prime}\right),  \nonumber \\
& \int_0^{+\infty}\frac{kdk}{2\pi} f_{nm}(\omega;k) f_{n^{\prime}m^{\prime}}(\omega,k) = \delta_{nn^{\prime}}.
\end{align}
With the same discretization as Eq.~(\ref{discrete}), the integral equation reduces to  
\begin{align}
    \sum_{k^{\prime}} H_{m,kk^{\prime}}^{\omega} V_{nm,k^{\prime}}^{\omega}= \xi_{nm}(\omega) V_{nm,k}^{\omega}, \label{2d-diag}
\end{align}
and  
\begin{align}
   & H_{m,kk^{\prime}}^{\omega} \equiv D(\omega;k)\delta_{kk^{\prime}} + \frac{\sqrt{kk^{\prime\prime}}}{L}F_m(k,k^{\prime}), \nonumber \\
   & V_{nm,k}^{\omega} \equiv  \sqrt{\frac{k}{L}} f_{nm}(\omega;k), \label{HandV-2d}
\end{align}
where $\sum\limits_{k} V_{nm,k}^{\omega}V_{n^{\prime}m,k}^{\omega} = \delta_{nn^{\prime}}$ and $\sum\limits_{n} V_{nm,k}^{\omega}V_{nm,k^{\prime}}^{\omega} = \delta_{kk^{\prime}}$. 
$F_{m}(k,k^{\prime})$ in Eq.~(\ref{Fm}) is evaluated numerically. 

In the numerical 
diagonalization of Eqs.~(\ref{3d-diag},\ref{2d-diag}), we set a large value of 
$L$ ($\sim 300 \times 2\pi$) and a large cutoff of $k$ 
($>15\times \text{max}(K_{F,a}, K_{F,b})$) so that the numerical solutions 
of the eigenvalues are convergent.



\subsection{Band curvatures of exciton and antiexciton bands around the $\Gamma$ point}

The previous section explains how we evaluate exciton and antiexciton energies at the $\Gamma$ point. This section explains 
how we evaluate energy-band curvatures of the exciton and antiexciton bands around the $\Gamma$ point. Let us begin with Eq.~(\ref{bs4}), 
\begin{align}
    \sum_{{\bm k}^{\prime}}\big[\widetilde{G}^{ex}({\bm q},\omega)^{-1}\big]_{{\bm k}{\bm k}^{\prime}} \langle {\bm k}^{\prime}|\phi_j({\bm q},\omega)\rangle 
    = \xi_{j}({\bm q},\omega)\langle{\bm k}|\phi_j({\bm q},\omega)\rangle. \label{bs5}
\end{align}
Suppose that the eigenvalue problem at ${\bm q}=0$ are 
solved for the 3D and 2D case respectively as in the previous section;
\begin{align}
        \sum_{{\bm k}^{\prime}}\big[\widetilde{G}^{ex}({\bm 0},\omega)^{-1}\big]_{{\bm k}{\bm k}^{\prime}} \langle {\bm k}^{\prime}|\phi_j({\bm 0},\omega)\rangle 
    = \xi_{j}({\bm 0},\omega)\langle{\bm k}|\phi_j({\bm 0},\omega)\rangle,  \label{bs5-0}
\end{align}
with a normalization, 
\begin{align}
   &\langle {\bm k}|\phi_j({\bm q}=0,\omega)\rangle = \nonumber \\
 & \ \ \left\{\begin{array}{lll} 
     \frac{1}{\sqrt{\Omega}} f_{nl}(\omega;k)Y_{lm}(\theta,\varphi), & j=(nlm), & {\rm in} \!\ \!\ {\rm 3D}, \\
     \frac{1}{\sqrt{\Omega}} f_{nm}(\omega;k)e^{im\varphi}, & j=(nm), & {\rm in} \!\ \!\ {\rm 2D}. \\
     \end{array}\right. \label{from-sum-to-integral}
\end{align}
The normalization gives a proper completeness relation, 
\begin{align}
    & \sum_{j}\langle {\bm k}|\phi_{j}({\bm 0},\omega)\rangle 
    \langle \phi_j({\bm 0},\omega)|{\bm k}^{\prime}\rangle = \delta_{{\bm k}{\bm k}^{\prime}}, 
\nonumber \\ 
 & \sum_{{\bm k}}
    \langle \phi_j({\bm 0},\omega)|{\bm k}\rangle \langle {\bm k}|\phi_{j^{\prime}}({\bm 0},\omega)\rangle  = \delta_{jj^{\prime}}. 
\end{align}
To obtain the band curvature of exciton and antiexciton bands 
around the $\Gamma$ point, 
we consider $|{\bm q}|$ as a small quantity, expand 
$\widetilde{G}^{ex}({\bm q},\omega)^{-1}$ in ${\bm q}$ up to the second order, and evaluate the second-order energy correction of $\xi_j({\bm q},\omega)$; 
$\xi_j({\bm q},\omega)=\xi_j({\bm 0},\omega) + b(\omega)q^2$. By the rotational symmetry, a $q$-linear energy correction is zero, 
while the $q$-quadratic energy correction depends only on the norm 
of ${\bm q}$, $q\equiv |{\bm q}|$.
\begin{figure*}[t]
\subfigure[ ]{\includegraphics[width= 2.2in]{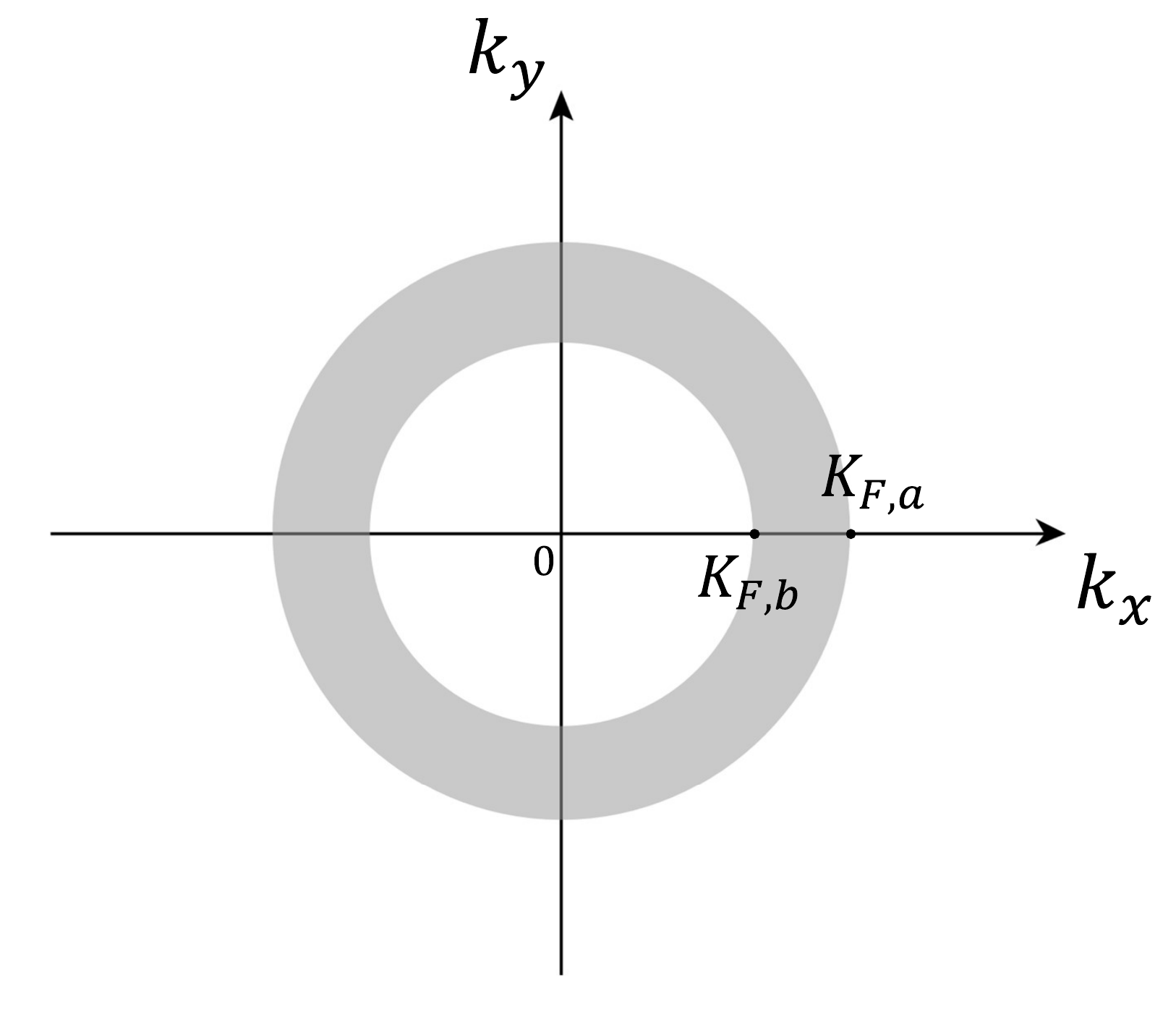}}
\subfigure[ ]{\includegraphics[width= 2.2in]{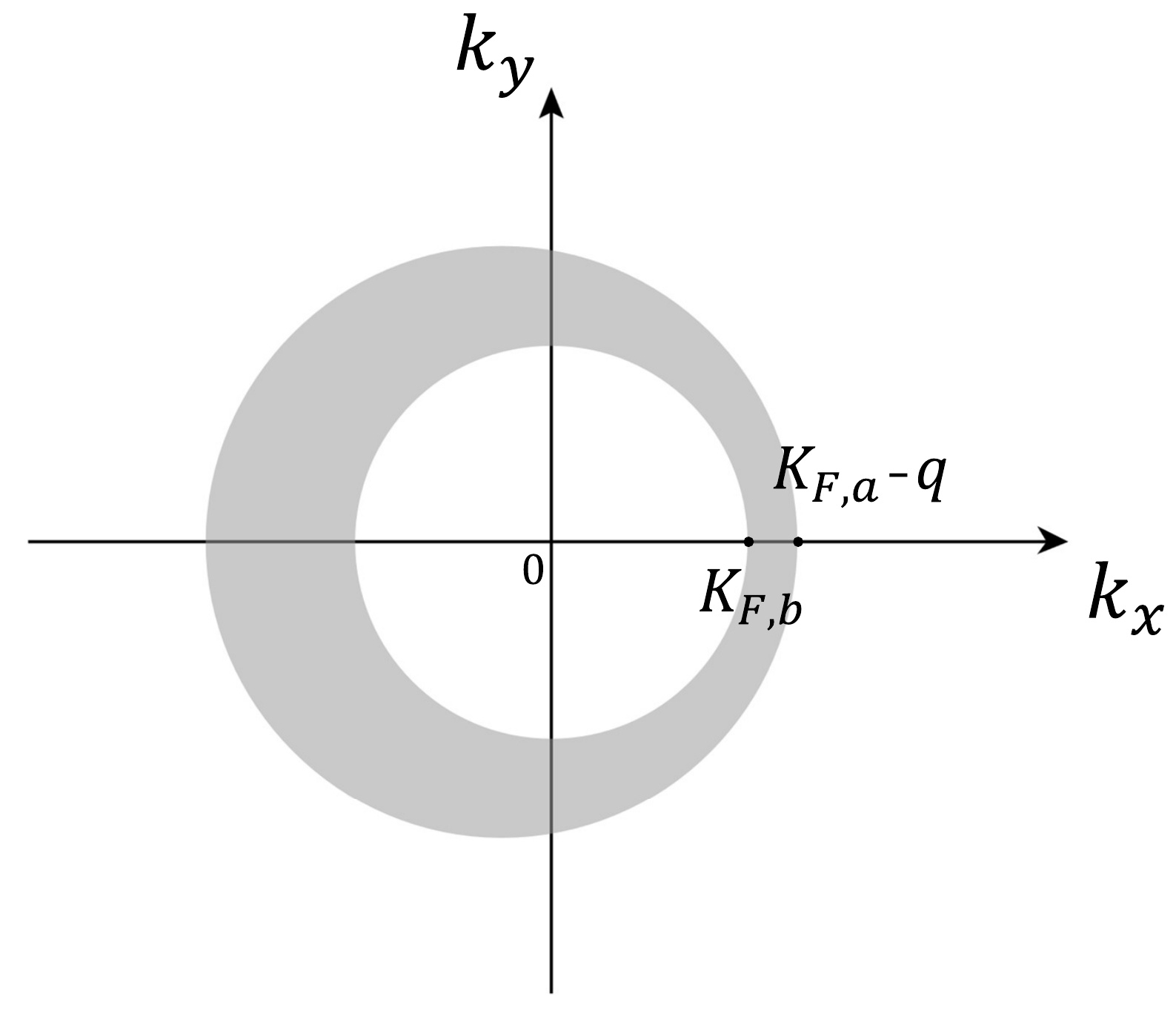}}
\subfigure[ ]{\includegraphics[width= 2.2in]{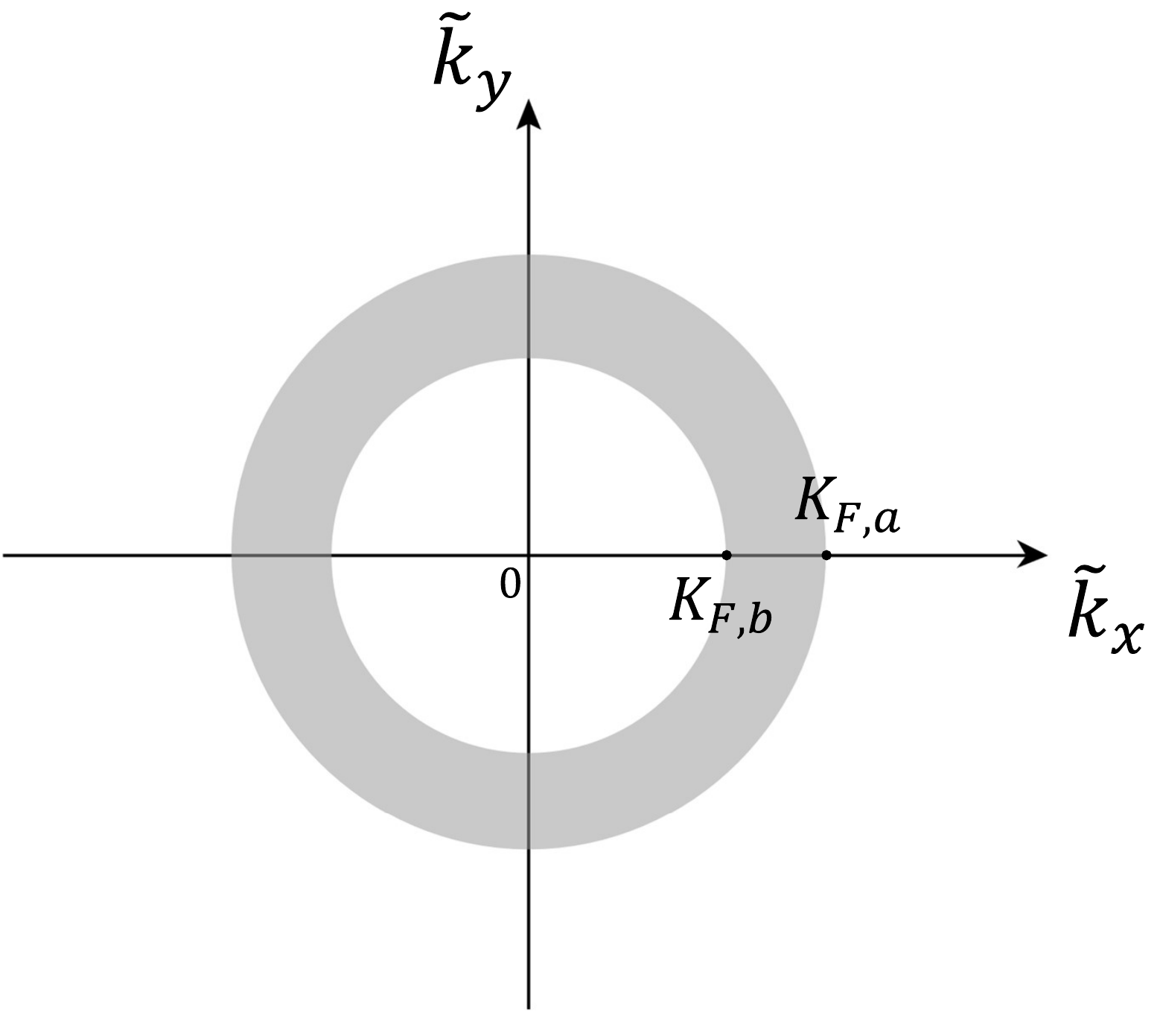}}
\caption{Domains of $\widetilde{G}^{ex}(\bm{0},\omega)^{-1}$ and $\widetilde{G}^{ex}(\bm{q},\omega)^{-1}$. 
The shaded regions are excluded for the domains of ${\bm k}$ and ${\bm k}^{\prime}$ in $[\widetilde{G}^{ex}(\bm{0},\omega)^{-1}]_{{\bm k}{\bm k}^{\prime}}$ (a), for the domains of ${\bm k}$ and ${\bm k}^{\prime}$ in $[\widetilde{G}^{ex}(\bm{q},\omega)^{-1}]_{{\bm k}{\bm k}^{\prime}}$ (b), and, for the domains of 
$\Tilde{\bm k}$ and $\Tilde{\bm k}^{\prime}$ in $[\widetilde{G}^{ex}(\bm{q},\omega)^{-1}]_{\Tilde{\bm k}\Tilde{\bm k}^{\prime}}$ (c). We take  $\bm{q}=q\bm{e}_x$ in the figure. The figures are plotted in the $xy$ plane. For 3D, the $y$ and $z$ directions are symmetric.}
\label{SMdomain}
\end{figure*}

Note that the domain of ${\bm k}$ and ${\bm k}^{\prime}$ for $[\widetilde{G}^{ex}(\bm{0},\omega)^{-1}]_{{\bm k}{\bm k}^{\prime}}$ (Fig.~\ref{SMdomain}(a))
and that for $[\widetilde{G}^{ex}({\bm q},\omega)^{-1}]_{{\bm k}{\bm k}^{\prime}}$ (Fig.~\ref{SMdomain}(b)) are different from each other. To rewrite this difference into a difference in the matrix elements, we adjust the domain of 
$[\widetilde{G}^{ex}({\bm q},\omega)^{-1}]$ into the domain of $[\widetilde{G}^{ex}({\bm 0},\omega)^{-1}]$ 
by defining the following two new variables as replacements of $\bm{k}$ and ${\bm k}^{\prime}$, respectively,
\begin{align}
&\Tilde{\bm{k}} \equiv \begin{cases}
   \bm{k}+\bm{q}, \   \  & |\bm{k}+\bm{q}| > K_{F,a} \\
    \bm{k}, \   \  & |\bm{k}|< K_{F,b} \\
  \end{cases}, \nonumber \\
&\Tilde{\bm{k^{\prime}}} \equiv \begin{cases}
   \bm{k^{\prime}}+\bm{q}, \   \  & |\bm{k^{\prime}}+\bm{q}| > K_{F,a} \\
    \bm{k^{\prime}}, \   \  & |\bm{k^{\prime}}|< K_{F,b} \\
  \end{cases}.
\end{align}
In terms of the new variables, 
$[\widetilde{G}^{ex}(\bm{q},\omega)^{-1}]_{\Tilde{\bm k}\Tilde{\bm k}^{\prime}}$ 
shares the identical domain (Fig.~\ref{SMdomain}(c)) with  $[\widetilde{G}^{ex}(0,\omega)^{-1}]_{\Tilde{\bm k}\Tilde{\bm k}^{\prime}}$; (i) $|\Tilde{\bm k}|>K_{F,a}$ or $K_{F,b}>|\Tilde{\bm k}|$ 
and (ii) $|\Tilde{\bm k}^{\prime}|>K_{F,a}$ or $K_{F,b}>|\Tilde{\bm k}^{\prime}|$. Let us compare their matrix elements in the domain. 
In terms of the new variables, $\widetilde{G}^{ex,-1}
=\widetilde{G}^{ex,-1}_0+W$ is given by   
\begin{align}
& [\widetilde{G}^{ex,-1}_0(\bm{q},\omega)]_{\Tilde{\bm{k}}\Tilde{\bm{k^{\prime}}}} = \nonumber \\
 & \delta_{\Tilde{\bm{k}}\Tilde{\bm{k^{\prime}}}}
     \begin{cases}
    \omega - [\epsilon_a(\Tilde{\bm{k}}) - \epsilon_b(\Tilde{\bm{k}} - \bm{q})],            & |\Tilde{\bm{k}}| > K_{F,a} \\
    -\{\omega - [\epsilon_a(\Tilde{\bm{k}} + \bm{q}) - \epsilon_b(\Tilde{\bm{k}})]\},                      & |\Tilde{\bm{k}}| < K_{F,b}
     \end{cases} ,
\end{align}
\begin{align}
    [W]_{\Tilde{\bm{k}}\Tilde{\bm{k^{\prime}}}} &= \begin{cases}
    \frac{w(\Tilde{\bm{k}}-\Tilde{\bm{k^{\prime}}})}{\Omega} & \   \ |\Tilde{\bm{k}}| > K_{F,a},\ \ |\Tilde{\bm{k^{\prime}}}| > K_{F,a} \\
    \frac{w(\Tilde{\bm{k}}-\Tilde{\bm{k^{\prime}}})}{\Omega} & \   \ |\Tilde{\bm{k}}| < K_{F,b},\ \ |\Tilde{\bm{k^{\prime}}}| < K_{F,b} \\
    \frac{w(\Tilde{\bm{k}} - \bm{q} -\Tilde{\bm{k^{\prime}}})}{\Omega} & \   \ |\Tilde{\bm{k}}| > K_{F,a},\ \ |\Tilde{\bm{k^{\prime}}}| < K_{F,b} \\
    \frac{w(\Tilde{\bm{k}} -\Tilde{\bm{k^{\prime}}} + \bm{q})}{\Omega} & \   \ |\Tilde{\bm{k}}| < K_{F,b},\ \ |\Tilde{\bm{k^{\prime}}}| > K_{F,a} \\
     \end{cases}. 
\end{align}
In the following, we take ${\bm q}=q{\bm e}_i$ ($\bm{e}_i$ is a unit vector along $i$, $i=x,y,z$) and take $q$-derivatives of the $\widetilde{G}^{ex}(\bm{q},\omega)_{\Tilde{\bm{k}}\Tilde{\bm{k^{\prime}}}}^{-1}$ matrix around 
the $\bm{q}=0$ point.

For simplicity of the presentation, let us call the new variables 
$\Tilde{\bm k}$ and $\Tilde{\bm k}^{\prime}$ as  ${\bm k}$ and 
${\bm k}^{\prime}$. The first $q$-derivative of $\widetilde{G}^{ex}(\bm{q},\omega)_{\bm{k}\bm{k^{\prime}}}^{-1}$ 
is given by   
\begin{align}
& [{\bm F}_{i}]_{\bm{k}\bm{k^{\prime}}} \equiv 
    [\frac{\partial \widetilde{G}^{ex}(q\bm{e_i},\omega)^{-1}}{ \partial q}|_{q=0}]_{\bm{k}\bm{k^{\prime}}} =  \nonumber \\
 & \begin{cases}
    \frac{k_i}{m_b}       \   \     & |\bm{k} | > K_{F,a}, \  \ \bm{k} = \bm{k^{\prime}} \\
    \frac{k_i}{m_a}      \   \       & |\bm{k}| < K_{F,b} , \  \ \bm{k} = \bm{k^{\prime}} \\
    0                     \   \    &  |\bm{k}| < K_{F,b}, \  \ |\bm{k^{\prime}}| < K_{F,b}, \  \ \bm{k} \ne \bm{k^{\prime}} \\
    0                     \   \    &  |\bm{k} | > K_{F,a}, \  \ |\bm{k^{\prime}} | > K_{F,a}, \  \ \bm{k} \ne \bm{k^{\prime}} \\
    \frac{1}{\Omega} \frac{\partial w(\bm{k} - \bm{k^{\prime}} + q\bm{e_i})}{\partial q}|_{q=0} \   \ &|\bm{k}| < K_{F,b}, \  \  |\bm{k^{\prime}} | > K_{F,a} \\
    \frac{1}{\Omega} \frac{\partial w(\bm{k} - q\bm{e_i} - \bm{k^{\prime}} )}{\partial q}|_{q=0} \   \ &|\bm{k} | > K_{F,a}, \   \ |\bm{k^{\prime}}| < K_{F,b} \\
     \end{cases},
\end{align}
with 
\begin{align}
  & \frac{\partial w(\bm{k} - \bm{k^{\prime}} + q\bm{e_i})}{\partial q}|_{q=0} = \nonumber \\
   &  \hspace{1cm} \begin{cases}
     -\frac{4\pi \times 2(k_i - k^{\prime}_i)}{(|\bm{k}-\bm{k^{\prime}}|^2 + k_{TF}^2)^2}  \   \ & \text{in 3D,} \\
     -\frac{2\pi}{(|\bm{k}-\bm{k^{\prime}}| + k_{TF})^2} \frac{k_i - k^{\prime}_i}{|\bm{k}-\bm{k^{\prime}}|} \   \ & \text{in 2D.} \\
    \end{cases} \label{dw/dq}
\end{align}
The second $q$-derivative of 
$\widetilde{G}^{ex}(\bm{q},\omega)_{\bm{k}\bm{k^{\prime}}}^{-1}$ 
is given by 
\begin{align}
& 2[{\bm S}_i]_{\bm{k}\bm{k^{\prime}}} \equiv 
    [\frac{\partial^2 \widetilde{G}^{ex}(q\bm{e_i},\omega)^{-1}}{ \partial q^2}|_{q=0}]_{\bm{k}\bm{k^{\prime}}} =\nonumber \\
& \begin{cases}
    -\frac{1}{m_b}       \   \     & |\bm{k} | > K_{F,a}, \  \ \bm{k} = \bm{k^{\prime}} \\
    \frac{1}{m_a}      \   \       & |\bm{k}| < K_{F,b} , \  \ \bm{k} = \bm{k^{\prime}} \\
    0                     \   \    &  |\bm{k}| < K_{F,b}, \  \ |\bm{k^{\prime}}| < K_{F,b}, \  \ \bm{k} \ne \bm{k^{\prime}} \\
    0                     \   \    &  |\bm{k} | > K_{F,a}, \  \ |\bm{k^{\prime}} | > K_{F,a}, \  \ \bm{k} \ne \bm{k^{\prime}} \\
    \frac{1}{\Omega} \frac{\partial^2 w(\bm{k} - \bm{k^{\prime}} + q\bm{e_i})}{\partial q^2}|_{q=0} \   \ &|\bm{k}| < K_{F,b}, \  \  |\bm{k^{\prime}} | > K_{F,a} \\
    \frac{1}{\Omega} \frac{\partial^2 w(\bm{k} - q\bm{e_i} - \bm{k^{\prime}} )}{\partial q^2}|_{q=0} \   \ &|\bm{k} | > K_{F,a}, \   \ |\bm{k^{\prime}}| < K_{F,b} \\
     \end{cases}, 
\end{align}
with 
\begin{align}
    \frac{\partial^2 w(\bm{k} - \bm{k^{\prime}} + q\bm{e_i})}{\partial q^2}|_{q=0} =& 
    -\frac{8\pi}{(|\bm{k}-\bm{k^{\prime}}|^2 + k_{TF}^2)^2} \nonumber \\
& + \frac{32\pi(k_i - k^{\prime}_i)^2 }{(|\bm{k}-\bm{k^{\prime}}|^2 + k_{TF}^2)^3}  \label{d^2w/dq^2-3d}
\end{align}
in 3D and 
\begin{align}
&\frac{\partial^2 w(\bm{k} - \bm{k^{\prime}} + q\bm{e_i})}{\partial q^2}|_{q=0} = 
     -\frac{2\pi}{(|\bm{k}-\bm{k^{\prime}}| + k_{TF})^2}\frac{1}{|\bm{k}-\bm{k^{\prime}}|} \nonumber \\
& \hspace{0.5cm} + \frac{4\pi}{(|\bm{k}-\bm{k^{\prime}}|+k_{TF})^3}\frac{(k_i - k^{\prime}_i)^2}{|\bm{k}-\bm{k^{\prime}}|^2} \nonumber \\
&  \hspace{2cm} + \frac{2\pi}{(|\bm{k}-\bm{k^{\prime}}| + k_{TF})^2} \frac{(k_i - k^{\prime}_i)^2}{|\bm{k}-\bm{k^{\prime}}|^3} 
\label{d^2w/dq^2-2d}
\end{align}
in 2D. 
Now that $\widetilde{G}^{ex}(q\bm{e_i},\omega)^{-1}$ is expanded in $q$,
\begin{align}
    [\widetilde{G}^{ex}(q\bm{e_i},\omega)^{-1}] = 
    [\widetilde{G}^{ex}(\bm{0},\omega)^{-1}] +  
    [{\bm F}_i] \!\ q + [{\bm S}_i] \!\ q^2 + {\cal O}(q^3),  
    \label{G_expand}
\end{align}
the second-order perturbation theory gives the second-order energy correction,
\begin{align}
     \xi_j(q\bm{e}_i,\omega) = \xi_j({\bm 0},\omega) + b_j(\omega) \!\ q^2 + {\cal O}(q^3). 
\end{align}
Here $b_{j}(\omega)$ is given by the 
eigenvectors and eigenvalues at ${\bm q}=0$ (See Eqs.~(\ref{bs5-0},\ref{from-sum-to-integral})), 
\begin{align}
  &  b_j(\omega) = \nonumber \\
& \sum_{j^{\prime} \ne j} \frac{\langle \phi_{j}(\bm{0},\omega) | \bm{F_i} |\phi_{j^{\prime}}(\bm{0},\omega)\rangle \langle \phi_{j^{\prime}}(\bm{0},\omega) | \bm{F_i} |\phi_{j}(\bm{0},\omega)\rangle}{\xi_j(\bm{0},\omega) - \xi_{j^{\prime}}(\bm{0},\omega)} \nonumber \\
& \ \ \ + \langle \phi_{j}(\bm{0},\omega) | \bm{S_i} |\phi_{j}(\bm{0},\omega)\rangle \equiv F_2 + S_1. \label{f2s1}
\end{align}
The previous section describes how to calculate the 
eigenvectors and eigenvalues of the lowest-energy $s$-wave exciton and 
antiexciton at ${\bm q}=0$ 
in the 3D case ($j \equiv (nlm)=(100)$) and 2D case ($j=(nm)=(10)$). In the following, 
we will describe how to calculate $b_{(100)}(\omega=\pm \omega_{\pm})$ 
in the 3D case and $b_{(10)}(\omega=\pm \omega_{\pm})$ in the 2D case.

\subsubsection{3D case} 
Consider the 3D case with $j \equiv (nlm)$ and take $(nlm)=(100)$ and 
$\bm{e_i} = \bm{e_z}$. Since $\xi_{j}({\bm q}={\bm 0},\omega=\pm \omega_{\pm})=0$, $F_2$ is 
given by 
\begin{align}
    F_2 = & - \sum_{nlm \ne (100)} \sum_{{\bm k}_1,{\bm k}_2,{\bm k}_3,{\bm k}_4} \xi^{-1}_{nl}({\bm 0},\omega) 
\times \nonumber \\
& \langle \phi_{100}(\bm{0},\omega) |{\bm k}_1\rangle [\bm{F_z}]_{{\bm k}_1{\bm k}_2} \langle {\bm k}_2|\phi_{nlm}(\bm{0},\omega)\rangle \times \nonumber \\
&  \langle \phi_{nlm}(\bm{0},\omega) |{\bm k}_3\rangle [\bm{F_z}]_{{\bm k}_3{\bm k}_4} \langle{\bm k}_4|\phi_{100}(\bm{0},\omega)\rangle,  
\end{align}
at $\omega= \pm \omega_{\pm}$. 
In terms of Eq.~(\ref{from-sum-to-integral}) and $Y_{00}(\theta,\varphi)=\frac{1}{2\sqrt{\pi}}$, we evaluate matrix elements of ${\bm F}_z$ by three-dimensional momentum integrals;
\begin{align}
&\sum_{{\bm k}_1,{\bm k}_2} \langle \phi_{100}(\bm{0},\omega) |{\bm k}_1\rangle [\bm{F_z}]_{{\bm k}_1{\bm k}_2} \langle{\bm k}_2|\phi_{nlm}(\bm{0},\omega)\rangle =  \nonumber \\
& \ \  \Big( \prod_{i=1}^{2} \int_0^{+\infty} \frac{dk_i}{(2\pi)^3}\ \ k_i^2 \int_{-1}^1 d \cos \theta_i \int_0^{2\pi} d\varphi_i  \Big)   \nonumber \\
&  \hspace{1cm} f_{10}(\omega;k_1) \frac{1}{2\sqrt{\pi}} \!\  \Omega[\bm{F_z}]_{\bm{k_1}\bm{k_2}} \!\    f_{nl}(\omega;k_2)Y_{lm}(\theta_2,\varphi_2), \label{select} 
\end{align}
with
\begin{align}
& \Omega[\bm{F_z}]_{\bm{k_1}\bm{k_2}}  = \nonumber \\
& \Big( \theta(k_1-K_{F,a}) \frac{k_1\cos \theta_1}{m_b} + \theta(K_{F,b}-k_1)\frac{k_1\cos \theta_1}{m_a} \Big) 
\nonumber \\  
& \times \frac{(2\pi)^3}{k^2_1} \delta(k_1-k_2) \sum_{lm} Y_{lm}(\theta_1,\varphi_1)Y^{*}_{lm}(\theta_2, \varphi_2)  \nonumber \\
& \hspace{0.4cm} + 2\sum_{lm} b_l(k_1,k_2) \frac{4\pi}{2l+1} Y_{lm}(\theta_1,\varphi_1)  Y_{lm}^*(\theta_2,\varphi_2) 
\nonumber \\
& \Big( \theta(k_1 - K_{F,a})\theta(K_{F,b} - k_2) (k_1\cos \theta_1 - k_2\cos \theta_2) \nonumber \\
& \ \  + \theta(k_2 - K_{F,a})\theta(K_{F,b} - k_1) (k_2\cos \theta_2 - k_1\cos \theta_1) \Big). \label{select2}
\end{align}
Here ${\bm k}_j \equiv k_j(\sin\theta_j\cos\varphi_j,\sin\theta_j\sin\varphi_j,\cos\theta_j)$ ($j=1,2,3,4$). We used the spherical expansion of Eq.~(\ref{dw/dq});  
\begin{align}
   & \frac{4\pi }{(|\bm{k_1}-\bm{k_2}|^2 + k_{TF}^2)^2} = \sum_l b_l(k_1,k_2)P_l (\cos \gamma) \nonumber \\
&\ \ = \sum_{lm} b_l(k_1,k_2) \frac{4\pi}{2l+1} Y_{lm}(\theta_1,\varphi_1)  Y_{lm}^*(\theta_2,\varphi_2).
\end{align}
The lowest and second lowest expansion coefficients are calculated as 
\begin{align}
    b_0(k_1, k_2) = & \frac{\pi}{k_1 k_2} \Big( \frac{1}{(k_1 - k_2)^2 + k_{TF}^2} - \frac{1}{(k_1 + k_2)^2 + k_{TF}^2} \Big), \\
    b_1(k_1, k_2) =& \frac{3\pi}{2 (k_1 k_2)^2} \Big\{ \ln [\frac{(k_1 - k_2)^2 + k_{TF}^2}{(k_1 + k_2)^2 + k_{TF}^2}] \nonumber \\
& + (k_1^2 + k_2^2 + k_{TF}^2)[\frac{1}{(k_1 - k_2)^2 + k_{TF}^2} \nonumber \\
& \ \  - \frac{1}{(k_1 + k_2)^2 + k_{TF}^2}]\Big\}. 
\end{align}
After the momentum integrals in Eq.~(\ref{select}), 
only $(lm)=(10)$ term remains finite in Eq.~(\ref{select}). 
With $\cos\theta = 2\sqrt{\frac{\pi}{3}} Y_{10}^{*}(\theta, \varphi)$, 
the non-zero term is evaluated as
\begin{align}
 & \langle \phi_{100}(\bm{0},\omega) | \bm{F_z} |\phi_{nlm}(\bm{0},\omega)\rangle 
=  \delta_{l1}\delta_{m0} \frac{4}{\sqrt{3}} \times \nonumber \\
 & \Big( \prod_{i=1}^{2} \int_0^{+\infty} \frac{dk_i}{(2\pi)^3}\ \ k_i^2 \Big) 
f_{10}(\omega;k_1)  \rho(k_1,k_2)   f_{n1}(\omega;k_2),
\end{align}
with 
\begin{align}
   & \rho(k_1,k_2) = \Big( \frac{\theta(k_1-K_{F,a})}{m_b} + \frac{\theta(K_{F,b}-k_1)}{m_a} \Big) (2\pi)^3\frac{\delta(k_1-k_2)}{k_1} \nonumber \\
                 &+ 8\pi \Big( \theta(k_1 - K_{F,a})\theta(K_{F,b} - k_2) (\frac{1}{3} b_1(k_1,k_2)k_1 - b_0(k_1,k_2) k_2) \nonumber \\
                 &+ \theta(k_2 - K_{F,a})\theta(K_{F,b} - k_1) ( b_0(k_1,k_2)k_2-\frac{1}{3} b_1(k_1,k_2)k_1) \Big).
\end{align}
Then we finally have
\begin{align}
     & F_2 =  -\frac{1}{3}\Big( \prod_{i=1}^{4} \int_0^{+\infty} \frac{dk_i}{(2\pi)^3}\ \ k_i^2 \Big) 
\sum_n \frac{1}{ \xi_{n1}(\bm{0},\omega)}f_{10}(\omega;k_1)  \nonumber \\
&      \rho(k_1,k_2) f_{n1}(\omega;k_2) f_{n1}(\omega;k_3) \rho(k_3,k_4)  f_{10}(\omega;k_4). \label{3D-F2} 
\end{align}

To numerically evaluate Eq.~(\ref{3D-F2}), we use 
the same discretization of $k$ as in 
Eq.~(\ref{discrete}),
\begin{align}
F_2 &=  -\frac{1}{3} \sum_{k_1,k_2,k_3,k_4,n}    \frac{V^{\omega}_{10,k_1}P_{k_1 k_2}  V^{\omega}_{n1,k_2} V^{\omega}_{n1,k_3}  P_{k_3 k_4} V^{\omega}_{10,k_4}}{ \xi_{n1}(\bm{0},\omega)}   \nonumber \\
&= -\frac{1}{3}  \sum_{k_1,k_2,k_3,k_4}    V^{\omega}_{10,k_1}P_{k_1 k_2}  H^{\omega,-1}_{1,k_2k_3}  
P_{k_3 k_4} V^{\omega}_{10,k_4}.  
        \label{numF2}
\end{align}
$V^{\omega}_{nl,k}$ and $H^{\omega}_{1,kk^{\prime}}$ are 
defined in Eq.~(\ref{HandV}) and 
\begin{align}
    & P_{k_1 k_2} = \frac{k_1 k_2}{L(2\pi)^2} \rho(k_1,k_2) = \nonumber \\
    & \Big( \theta(k_1-K_{F,a}) \frac{k_1}{m_b} + \theta(K_{F,b}-k_1)\frac{k_1}{m_a} \Big) \delta_{k_1 k_2} 
+ \frac{8\pi k_1 k_2}{L(2\pi)^2} \times \nonumber \\
 & \Big( \theta(k_1 - K_{F,a})\theta(K_{F,b} - k_2) (\frac{1}{3} b_1(k_1,k_2)k_1 - b_0(k_1,k_2) k_2) \nonumber \\
& + \theta(k_2 - K_{F,a})\theta(K_{F,b} - k_1) (b_0(k_1,k_2)k_2-\frac{1}{3} b_1(k_1,k_2)k_1) \Big).
\end{align}

$S_1$ in Eq.~(\ref{f2s1}) is given by
\begin{align}
    & S_1 = \sum_{{\bm k}_1,{\bm k}_2} 
    \langle \phi_{100}(\bm{0},\omega) |{\bm k}_1\rangle 
    [\bm{S_z}]_{{\bm k}_1{\bm k}_2} \langle {\bm k}_2|\phi_{100}(\bm{0},\omega)\rangle \nonumber \\ 
   & =  \frac{1}{3}\sum_{{\bm k}_1,{\bm k}_2}
        \langle \phi_{100}(\bm{0},\omega) |{\bm k}_1\rangle 
    [\bm{S_z}+{\bm S_x} +{\bm S_y}]_{{\bm k}_1{\bm k}_2} \langle {\bm k}_2|\phi_{100}(\bm{0},\omega)\rangle  \nonumber \\
        &=  \Big( \prod_{i=1}^{2} \int_0^{+\infty} \frac{dk_i}{(2\pi)^3}\ \ k_i^2 \int_{-1}^1 d \cos \theta_i \int_0^{2\pi} d\varphi_i  \Big)  \nonumber \\
& \hspace{1cm} \frac{1}{4\pi}  f_{10}(\omega;k_1)  \Omega\bm{S}(\bm{k_1},\bm{k_2})   f_{10}(\omega;k_2) , 
        \label{S1}
\end{align}
with 
\begin{align}
& \Omega\bm{S}(\bm{k_1},\bm{k_2})  = \Big( - \frac{\theta(k_1-K_{F,a}) }{2m_b} + \frac{\theta(K_{F,b}-k_1)}{2m_a} \Big) \nonumber \\ 
& \ \ \times \frac{(2\pi)^3}{k^2_1} \delta(k_1-k_2) \sum_{lm} Y_{lm}(\theta_1,\varphi_1)Y^{*}_{lm}(\theta_2, \varphi_2) \nonumber \\
& + \sum_{lm} \beta_l(k_1,k_2) \frac{4\pi}{2l+1} Y_{lm}(\theta_1,\varphi_1)  Y_{lm}^*(\theta_2,\varphi_2) \nonumber \\
& \ \ \Big( \theta(k_1 - K_{F,a})\theta(K_{F,b} - k_2) 
+ \theta(k_2 - K_{F,a})\theta(K_{F,b} - k_1)\Big). \label{S1a}
\end{align}
Here we used the spherical expansion of Eq.~(\ref{d^2w/dq^2-3d}); 
\begin{align}
   & -\frac{4\pi }{(|\bm{k_1}-\bm{k_2}|^2 + k_{TF}^2)^2} + \frac{4\pi\times 4|\bm{k_1}-\bm{k_2}|^2/3}{(|\bm{k_1}-\bm{k_2}|^2 + k_{TF}^2)^3}  \nonumber \\
   & = \sum_{lm} \beta_l(k_1,k_2) \frac{4\pi}{2l+1} Y_{lm}(\theta_1,\varphi_1)  Y_{lm}^*(\theta_2,\varphi_2).
\end{align}
The lowest order expansion coefficient is calculated as 
\begin{align}
 &\beta_0(k_1, k_2) = \frac{\pi}{3k_1 k_2}\Big( \frac{1}{(k_1 - k_2)^2 + k_{TF}^2} - \frac{1}{(k_1 + k_2)^2 + k_{TF}^2}\Big) \nonumber \\
&  - \frac{2\pi k_{TF}^2}{3k_1 k_2} \Big( \frac{1}{((k_1 - k_2)^2 + k_{TF}^2)^2} - \frac{1}{((k_1 + k_2)^2 + k_{TF}^2)^2}\Big). 
\end{align}
After the integration in Eq.~(\ref{S1}), only the $(lm) = (00)$ term in Eq.~(\ref{S1a}) remains finite;
\begin{align}
    S_1 &=  \Big( \prod_{i=1}^{2} \int_0^{+\infty} \frac{dk_i}{(2\pi)^3}\ \ k_i^2  \Big) f_{10}(\omega;k_1)   c(k_1, k_2)  f_{10}(\omega;k_2), \label{S1b}
\end{align}
with 
\begin{align}
&     c(k_1, k_2)  = \Big( -\frac{\theta(k_1-K_{F,a})}{2m_b} + \frac{\theta(K_{F,b}-k_1)}{2m_a} \Big) \times \nonumber \\
&(2\pi)^3\frac{\delta(k_1-k_2)}{k_1^2} +  \beta_0(k_1,k_2) (4\pi) \times \nonumber \\
&\ \  \Big( \theta(k_1 - K_{F,a})\theta(K_{F,b} - k_2) + \theta(k_2 - K_{F,a})\theta(K_{F,b} - k_1)  \Big). \label{S1c}
\end{align}
To numerically evaluate Eq.~(\ref{S1b}), we discretize 
$k_1$ and $k_2$ as in Eq.~(\ref{discrete}),
\begin{align}
S_1 &=  \sum_{k_1,k_2} V^{\omega}_{10,k_1} C_{k_1k_2}  V^{\omega}_{10,k_2}, \label{numS1}
\end{align}
with
\begin{align}
 & C_{k_1 k_2} \equiv \frac{k_1 k_2}{L(2\pi)^2} c(k_1,k_2) \nonumber \\
& = \Big( -\frac{\theta(k_1-K_{F,a})}{2m_b} + \frac{\theta(K_{F,b}-k_1)}{2m_a} \Big) \delta_{k_1 k_2} \nonumber \\
                 & \ \  + \frac{k_1 k_2}{L(2\pi)^2}  \beta_0(k_1,k_2) (4\pi)\times \nonumber \\
& \ \ \   \Big( \theta(k_1 - K_{F,a})\theta(K_{F,b} - k_2) + \theta(k_2 - K_{F,a})\theta(K_{F,b} - k_1)  \Big).
\end{align}

In summary, $b_j(\omega)$ is given by 
Eqs.~(\ref{f2s1},\ref{numF2},\ref{numS1}) at $\omega=\pm \omega_{\pm}$. The energy-band curvatures of 
the lowest $s$-wave ($j=(100)$) exciton and antiexciton bands at the $\Gamma$ point can be determined by $b_j(+\omega_+)$ and $b_{j}(-\omega_{-})$ respectively;
\begin{widetext}
\begin{align}
 \left\{\begin{array}{ll}
 \xi_{j}({\bm q},\omega_{+}(\bm{q})) = (\omega_{+}(\bm{q})-\omega_{+}) 
 \Big[\frac{d\xi_{j}({\bm 0},\omega)}{d\omega}\Big]_{|\omega=\omega_+} + 
 b_j(\omega_{+}) q^2 = 0, &  {\rm for}\!\ \!\ {\rm exciton,} \\ 
 \xi_{j}({\bm q},-\omega_{-}(\bm{q})) =  (-\omega_{-}(\bm{q})+\omega_{-}) 
 \Big[\frac{d\xi_{j}({\bm 0},\omega)}{d\omega}\Big]_{|\omega=-\omega_-} + 
 b_j(-\omega_{-}) q^2 = 0, & {\rm for} \!\ \!\ {\rm antiexciton.} \\
 \end{array}\right.  \label{mass1}
\end{align} 
From this, we obtain the energy-band curvatures of
the exciton and antiexciton around the $\Gamma$ point,  
\begin{align}
 \left\{\begin{array}{ll}
 \omega_{+}(\bm{q})=\omega_{+} -  
 b_j(\omega_{+}) \!\ \Big[\frac{d\xi_{j}({\bm 0},\omega)}{d\omega}\Big]^{-1}_{|\omega=\omega_+}
 q^2+ \cdots, &  {\rm for}\!\ \!\ {\rm exciton,} \\ 
 \omega_{-}(\bm{q}) = \omega_{-} + 
 b_j(-\omega_{-}) \!\ \Big[\frac{d\xi_{j}({\bm 0},\omega)}{d\omega}\Big]^{-1}_{|\omega=-\omega_-}
  q^2 + \cdots, & {\rm for} \!\ \!\ {\rm antiexciton.} \\
 \end{array}\right. \label{mass2}
\end{align}
\end{widetext}

\subsubsection{2D case}
Consider the 2D case with $j=(nm)$ and take $(nm)=(10)$ and  
$\bm{e_i} = \bm{e_x}$. Since $\xi_{j}({\bm 0},\pm\omega_{\pm})=0$, $F_2$ in Eq.~(\ref{f2s1}) is given by
\begin{align}
    & F_2 = -\sum_{nm \ne 10} \sum_{{\bm k}_1,{\bm k}_2,{\bm k}_3,{\bm k}_4} 
\xi^{-1}_{nm}(\bm{0},\omega) \nonumber \\
&\ \  \langle \phi_{10}(\bm{0},\omega) |{\bm k}_1\rangle [\bm{F_x}]_{{\bm k}_1{\bm k}_2} \langle{\bm k}_2|\phi_{nm}(\bm{0},\omega)\rangle \nonumber \\
& \ \ \ \  \langle \phi_{nm}(\bm{0},\omega) |{\bm k}_3\rangle 
    [\bm{F_x}]_{{\bm k}_3{\bm k}_4} \langle {\bm k}_4|\phi_{10}(\bm{0},\omega)\rangle. \label{F2-2D}  
\end{align}
In terms of Eq.~(\ref{from-sum-to-integral}), we evaluate the matrix elements in $F_2$ by the 
2D momentum integrals;
\begin{align}
 & \sum_{{\bm k}_1,{\bm k}_2}\langle \phi_{10}(\bm{0},\omega) |{\bm k}_1\rangle [\bm{F_x}]_{{\bm k}_1{\bm k}_2} \langle {\bm k}_2|\phi_{nm}(\bm{0},\omega)\rangle \nonumber \\
& =  \Big( \prod_{i=1}^{2} \int_0^{+\infty} \frac{k_i dk_i}{2\pi} \int_0^{2\pi} \frac{d\varphi_i}{2\pi}  \Big) \nonumber \\
&\hspace{1cm} f_{10}(\omega;k_1)  \Omega\bm{F_x}(\bm{k_1},\bm{k_2})  f_{nm}(\omega;k_2)e^{im\varphi_2}. \label{F2-2D-c}
\end{align}
Eq.~(\ref{dw/dq}) is decomposed in terms of the trigonometric functions, 
\begin{align}
    \frac{2\pi}{(|\bm{k_1}-\bm{k_2}| + k_{TF})^2} \frac{1}{|\bm{k_1}-\bm{k_2}|} &= \sum_m J_m(k_1,k_2)e^{im(\varphi_1-\varphi_2)}, \nonumber 
\end{align}
with
\begin{align}
& J_m(k_1,k_2) =  \int _0^{2\pi} \frac{d(\varphi_1-\varphi_2)}{2\pi} e^{-im(\varphi_1-\varphi_2)} \nonumber \\
& \ \ \frac{2\pi}{(\sqrt{k_1^2 + k_2^2 -2k_1 k_2 \cos(\varphi_1 -\varphi_2)} + k_{TF})^2}  \nonumber \\
& \ \ \ \frac{1}{\sqrt{k_1^2 + k_2^2 -2k_1 k_2 \cos(\varphi_1 -\varphi_2)}}.
\end{align}
Eq.~(\ref{F2-2D}) is calculated in terms of the expansion, 
\begin{align}
    &\langle \phi_{10}(\bm{0},\omega) | \bm{F_x} |\phi_{nm}(\bm{0},\omega)\rangle =  \nonumber \\
    &\Big( \prod_{i=1}^{2} \int_0^{+\infty} \frac{dk_i}{2\pi}\ \ k_i \int_0^{2\pi} \frac{d\varphi_i}{2\pi}  \Big) f_{10}(\omega;k_1) f_{nm}(\omega;k_2)e^{im\varphi_2}  \nonumber\\
    &\  \ \bigg\{ \Big(\frac{\theta(k_1 - K_{F,a})}{m_b} + \frac{\theta(K_{F,b}-k_1)}{m_a}\Big) \nonumber \\
& \ \  \times \pi\delta\left(k_1-k_2\right)(e^{i\varphi_1}+e^{-i\varphi_1})\sum_{m^{\prime}} e^{i m^{\prime}\left(\varphi_1-\varphi_2\right)} \nonumber \\
    &+\theta(k_1 - K_{F,a})\theta(K_{F,b} - k_2)\Big(\frac{k_1 (e^{i\varphi_1}+e^{-i\varphi_1})}{2} \nonumber \\
& \hspace{0.3cm} - \frac{k_2 (e^{i\varphi_2}+e^{-i\varphi_2})}{2}\Big) \sum_{m^{\prime}} J_{m^{\prime}}(k_1,k_2)e^{im^{\prime}\varphi_1}e^{-im^{\prime}\varphi_2}  \nonumber\\
    &+\theta(k_2 - K_{F,a})\theta(K_{F,b} - k_1)\Big(\frac{k_2 (e^{i\varphi_2}+e^{-i\varphi_2})}{2}  \nonumber \\
& \hspace{0.3cm} 
 - \frac{k_1 (e^{i\varphi_1}+e^{-i\varphi_1})}{2}\Big) \sum_{m^{\prime}} J_{m^{\prime}}(k_1,k_2)e^{im^{\prime}\varphi_1}e^{-im^{\prime}\varphi_2}\bigg\}. \label{F2-2D-d}
\end{align}
After the momentum integration in Eq.~(\ref{F2-2D-c}), only $m = \pm1$ terms remain finite in Eq.~(\ref{F2-2D-c}). Since $f_{nm}(\omega;k)=f_{n(-m)}(\omega;k)$, $\xi_{nm}(\omega;k)=\xi_{n(-m)}(\omega;k)$, and $J_{m}(k,k^{\prime})=J_{-m}(k,k^{\prime})$, the matrix elements for $m=+1$ and those for $m=-1$ 
are the same,
\begin{align}
&  \langle \phi_{10}(\bm{0},\omega) | \bm{F_x} |\phi_{n(\pm 1)}(\bm{0},\omega)\rangle   
    = \Big( \prod_{i=1}^{2} \int_0^{+\infty} \frac{dk_i}{2\pi}\ \ k_i  \Big) \nonumber  \\ 
& \hspace{2.5cm}  f_{10}(\omega;k_1) \tau(k_1, k_2)f_{n1}(\omega;k_2),
\end{align}
with
\begin{align}
 & \tau(k_1, k_2) = \Big( \frac{\theta(k_1 - K_{F,a})}{m_b} + \frac{\theta(K_{F,b}-k_1)}{m_a} \Big) 
\pi\delta\left(k_1-k_2\right) \nonumber\\
  &  +\Big( \theta(k_1 - K_{F,a})\theta(K_{F,b} - k_2) + \theta(k_2 - K_{F,a})\theta(K_{F,b} - k_1) \Big) \nonumber \\
 & \ \ \ \times \Big(\frac{k_1}{2} J_1(k_1,k_2) - \frac{k_2}{2}J_0(k_1,k_2)\Big). 
\end{align}
Then we finally have 
\begin{align}
    &  F_2 =  -2\Big( \prod_{i=1}^{4} \int_0^{+\infty} \frac{dk_i}{2\pi}\ \ k_i \Big) \sum_n \xi^{-1}_{n1}(\bm{0},\omega) 
f_{10}(\omega;k_1) \times \nonumber \\
&  \tau(k_1,k_2) f_{n1}(\omega;k_2) f_{n1}(\omega;k_3) \tau(k_3,k_4)  f_{10}(\omega;k_4). \label{F2-2D-a} 
\end{align}
To evaluate Eq.~(\ref{F2-2D-a}) numerically, we discretize $k_i$ ($i=1,2,3,4$) 
as in Eq.~(\ref{discrete}),  
\begin{align}
   &  F_2 =  -2\sum_{k_1,k_2,k_3,k_4,n}  \frac{V^{\omega}_{10,k_1}T_{k_1 k_2}  V^{\omega}_{n1,k_2} V^{\omega}_{n1,k_3}  T_{k_3 k_4} V^{\omega}_{10,k_4} }{ \xi_{n1}(\bm{0},\omega)} \nonumber \\ 
& \ \ =-2\sum_{k_1,k_2,k_3,k_4}
        V^{\omega}_{10,k_1}  T_{k_1 k_2} H^{\omega,-1}_{1,k_2k_3} T_{k_3k_4} V^{\omega}_{10,k_4}.  
        \label{numF2_2D}
\end{align}
Here $V^{\omega}_{nm,k}$ and $H^{\omega}_{m,kk^{\prime}}$ are defined in Eq.~(\ref{HandV-2d}) and
\begin{align}
    & T_{k_1 k_2} \equiv \frac{\sqrt{k_1 k_2}}{L} \tau(k_1,k_2) \nonumber \\
& = k_1 \Big(  \frac{\theta(k_1-K_{F,a})}{2m_b} + \frac{\theta(K_{F,b}-k_1)}{2m_a} \Big) \delta_{k_1 k_2} 
+ \frac{\sqrt{k_1 k_2}}{L} \times \nonumber \\
& \!\ \Big( \theta(k_1 - K_{F,a})\theta(K_{F,b} - k_2) + \theta(k_2 - K_{F,a})\theta(K_{F,b} - k_1) \Big) \nonumber \\
& \times \Big(\frac{k_1}{2} J_1(k_1,k_2) - \frac{k_2}{2}J_0(k_1,k_2)\Big). \nonumber 
\end{align}

$S_1$ in Eq.~(\ref{f2s1}) is given by the 2D momentum integrals, 
\begin{align}
  &  S_1 = \sum_{{\bm k}_1,{\bm k}_2}
    \langle \phi_{10}(\bm{0},\omega) |{\bm k}_1\rangle [\bm{S_x}]_{{\bm k}_1{\bm k}_2} \langle {\bm k}_2|\phi_{10}(\bm{0},\omega)\rangle \nonumber \\
& = \frac{1}{2} \sum_{{\bm k}_1,{\bm k}_2}
    \langle \phi_{10}(\bm{0},\omega) |{\bm k}_1\rangle [\bm{S_x} + \bm{S_y}]_{{\bm k}_1{\bm k}_2} \langle {\bm k}_2|\phi_{10}(\bm{0},\omega)\rangle  \nonumber \\
        &=  \Big( \prod_{i=1}^{2} \int_0^{+\infty} \frac{dk_i}{2\pi}\ \ k_i  \int_0^{2\pi} \frac{d\varphi_i}{2\pi}  \Big)f_{10}(\omega;k_1) \nonumber \\
& \hspace{2cm} \Omega\bm{S}(\bm{k_1},\bm{k_2}) f_{10}(\omega;k_2).  
        \label{S1-2D}
\end{align}
$\Omega\bm{S} = \frac{\Omega}{2}(\bm{S_x} + \bm{S_y})$ is further expanded 
in terms of the Fourier series, 
\begin{align}
\Omega\bm{S}(\bm{k_1},\bm{k_2}) & = \Big(-\frac{\theta(k_1-K_{F,a})}{m_b} + \frac{\theta(K_{F,b}-k_1)}{m_a} \Big) 
\nonumber \\
 & \times \frac{\pi\delta(k_1-k_2)}{k_1} \sum_{m} e^{i m\left(\varphi_1-\varphi_2\right)} +  \nonumber \\
& \hspace{-2cm}  \Big( \theta(k_1 - K_{F,a})\theta(K_{F,b} - k_2)
                                 + \theta(k_2 - K_{F,a})\theta(K_{F,b} - k_1)  \Big) \nonumber \\
& \times \sum_{m} K_m(k_1,k_2) e^{i m\left(\varphi_1-\varphi_2\right)}, \label{S-matrix}
\end{align}
with Fourier coefficients
\begin{align}
    &K_m(k_1,k_2) = \frac{1}{2}\int _0^{2\pi} 
    \frac{d(\varphi_1-\varphi_2)}{2\pi} \ \  e^{-im(\varphi_1-\varphi_2)} \nonumber \\
    & \bigg\{ \frac{2\pi}{(\sqrt{k_1^2 + k_2^2 -2k_1 k_2 \cos(\varphi_1 -\varphi_2)} + k_{TF})^3} \nonumber\\
    &\ \ -\frac{\pi}{(\sqrt{k_1^2 + k_2^2 -2k_1 k_2 \cos(\varphi_1 -\varphi_2)} + k_{TF})^2} \nonumber \\
& \ \ \ \times \frac{1}{\sqrt{k_1^2 + k_2^2 -2k_1 k_2 \cos(\varphi_1 -\varphi_2)}} \bigg\}. 
\end{align}
After the integrals over $\varphi_1$ and $\varphi_2$ in Eq.~(\ref{S1-2D}), 
only $m=0$ term in Eq.~(\ref{S-matrix}) remains finite, 
\begin{align}
    S_1 &=  \Big( \prod_{i=1}^{2} \int_0^{+\infty} \frac{dk_i}{2\pi}\ \ k_i  \Big) f_{10}(\omega;k_1)   \chi(k_1, k_2)  f_{10}(\omega;k_2), \label{S1-2D-a}
\end{align}
with 
\begin{align}
    & \chi(k_1, k_2)  = \nonumber \\
& \Big( -\frac{\theta(k_1-K_{F,a})}{m_b} + \frac{\theta(K_{F,b}-k_1)}{m_a} \Big) \frac{\pi\delta(k_1-k_2)}{k_1} + \nonumber  \\
     &\Big( \theta(k_1 - K_{F,a})\theta(K_{F,b} - k_2) + \theta(k_2 - K_{F,a})\theta(K_{F,b} - k_1)  \Big) \nonumber \\
& \times K_0(k_1,k_2).  
\end{align}
To evaluate Eq.~(\ref{S1-2D-a}) numerically, we put it 
in a discrete form 
\begin{align}
    S_1 &=   \sum_{k_1,k_2}  V^{\omega}_{10,k_1} X_{k_1k_2}  V^{\omega}_{10,k_2}, \label{numS1_2D}
\end{align}
where $V^{\omega}_{nm,k}$ are defined in Eq.~(\ref{HandV-2d}) and
\begin{align}
 &X_{k_1 k_2} = \frac{\sqrt{k_1 k_2}}{L} \chi(k_1,k_2) \nonumber \\ 
& =  \Big( -\frac{\theta(k_1-K_{F,a})}{2m_b} +\frac{\theta(K_{F,b}-k_1)}{2m_a} \Big) \delta_{k_1 k_2} 
+  \frac{\sqrt{k_1 k_2}}{L} \nonumber  \\
     & \!\ \times \Big( \theta(k_1 - K_{F,a})\theta(K_{F,b} - k_2)
         + \theta(k_2 - K_{F,a})\theta(K_{F,b} - k_1)  \Big)  \nonumber \\
 & \times K_0(k_1,k_2)  .
\end{align}
In summary, $b_{j}(\omega)$ is given by Eqs.~(\ref{f2s1},\ref{numF2_2D},\ref{numS1_2D}) evaluated at $\omega=\pm\omega_{\pm}$. The energy-band curvatures of 
exciton and antiexciton bands are obtained by Eqs.~(\ref{mass1},\ref{mass2}).

We can see from the expression in Eq.~(\ref{f2s1}) that the $F_2$ term is always positive. Our numerical calculation shows that in parameter regions we studied, $F_2$ term dominates over the $S_1$ term, which makes $b_j(\omega_{+})$ and $b_j(-\omega_{-})$ to be always positive. Thus, from  Eq.~(\ref{mass2}), we can see that the band curvatures around the $\Gamma$ point become negative. In Fig.~\ref{curv}, we show the values of the band curvatures as we vary the electron mass $\bar{m}$ ($m_a=m_b$) and fix $E_g$ and $\mu$.

\section{``Antiparticle of an exciton" in previous literature}
\label{appendix:b}
``Antiexciton" had been introduced in previous theoretical literature with different definitions~\cite{PhysRevB.44.1368,PhysRevB.44.3638,lerner1981two,PhysRevB.59.5627}. In the following 
two subsections, we clarify the physical difference between our concept of the antiexciton 
and those in the literature. 

\subsection{``Antiparticle of an exciton" that is an identical entity to its own counterpart exciton}
Unlike the antiexciton proposed 
in this paper, the ``antiexciton'' introduced in some literature~\cite{PhysRevB.44.1368,PhysRevB.44.3638} characterizes 
the {\it same} inter-band collective mode as its own counterpart 
exciton. Ref.~\cite{PhysRevB.44.1368,PhysRevB.44.3638} studied generic semiconductors without 
the U(1)$\times$U(1) symmetry. 
An effective exciton Hamiltonian in such a system has no U(1) symmetry and  
the two-particle Green's function has pairs of a 
positive-energy pole and a negative-energy 
pole that are related to each other by a generic 
particle-hole symmetry of the effective exciton Hamiltonian. 
The literature~\cite{PhysRevB.44.1368,PhysRevB.44.3638} defines such
pairs of a positive-energy pole and a negative energy pole as excitons and 
their antiparticle counterparts (``antiexcitons''), respectively. 
In this section, we will show that the negative-energy pole 
related to the positive-energy pole by 
the generic particle-hole symmetry is redundant and the two  poles
characterize an identical physical excitation.  

In the following, we first explain a generic particle-hole symmetry 
of a free boson Hamiltonian. The symmetry relates 
a pair of a positive-energy eigenstate and a 
negative-energy eigenstate, while the two ``states'' 
actually characterize an identical physical state. 
To see this point in a simple way, let us begin with a general 
Bogoliubov-de Gennes (BdG) Hamiltonian for free boson systems. 
For a later comparison to the Bethe-Salpeter equation for excitons, 
we assume the spatial 
translational symmetry, while the following argument can be 
easily generalized into the other cases without 
the translational symmetry. The Hamiltonian in the 
momentum space 
reads,  
\begin{align}
{\cal H} &= \frac{1}{2} \sum_{\bm q} \left(\begin{array}{cc} 
{\bm \gamma}^{\dagger}_{\bm q} & {\bm \gamma}_{-{\bm q}} \\
\end{array}\right) \left(\begin{array}{cc} 
{\bm A}({\bm q}) & {\bm B}({\bm q}) \\
{\bm B}^*({-{\bm q}}) & {\bm A}^{*}({-{\bm q}}) \\
\end{array}\right) \left(\begin{array}{c} 
{\bm \gamma}_{\bm q} \\
{\bm \gamma}^{\dagger}_{-{\bm q}} \\
\end{array}\right)\nonumber \\
 &\equiv \frac{1}{2} \sum_{\bm q} \left(\begin{array}{cc} 
{\bm \gamma}^{\dagger}_{\bm q} & {\bm \gamma}_{-{\bm q}} \\
\end{array}\right) {\bm H}_{\rm BdG}({\bm q}) 
\left(\begin{array}{c} 
{\bm \gamma}_{\bm q} \\
{\bm \gamma}^{\dagger}_{-{\bm q}} \\
\end{array}\right). 
\end{align}
Here ${\bm \gamma}^{\dagger}_{\bm q}$ (${\bm \gamma}_{-\bm q}$)
is an $m$-component vector of boson creation 
(annihilation) operators with a momentum 
$\bm q$ ($-{\bm q}$), e.g. 
\begin{align}
\left(\begin{array}{cc} 
{\bm \gamma}^{\dagger}_{\bm q} & {\bm \gamma}_{-{\bm q}} \\
\end{array}\right) \equiv \left(\begin{array}{cccccc} 
\gamma^{\dagger}_{1,\bm q} & \cdots & \gamma^{\dagger}_{m,\bm q}
& {\bm \gamma}_{1,-{\bm q}} & \cdots & 
{\bm \gamma}_{m,-{\bm q}}\\
\end{array}\right). 
\end{align}
Each component of the creation and annihilation operators obey the commutation 
relations, e.g. $[\gamma_{j,{\bm q}},{\gamma}^{\dagger}_{m,{\bm q}^{\prime}}]
=\delta_{j,m} \delta_{{\bm q},{\bm q}^{\prime}}$. The boson's 
commutation relation leads to 
${\bm B}^T({-{\bm q}}) = {\bm B}({\bm q})$. The Hermiticity of ${\cal H}$ 
results in the Hermiticity of the $m$ by $m$ matrix ${\bm A}({\bm q})$ 
$({\bm A}({\bm q})={\bm A}^{\dagger}({\bm q}))$ 
as well as a following generic symmetry of the $2m$ by $2m$ 
matrix ${\bm H}_{\rm BdG}({\bm q})$,  
\begin{align}
    \sigma_1 \!\ {\bm H}_{\rm BdG}({\bm q}) \!\ \sigma_1 &= 
    \left(\begin{array}{cc} 
    {\bm A}^*({-{\bm q}}) & {\bm B}^{*}({-{\bm q}}) \\
    {\bm B}({\bm q}) & {\bm A}({\bm q}) \\
    \end{array}\right) \nonumber \\
&= {\bm H}^*_{\rm BdG}(-{\bm q}). \label{ph}
\end{align}
Here $\sigma_1$ exchanges ${\bm \gamma}^{\dagger}_{-{\bm q}}$ (particle) and 
${\bm \gamma}_{{\bm q}}$ (hole), so that we dub this symmetry as the
particle-hole symmetry. Unlike the particle-hole symmetry for a many-body Hamiltonian~\cite{srednicki_2007}, 
the particle-hole symmetry here 
is only due to the Hermiticity of ${\cal H}$ and the boson's 
commutation relation. 
Thus, any free boson Hamiltonians have this particle-hole symmetry 
in some bases (e.g. see below for an effective exciton Hamiltonian). 
To emphasize this point, we call this symmetry as 
a {\it generic} particle-hole symmetry. 

The BdG  Hamiltonian is 
diagonalized by a Bogoliubov transformation ${\bm T}$, 
\begin{align}
  &  {\bm T}^{\dagger} \!\ {\bm H}_{\rm BdG}({\bm q}) \!\ 
    {\bm T} = \left(\begin{array}{cc}
        {\bm E}({\bm q}) & \\
        & {\bm E}(-{\bm q}) \\
        \end{array}\right), \nonumber \\
&  \left(\begin{array}{c}
        {\bm \gamma}_{\bm q} \\
        {\bm \gamma}^{\dagger}_{-{\bm q}} \\
        \end{array}\right) = {\bm T} \left(\begin{array}{c}
        {\bm \eta}_{\bm q} \\
        {\bm \eta}^{\dagger}_{-{\bm q}} \\
        \end{array}\right).  \label{bogo}
\end{align}
${\bm E}({\bm q})$ is a diagonal matrix whose diagonal elements are eigenenergies 
of new boson fields ${\bm \eta}_{\bm q}$ with the momentum ${\bm q}$, 
\begin{align}
    {\cal H} = \frac{1}{2} 
    \sum_{{\bm q}} \sum^m_{j=1} \Big( 
    {\bm E}_{j}({\bm q}) \!\ {\bm \eta}^{\dagger}_{j,{\bm q}} {\bm \eta}_{j,{\bm q}} 
    +  {\bm E}_{j}(-{\bm q}) \!\  
    {\bm \eta}_{j,-{\bm q}} {\bm \eta}^{\dagger}_{j,-{\bm q}} \Big). \label{dia-2}
\end{align}
To connect the commutation relations among the old boson fields with the commutation relations among the new boson fields, the Bogoliubov transformation 
must satisfy the paraunitarity condition, 
\begin{align}
    {\bm T}^{\dagger} \sigma_3 {\bm T} = {\bm T} \sigma_3 {\bm T}^{\dagger} = \sigma_3.  \label{paraunitarity}
\end{align}
$\sigma_3$ in the equation is a 2-by-2 diagonal Pauli matrix, 
taking $+1$ in the hole space (${\bm \gamma}_{{\bm q}}$ or ${\bm  \eta}_{{\bm q}}$) and $-1$ in the particle 
space (${\bm \gamma}^{\dagger}_{-{\bm q}}$ or ${\bm \eta}^{\dagger}_{-{\bm q}}$). With the commutation relation of the new boson fieds, we have 
${\cal H}= \sum_{{\bm q},j} {\bm E}_{j}({\bm q}) \!\ {\bm \eta}^{\dagger}_{j,{\bm q}} {\bm \eta}_{j,{\bm q}} + {\rm const.}$.   

To fasciliate a later comparison to 
the Bethe-Salpeter equation for the inter-band excitations, 
let us next derive an equation of motion 
of the free boson systems. The equation of motion for the 
free boson fields takes a 
form of a generalized Hermitian eigenvalue problem, 
\begin{align}
    i\sigma_3 \!\ \partial_t \left(\begin{array}{c} 
    {\bm \gamma}_{\bm q} \\
    {\bm \gamma}^{\dagger}_{-{\bm q}} \\
    \end{array}\right) = 
    {\bm H}_{\rm BdG}({\bm q}) \left(\begin{array}{c} 
    {\bm \gamma}_{\bm q} \\
    {\bm \gamma}^{\dagger}_{-{\bm q}} \\
    \end{array}\right),   \label{g-eigenvalue}
\end{align}
with ${\bm H}_{\rm BdG}({\bm q})={\bm H}^{\dagger}_{\rm BdG}({\bm q})$. 
The diagonal Pauli matrix $\sigma_3$ in the left hand side comes 
from the boson's commutation relations, $[\gamma_{{\bm q},i},\gamma^{\dagger}_{{\bm q},j}] = - [\gamma^{\dagger}_{{\bm q},i},\gamma_{{\bm q},j}] = \delta_{i,j}$. 

The generic particle-hole symmetry of the BdG Hamiltonian (Eq.~(\ref{ph}))
formally relates a pair of positive-energy and negative-energy eigenmodes 
of the equation of motion. Suppose that the equation of motion has a 
solution ${\bm \Psi}$ of a positive energy $\omega$ and a momentum ${\bm q}$ as, 
\begin{align} 
\left(\begin{array}{c}
 {\bm \gamma}_{\bm q} \\
 {\bm \gamma}^{\dagger}_{-{\bm q}} \\
 \end{array}\right)
={\bm \Psi} e^{-i\omega t} \label{p-state}
\end{align}
with $\omega \sigma_3 \Psi = {\bm H}_{\rm BdG}({\bm q}) \Psi$. Such  
${\bm \Psi}$ is a column vector of 
${\bm T}$ for ${\bm \eta}_{\bm q}$ in Eq.~(\ref{bogo}). Then the 
particle-hole symmetry in Eq.~(\ref{ph}) formally relates this solution  
with a solution of the equation of motion at $-{\bm q}$ 
with a negative energy $-\omega$, 
\begin{align}
\left(\begin{array}{c}
{\bm \gamma}_{-{\bm q}} \\  
{\bm \gamma}^{\dagger}_{{\bm q}} \\
\end{array}\right) 
=\sigma_1 {\bm \Psi}^* e^{i\omega t}. \label{h-state}
\end{align} 
Namely, $-\omega \sigma_3 (\sigma_1 \Psi^*) = {\bm H}_{\rm BdG}(-{\bm q})  (\sigma_1\Psi^*)$. 
These two solutions clearly describe an identical inter-band collective 
mode, because the Hermitian conjugate of Eq.~(\ref{p-state}) is nothing but 
Eq.~(\ref{h-state}). 

In the following, we will review previous theories, where the 
Bethe-Salpeter equation for the two-particle Green's function 
always reduce to a generalized Hermitian eigenvalue problem of 
Eq.~(\ref{g-eigenvalue}) together with a BdG-type Hamiltonian~\cite{PhysRevB.62.4927,martin_reining_ceperley_2016,PhysRevB.44.1368,PhysRevB.44.3638}. 
The BdG Hamiltonian is nothing but a free boson Hamiltonian of 
excitons (more generally, inter-band excitations), and it has 
a generic particle-hole symmetry in a certain basis. The symmetry comes from the Hermiticity of an 
original many-body Hamiltonian of electrons and 
bosonic nature of excitons. 
In this sense, it is equivalent to Eq.~(\ref{ph}). 
Due to the generic particle-hole symmetry, the Green's function 
formally has pairs of a positive-energy pole ($\omega$) at 
${\bm q}$ and a negative-energy pole ($-\omega$) at $-{\bm q}$ 
that are connected to each other by the symmetry. Some literature~\cite{PhysRevB.44.1368,PhysRevB.44.3638} define 
the positive-energy 
pole at $\omega$ and the negative-energy pole ($-\omega$) as exciton 
and and its antiparticle counterpart (``antiexciton''), respectively.
However, the two poles characterize exactly an identical 
physical eigenmode as in Eqs.~(\ref{p-state},\ref{h-state}). 

To show this more explicitly, we follow Glutsch and Bechstedt's papers~\cite{PhysRevB.44.1368,PhysRevB.44.3638} and 
add an inter-band hopping term into a two-band system; 
\begin{align}
     \hat{H} &= \hat{H_0} + \hat{V}, \nonumber \\
    \hat{H_0} &= \sum_{\bm{k}} \Big[\epsilon_a(\bm{k}) a^{\dagger}_{\bm{k}}a_{\bm{k}} + \epsilon_b(\bm{k})b^{\dagger}_{\bm{k}}b_{\bm{k}} \nonumber \\
 & \hspace{2cm} + \Delta(\bm{k})a^{\dagger}_{\bm{k}}b_{\bm{k}} + \Delta^{*}(\bm{k})b^{\dagger}_{\bm{k}}a_{\bm{k}} \Big].   \label{Hami-3}
\end{align}
Where $\hat{V}$ is the same as Eq.~(\ref{Hint}). The inter-band hopping 
term $\Delta({\bm k})$ can come from an intrinsic band 
hybridization~\cite{PhysRevB.62.4927, martin_reining_ceperley_2016}, 
an external pump field~\cite{PhysRevB.44.1368,PhysRevB.44.3638}, or 
spontaneous symmetry breaking by exciton 
condensation~\cite{lerner1981two,PhysRevB.59.5627}. When a Fermi level 
is placed inside a band gap (see below), the Hamiltonian describes 
a generic semiconductor without the U(1)$\times$U(1) symmetry. 
$a_{\bm{k}} = u(\bm{k})\alpha_{\bm{k}} + v(\bm{k})\beta_{\bm{k}}$, $b_{\bm{k}} = -v^{*}(\bm{k})\alpha_{\bm{k}} + u(\bm{k})\beta_{\bm{k}}$ with 
$|u(\bm{k})|^2 + |v(\bm{k})|^2 = 1$, 
\begin{align}
    \hat{H_0} &= \sum_{\bm{k}} \Big[ \Omega_{+}(\bm{k})\alpha^{\dagger}_{\bm{k}}\alpha_{\bm{k}}   + \Omega_{-}(\bm{k})\beta^{\dagger}_{\bm{k}}\beta_{\bm{k}} \Big], \nonumber 
\end{align} 
with 
\begin{align}
\Omega_{\pm}(\bm{k}) \equiv \frac{(\epsilon_a(\bm{k}) + \epsilon_b(\bm{k})}{2} \pm \frac{\sqrt{(\epsilon_a(\bm{k}) - \epsilon_b(\bm{k}))^2 + 4|\Delta(\bm{k})|^2}}{2} .
\end{align} 
In terms of interaction between $\alpha$ and $\beta$ fermions, the interaction part is given by 
\begin{widetext}
\begin{align}
    \hat{V} 
    = & \frac{1}{2\Omega} \sum_{{\bm q},{\bm k},{\bm k}^{\prime}} \bigg( 
    2{\bm A}_{{\bm k},{\bm k}^{\prime}}({\bm q}) \!\  \alpha^{\dagger}_{{\bm k}+{\bm q}} 
    \beta^{\dagger}_{{\bm k}^{\prime}} \beta_{\bm k} \alpha_{{\bm k}^{\prime}+{\bm q}} 
    + {\bm B}_{{\bm k},{\bm k}^{\prime}}({\bm q}) \!\  \alpha^{\dagger}_{{\bm k}+{\bm q}} 
    \alpha^{\dagger}_{-{\bm k}^{\prime}-{\bm q}} \beta_{\bm k} \beta_{-{\bm k}^{\prime}}
    + {\bm B}^*_{-{\bm k},-{\bm k}^{\prime}}(-{\bm q}) \!\  \beta^{\dagger}_{-{\bm k}} 
    \beta^{\dagger}_{{\bm k}^{\prime}} \alpha_{-{\bm k}-{\bm q}} 
    \alpha_{{\bm k}^{\prime}+{\bm q}} \bigg) + \cdots \nonumber \\ 
    = & \frac{1}{2\Omega} \sum_{{\bm q},{\bm k},{\bm k}^{\prime}} \bigg( 
    2{\bm A}^*_{-{\bm k},-{\bm k}^{\prime}}(-{\bm q}) \!\  \beta^{\dagger}_{-{\bm k}} 
    \alpha^{\dagger}_{-{\bm k}^{\prime}-{\bm q}} \alpha_{-{\bm k}-{\bm q}} \beta_{{\bm k}^{\prime}} 
     + {\bm B}_{{\bm k},{\bm k}^{\prime}}({\bm q}) \!\  \alpha^{\dagger}_{\cdots} 
    \alpha^{\dagger}_{\cdots} \beta_{\cdots} \beta_{\cdots}
     + {\bm B}^*_{-{\bm k},-{\bm k}^{\prime}}(-{\bm q}) \!\  \beta^{\dagger}_{\cdots} 
    \beta^{\dagger}_{\cdots} \alpha_{\cdots} 
    \alpha_{\cdots} \bigg) + \cdots, \label{int-V}
\end{align}
where matrix elements of 
${\bm A}({\bm q})$ and ${\bm B}({\bm q})$ 
are given by the unitary transformation. ``$\cdots$'' in the right hand sides are those terms that take forms of either $\alpha^{\dagger}\alpha^{\dagger}\alpha\alpha$ or  $\beta^{\dagger}\beta^{\dagger}\beta\beta$, while they do not contribute to the BS equation in a semiconductor region considered below. Note that 
the Hermiticity of $\hat{V}$ requires 
${\bm A}_{{\bm k},{\bm k}^{\prime}}({\bm q})={\bm A}^*_{{\bm k}^{\prime},{\bm k}}({\bm q})$, and relates the matrix elements of the second term  $(\alpha^{\dagger}\alpha^{\dagger}\beta\beta)$ and the third term $(\beta^{\dagger}\beta^{\dagger}\alpha\alpha)$ as in the equation. With the sum over ${\bm q}$, ${\bm k}$ and ${\bm k}^{\prime}$, we can always symmetrize ${\bm B}({\bm q})$ such that ${\bm B}_{{\bm k},{\bm k}^{\prime}}({\bm q})={\bm B}_{-{\bm k}^{\prime},-{\bm k}}(-{\bm q})$.

Following the papers~\cite{PhysRevB.44.1368,PhysRevB.44.3638}, 
we consider that $\Delta({\bm k})$ results in 
a direct gap between the $\alpha$ band and $\beta$ 
band, and we place a chemical potential $\mu$ inside the energy 
gap; $\Omega_{+}({\bm k})> \mu > \Omega_{-}({\bm k})$ for all the ${\bm k}$. 
A many-body ground state $|0\rangle$ is considered to be adiabatically connected to 
this semiconductor ground state. The 
inter-band collective modes above the many-body ground state can be analyzed by 
an inter-band two-particle Green's function, which generally takes 
a two-by-two matrix form, 
\begin{align}
\hat{\cal G}^{ex}({\bm x}-{\bm x}^{\prime},t-t^{\prime})_{{\bm y}{\bm y}^{\prime}} 
\equiv -(-i)^2 \left(\begin{array}{cc}
\langle 0 | \mathcal{T} \{ \gamma({\bm x},{\bm y};t) \gamma^{\dagger}({\bm x}^{\prime},{\bm y}^{\prime};t^{\prime}) \} | 0 \rangle
& \langle 0 | \mathcal{T} \{ \gamma({\bm x},{\bm y},t) \gamma({\bm x}^{\prime},{\bm y}^{\prime};t^{\prime}) \} | 0 \rangle  \\ 
\langle 0 | \mathcal{T} \{ \gamma^{\dagger}({\bm x},{\bm y},t) \gamma^{\dagger}({\bm x}^{\prime},{\bm y}^{\prime},t^{\prime}) \} | 0 \rangle
& \langle 0 | \mathcal{T} \{ \gamma^{\dagger}({\bm x},{\bm y},t) \gamma({\bm x}^{\prime},{\bm y}^{\prime},t^{\prime}) \} | 0 \rangle \\ 
\end{array}\right), 
\end{align}
with $\gamma({\bm x},{\bm y};t) \equiv \alpha_{{\bm x}}(t) \beta^{\dagger}_{{\bm x} + {\bm y}}(t)$ 
and $\gamma^{\dagger}({\bm x},{\bm y};t) \equiv  \beta_{{\bm x} + {\bm y}}(t) \alpha^{\dagger}_{{\bm x}}(t)$. 
The $t$-dependence of the operators is in the Heisenberg picture. 
$\gamma({\bm x},{\bm y})$ and 
$\gamma^{\dagger}({\bm x},{\bm y})$ can be regarded as  
annihilation and creation of excitons, respectively. Note that  
the $\alpha^{\dagger}\alpha^{\dagger} \beta \beta$ 
and $\beta^{\dagger}\beta^{\dagger} \alpha \alpha$ terms in $\hat{V}$ 
break a U(1) symmetry of   
$\alpha \rightarrow \alpha e^{i\theta}$ and $\beta \rightarrow \beta 
e^{-i\theta}$. Thereby,    
the inter-band Green's function generally takes the two-by-two matrix form 
in particle-hole space of the exciton fields. 

After Fourier transforms of 
the spatial coordinates, the Green's function is given by a function of a momentum ${\bm q}$ of 
the center-of-mass coordinate between two fermions and 
momenta ${\bm k}$, ${\bm k}^{\prime}$ of the relative coordinates,  
\begin{align}
 &\hat{\cal G}^{ex}\left({\bm q}, t-t^{\prime}\right)_{{\bm k}{\bm k}^{\prime}} 
    \equiv  \int 
d^d(\bm{x}-{\bm x}^{\prime}) \int d^d {\bm y} \int d^d {\bm y}^{\prime} 
e^{ - i{\bm q}\cdot ({\bm x}-{\bm x}^{\prime}) + i{\bm k}\cdot{\bm y} - i{\bm k}^{\prime}\cdot{\bm y}^{\prime}} 
\hat{\cal G}^{ex}({\bm x} - {\bm x}^{\prime}, t-t^{\prime})_{{\bm y}{\bm y}^{\prime}}, \label{ft2} 
\end{align}
with 
\begin{eqnarray}
&& \hat{\cal G}^{ex}({\bm q},t-t^{\prime})_{{\bm k}{\bm k}^{\prime}} 
= -(-i)^2 \Omega \times \nonumber \\
&& \hspace{1cm}  \left(\begin{array}{cc} 
\langle 0 |{\cal T} \{  \alpha_{\bm{k}+\bm{q}}(t) \beta^{\dagger}_{\bm{k}}(t) \!\  
    \beta_{\bm{k^{\prime}}}(t^{\prime})\alpha^{\dagger}_{\bm{k^{\prime} } + \bm{q}}(t^{\prime}) \} | 0 \rangle & \langle 0 | {\cal T} \{  
    \alpha_{\bm{k}+\bm{q}}(t) \beta^{\dagger}_{\bm{k}}(t) \!\ 
    \alpha_{-\bm{k^{\prime}} - \bm{q}}(t^{\prime})\beta^{\dagger}_{-\bm{k^{\prime}}}(t^{\prime})    \} |0 \rangle \\
 \langle 0| {\cal T} \{ \beta_{-\bm{k}}(t) \alpha^{\dagger}_{-\bm{k}-\bm{q}}(t) \!\  \beta_{\bm{k^{\prime}}}(t^{\prime})\alpha^{\dagger}_{\bm{k^{\prime}}+\bm{q}}(t^{\prime})
    \} |0 \rangle & \langle 0| {\cal T} \{\beta_{-\bm{k}}(t) \alpha^{\dagger}_{-\bm{k}-\bm{q}}(t) \!\  \alpha_{-\bm{k^{\prime}}-\bm{q}}(t^{\prime})\beta^{\dagger}_{-\bm{k^{\prime}}}(t^{\prime}) \} |0 \rangle \\
\end{array}\right).
\end{eqnarray}
Here $\Omega$ is the system volume. 

The ground state in the non-interacting limit is a vacuum of $\alpha$ 
and $\beta^{\dagger}$, $\alpha |0\rangle_{V=0}= \beta^{\dagger}|0\rangle_{V=0}=0$, where the inter-band  
Green's function takes a diagonal form in the particle-hole space, 
\begin{align}
    \hat{\cal G}^{ex}_0({\bm q},t-t^{\prime})_{{\bm k}{\bm k}^{\prime}} 
= \Omega\delta_{{\bm k},{\bm k}^{\prime}}
    \left(\begin{array}{cc}
    \theta(t-t^{\prime}) e^{-i(\Omega_{+}({\bm k}+{\bm q})-\Omega_{-}({\bm k}))(t-t^{\prime})} & 0 \\
    0 &  \theta(t^{\prime}-t) e^{i(\Omega_{+}(-{\bm k}-{\bm q})-\Omega_{-}(-{\bm k}))(t-t^{\prime})} \\ 
    \end{array}\right). 
\end{align}
In the presence of $\hat{V}$ of Eq.~(\ref{int-V}), the inter-band two-particle 
Green's function is given by a solution of the following Bethe-Salpeter equation, 
\begin{align}
\hat{\cal G}^{ex}({\bm q},t-t^{\prime})_{{\bm k}{\bm k}^{\prime}} &= \hat{\cal G}^{ex}_0({\bm q},t-t^{\prime})_{{\bm k}{\bm k}^{\prime}} 
\nonumber \\
& \ \ \ +  \frac{i}{\Omega^2} 
\sum_{\overline{\bm k},\overline{\bm k}^{\prime}} 
\int d\overline{t} \!\ \hat{\cal G}^{ex}_0({\bm q},t-\overline{t})_{{\bm k}\overline{\bm k}} 
 \!\ \left(\begin{array}{cc} 
   {\bm A}_{\overline{\bm k},\overline{\bm k}^{\prime}}({\bm q}) & {\bm B}_{\overline{\bm k},\overline{\bm k}^{\prime}}({\bm q}) \\ 
    {\bm B}^{*}_{-\overline{\bm k},-\overline{\bm k}^{\prime}}(-{\bm q}) & {\bm A}^{*}_{-\overline{\bm k},-\overline{\bm k}^{\prime}}(-{\bm q}) \\
    \end{array}\right) \!\ 
    \hat{\cal G}^{ex}({\bm q},\overline{t}-t^{\prime})_{\overline{\bm k}^{\prime}{\bm k}^{\prime}}.
\end{align}
After the Fourier transform in time, 
the equation reduces to a generalized Hermitian eigenvalue problem with a BdG-type Hamiltonian 
${\bm H}_{\rm BdG}({\bm q})$; 
\begin{align}
    \sum_{\overline{\bm k}} \Big( \delta_{{\bm k},\overline{\bm k}} (\omega \!\ \sigma_3 
    + i0^{+} \sigma_0 ) 
    - {\bm H}_{\rm BdG}({\bm q})_{{\bm k},\overline{\bm k}} \Big) \!\ \widetilde{\cal G}^{ex}({\bm q},\omega)_{\overline{\bm k},{\bm k}^{\prime}} = \delta_{{\bm k},{\bm k}^{\prime}} \sigma_0,  \label{g-eigenvalue-2}
\end{align}
with ${\bm H}_{\rm BdG}({\bm q}) = 
{\bm H}^{\dagger}_{\rm BdG}({\bm q})$. Here 
the Green's function is normalized by $i\Omega$, 
$i\Omega \widetilde{\cal G}^{ex} \equiv \hat{\cal G}^{ex}$. 
The BdG Hamiltonian is a free boson Hamiltonian of excitons (inter-band excitations) 
and it takes a 2-by-2 matrix form 
in the particle-hole space of the exciton fields, 
\begin{align}
{\bm H}_{\rm BdG}({\bm q})_{{\bm k},{\bm k}^{\prime}} 
= \delta_{{\bm k},{\bm k}^{\prime}} \left(\begin{array}{cc} 
\Omega_+({\bm k}+{\bm q}) - \Omega_{-}({\bm k}) & \\
& \Omega_{+}(-{\bm k}-{\bm q}) - \Omega_{-}(-{\bm k}) \\
\end{array}\right) - \frac{1}{\Omega} \left(\begin{array}{cc} 
A_{{\bm k},{\bm k}^{\prime}}({\bm q}) & B_{{\bm k},{\bm k}^{\prime}}({\bm q}) \\
B^*_{-{\bm k},-{\bm k}^{\prime}}(-{\bm q}) & A^{*}_{-{\bm k},-{\bm k}^{\prime}}(-{\bm q}) \\ 
\end{array}\right). \label{bdg-Hami-1}
 \end{align}
\end{widetext}
$\sigma_3$ in the left hand side of Eq.~(\ref{g-eigenvalue-2}) 
is the diagonal Pauli matrix taking $+1$ for the hole space of the exciton 
field ($\gamma$) and $-1$ for the particle space of the exciton 
field ($\gamma^{\dagger}$). $\omega \!\ \sigma_3$ is nothing but 
a Frouier transform of $i\sigma_3 \partial_t$ in the left hand side of 
Eq.~(\ref{g-eigenvalue}), standing for the boson's commutation relations of 
excitons fields. 
Thus, solving this Bethe-Salpeter equation of Eq.~(\ref{g-eigenvalue-2}) is essentially equivalent to 
solving the equation of motion of Eq.~(\ref{g-eigenvalue}).

The eigenvalue problem can be solved in terms of the paraunitary transformation. 
Suppose that the BdG Hamiltonian is diagonalized by a paraunitary transformation of Eq.~(\ref{bogo}). Thereby, the diagonal elements in ${\bm E}({\bm q})$ of Eq.~(\ref{bogo}) are nothing but 
excitation energies of inter-band collective 
and individual excitations with the momentum ${\bm q}$. Column vectors in ${\bm T}$ of Eq.~(\ref{bogo}) stand for 
wavefunctions of these eigenmodes. 
Due to the particle-hole mixing nature of ${\bm H}_{\rm BdG}({\bm q})$, 
${\bm T}$ thus introduced comrpises of both the hole-type 
wavefunctions for ${\bm \eta}_{j,{\bm q}}$ and the particle-type wavefunctions for ${\bm \eta}^{\dagger}_{j,-{\bm q}}$. We dub them 
as ${\bm \Psi}_j$ and ${\bm \Phi}_j$ respectively ($j=1,2,\cdots$);  
\begin{align}
{\bm T} \equiv \left(\begin{array}{cccccc} 
{\bm \Psi}_1 & {\bm \Psi}_2 & \cdots {\bm \Phi}_1 & {\bm \Phi}_2 & \cdots \\
\end{array}\right). 
\end{align}
${\bm \Psi}_j$ for $\eta_{j,{\bm q}}$ here corresponds to 
${\bm \Psi}$ in Eq.~(\ref{p-state}). 
To preserve the boson statistics between ${\bm \eta}_{j,{\bm q}}$ 
and ${\bm \eta}^{\dagger}_{m,{\bm q}}$ of Eq.~(\ref{bogo}), these 
wavefunctions are normalized with the paraunitary condition of Eq.~(\ref{paraunitarity}). The orthnormalization 
is given by 
${\bm \Psi}^{\dagger}_m \sigma_3 {\bm \Psi}_j = \delta_{j,m}$, 
${\bm \Phi}^{\dagger}_m \sigma_3 {\bm \Phi}_j = - \delta_{j,m}$, and 
${\bm \Psi}^{\dagger}_m \sigma_3 {\bm \Phi}_j = 
{\bm \Phi}^{\dagger}_m \sigma_3 {\bm \Psi}_j = 0$ $(j,m=1,2,\cdots)$. The completeness relation is defined by
\begin{align}
    {\bm T} \sigma_3 {\bm T}^{\dagger} &= 
    \sum_{j} {\bm \Psi}_{j,{\bm k}}{\bm \Psi}^*_{j,{\bm k}^{\prime}} 
    -  \sum_{j} {\bm \Phi}_{j,{\bm k}}{\bm \Phi}^*_{j,{\bm k}^{\prime}}  \nonumber \\
    &= \sigma_3 \delta_{{\bm k},{\bm k}^{\prime}}. \label{complete}
\end{align}
With these orthogonality and completeness relations, 
Eq.~(\ref{bogo}) can be rewritten into 
\begin{align}
    \left\{\begin{array}{l}
    {\bm H}_{\rm BdG}({\bm q}) \!\ {\bm \Psi}_{j} = \sigma_3 
    {\bm \Psi}_{j} E_{j}({\bm q}), \\
    {\bm H}_{\rm BdG}({\bm q}) \!\ {\bm \Phi}_{j} = - \sigma_3 
    {\bm \Phi}_{j} E_{j}(-{\bm q}), \\
    \end{array}\right.
\end{align}
for $j=1,2,\cdots$. 
Given the paraunitary transformation, the inter-band Green's function 
at ${\bm q}$ is obtained as a solution of the Bethe-Salpeter equation, 
\begin{align}
    \widetilde{\cal G}^{ex}({\bm q},\omega)_{{\bm k},{\bm k}^{\prime}} =& \sum_{j}\frac{{\bm \Psi}_{j,{\bm k}}{\bm \Psi}^*_{j,{\bm k}^{\prime}}}{\omega - E_{j}(\bm q)+i0^{+}} \nonumber \\
& -  \sum_{j}\frac{{\bm \Phi}_{j,{\bm k}}{\bm \Phi}^*_{j,{\bm k}^{\prime}}}{\omega + E_{j}(-{\bm q})-i0^{+}}.  \label{sol-q}
\end{align}
Note that thanks to the completeness relation of Eq.~(\ref{complete}), 
the solution satisfies a sum rule, 
\begin{align}
    \int^{+\infty}_{-\infty} \frac{d\omega}{2\pi i} \!\ 
    \widetilde{\cal G}^{ex}({\bm q},\omega)_{{\bm k},{\bm k}^{\prime}} 
    \Big(e^{-i\omega 0^{+}} - e^{i\omega 0^{+}}\Big) 
    = \sigma_3 \delta_{{\bm k},{\bm k}^{\prime}}. \label{boundary} 
\end{align}
The sum rule is a boundary condition in time 
$i\widetilde{\cal G}({\bm q},t=0+)-i\widetilde{\cal G}({\bm q},t=0-)
= \sigma_3 \delta_{{\bm k},{\bm k}^{\prime}}$. One can 
obtain the boundary condition by noting 
$\langle 0 |\alpha^{\dagger}_{\bm k}\alpha_{\bm k}|0\rangle = 0$ 
and $\langle 0 |\beta^{\dagger}_{\bm k}\beta_{\bm k}|0\rangle = 1$. 
Here $|0\rangle$ is the many-body ground state. 

Like in the free boson system, the BdG Hamiltonian of 
excitons ${\bm H}_{\rm BdG}({\bm q})$ 
has a generic particle-hole symmetry in a certain basis;
\begin{align}
    \sigma_1 \!\ 
    \hat{\bm H}^*_{\rm BdG}({\bm q})_{{\bm k},{\bm k}^{\prime}} \!\ 
    \sigma_1 = \hat{\bm H}_{\rm BdG}(-{\bm q})_{-{\bm k},-{\bm k}^{\prime}}. 
\end{align}
As was explained below Eq.~(\ref{int-V}), the symmetry 
comes from the Hermiticity of the original many-body Hamiltonian 
of electrons and the bosonic nature of excitons. Due 
to this generic symmetry, the BdG Hamiltonian 
${\bm H}_{\rm BdG}(-{\bm q})$ at $-{\bm q}$ is diagonalized 
by the following paraunitary transformation, 
\begin{align}
\sigma_1 {T}^* \sigma_1 = \left(\begin{array}{cccccc} 
\sigma_1{\bm \Phi}^*_1 & \sigma_1{\bm \Phi}^*_2 & \cdots 
\sigma_1{\bm \Psi}^*_1 & \sigma_1{\bm \Psi}^*_2 & \cdots \\
\end{array}\right). 
\end{align}
This gives a solution of the inter-band 
Green's function at $-{\bm q}$;
\begin{align}
    \widetilde{\cal G}^{ex}(-{\bm q},\omega)_{{\bm k},{\bm k}^{\prime}} = &\sum_{j}\frac{\sigma_1{\bm \Phi}^*_{j,-{\bm k}}{\bm \Phi}_{j,-{\bm k}^{\prime}}\sigma_1}{\omega - E_{j}(-{\bm q})+i0^{+}} \nonumber \\
& -  \sum_{j}\frac{\sigma_1{\bm \Psi}^*_{j,-{\bm k}}{\bm \Psi}_{j,-{\bm k}^{\prime}}\sigma_1}{\omega + E_{j}({\bm q})-i0^{+}}. \label{sol--q}
\end{align}
Some previous works~\cite{PhysRevB.44.1368,PhysRevB.44.3638} in the literature~\cite{PhysRevB.62.4927,martin_reining_ceperley_2016,PhysRevB.44.1368,PhysRevB.44.3638} defined the positive-energy poles at $\omega=E_j({\bm q})$ as excitons and the 
negative-energy poles at $\omega=-E_j(-{\bm q})$ in 
Eq.~(\ref{sol-q}) as its counterpart ``antiexcitons", 
respectively. However, a comparison between Eq.~(\ref{sol-q}) and Eq.~(\ref{sol--q}) 
clearly shows that the negative-energy poles are completely redundant, 
because all the information of physical eigenmodes of the system are solely encoded in 
the positive-energy poles of {\it all} the ${\bm q}$. In fact, as shown explicitly 
in the comparison between Eq.~(\ref{p-state}) and Eq.~(\ref{h-state}), 
the negative energy poles at $-{\bm q}$ with $\omega=-E_j({\bm q})$ 
characterizes the same physical excitation as the positive 
energy poles at ${\bm q}$ with $\omega=E_j({\bm q})$. The equivalence 
is nothing but the equivalence between ${\eta}_{j,{\bm q}}$ in the 
first term of Eq.~(\ref{dia-2}) at ${\bm q}$ and ${\eta}^{\dagger}_{j,{\bm q}}$ 
in the second term of Eq.~(\ref{dia-2}) at $-{\bm q}$. 

Unlike ``antiexciton'' introduced in these literature~\cite{PhysRevB.44.1368,PhysRevB.44.3638}, $|n\rangle$ state 
(exciton) and $|n^{\prime}\rangle$ state (antiexciton) in Eq.~(\ref{eqn_Lehmann}) describes two {\it different} excited eigenstates of the two-band semimetal model. In fact, in the presence of the U(1) $\times$ U(1) 
symmetry, where the two-band semimetal model commutes with total electron 
numbers of the $a$ band and the $b$ band, the excitons state lives 
in a Hilbert space of $|N_a+1,N_b-1\rangle$,   
the antiexcitons state lives in a Hilbert space of $|N_a-1,N_b+1\rangle$, 
and these two spaces are decoupled from each other in an exact 
diagonalization of the many-body Hamiltonian for the semimetal model. 
Here $N_a$ and $N_b$ are the total electron numbers of the $a$ band and the $b$ band, 
respectively, and the semimetal ground state $|0\rangle$ 
is in a Hilbert space of $|N_a,N_b\rangle$. 
Since $a^{\dagger} b$ creates such an exciton and annihilates
such an antiexciton in 
the semimetal ground state, the Green's function of 
$G^{ex}({\bm x}-{\bm x}^{\prime},t-t^{\prime})_{{\bm y}{\bm y}^{\prime}} \equiv \langle 0 | {\cal T} \{ a_{{\bm x}}(t)b^{\dagger}_{{\bm x}+{\bm y}}(t) b_{{\bm x}^{\prime}+{\bm y}^{\prime}}(t^{\prime}) a^{\dagger}_{{\bm x}^{\prime}}(t^{\prime}) \} |0\rangle$  
has a pole for the exciton state in the positive-$\omega$ region and a pole for the antiexciton state in the negative-$\omega$ region. 

In summary, our manuscript proposed a pair of the exciton and antiexciton as two {\it distinct} excited eigenstates above the semimetal ground state. Physically speaking, our concept of the antiexciton should be 
distinguished from ``antiexciton'' in the literature~\cite{PhysRevB.44.1368,PhysRevB.44.3638} 
which actually describes an identical excited eigenstate as its 
counterpart exciton state.  

In the presence of single-particle hybridization between 
the two bands, $\Delta a^{\dagger} b$, $N_a-N_b$ 
becomes no longer a quantum number, 
and excitons (inter-band collective modes in the Hilbert space of $|N_a+1,N_b-1\rangle$) and antiexcitons (inter-band collective modes in the Hilbert space of $|N_a-1,N_b+1\rangle$)  shall be hybridized in general. Thus, inter-band collective modes in such two-band models are no longer classified in terms of the exciton or antiexciton proposed in our paper. Nonetheless, when 
the hybridization $\Delta$ is much smaller than an energy 
difference between original exciton and antiexciton states 
at $\Delta=0$, two weakly-hybridized inter-band collective 
modes can be approximately regarded as an exciton mode and
an antiexicton mode. In this sense, 
our concept of the exciton and antiexciton still provides 
a useful picture for distinguishing the two types of 
inter-band collective modes in two-bands semimetals 
even with the hybridization.

\subsection{``Antiparticle of an exciton" that does not coexist with its counterpart exciton}
Ref.~\cite{lerner1981two} introduced a concept of ``antiexciton"
in two-dimensional (2D) electron-hole gas (EHG) under high magnetic field, which one could consider to share the similar physical picture as the antiexciton proposed in this paper. The concept was further cited in Ref.~\cite{PhysRevB.59.5627}. Under the magnetic field, the 2D EHG forms Landau levels (LLs) of an 
electron-type band and a hole-type band. Though Ref.~\cite{lerner1981two} considers an effective inter-band hybridization by exciton condensation, the classification of inter-band collective modes for the U(1)$\times$U(1) symmetric case at zero temperature is approximately applicable for their case. 

When the lowest LL (LLL) of the electron band ($a$ band) is higher than the LLL of the hole band ($b$ band) in energy and the Fermi level is set to the middle between the two LLLs (``small $\rho$ limit''), an inter-band collective mode 
in such a positive band gap semiconductor regime is a bound 
state of a particle in the $a$ band and a hole in the $b$ band; 
the inter-band collective mode lives in a Hilbert space of $|N_a + 1,N_b-1\rangle$ (Here we consider that the ground is in a Hilbert space of $|N_a,N_b\rangle$ with $N_a \ll N_b$). When the LLL of the electron band is lower than the LLL of the hole band and the Fermi level is placed into the middle of the two LLLs 
(``small $1-\rho$ limit''), 
an inter-band collective mode in such a negative band gap semiconductor regime is a bound state of a hole in the $a$-band and a particle in the $b$-band; the collective mode lives in a Hilbert space of $|N_a-1,N_b + 1\rangle$ (Here we consider that the ground is in a Hilbert space of $|N_a,N_b\rangle$ with $N_a \gg N_b$). Lerner and Lozovik called the inter-band collective modes in these 
positive band gap and negative band gap semiconductor regimes as excitons and antiexcitons respectively, so that the concept of the antiexciton proposed in our manuscript shares 
the similar physical picture as Lerner and Lozovik's. 

Nonetheless, unlike in the semimetal model studied in this manuscript, 
the excitons and antiexcitons in the 2D EHG under the field {\it do not coexist} inside the same 2D bulk. Excitons exist only in the small $\rho$ limit and antiexcitons exist in the small $1-\rho$ limit. Physically speaking, the band inversion can be induced by changing the magnetic field, so that these two 
limits are realized in two different regions of the magnetic field. 
To our best knowledge, 
the coexistence of excitons and antiexcitons is demonstrated for the first time in the semimetal model
by our manuscript. The coexistence of a pair of excitons and antiexcitons leads to two distinct absorption peaks in optical spectroscopy experiments as well as an fertile excitonic-analog of the electron-positron pair annihilation phenomenon. These physical consequences cannot be realized by the 2D EHG under the
high magnetic field.

\bibliography{antiexc}

\providecommand{\noopsort}[1]{}\providecommand{\singleletter}[1]{#1}%
\begin{thebibliography}{47}%
\makeatletter
\providecommand \@ifxundefined [1]{%
 \@ifx{#1\undefined}
}%
\providecommand \@ifnum [1]{%
 \ifnum #1\expandafter \@firstoftwo
 \else \expandafter \@secondoftwo
 \fi
}%
\providecommand \@ifx [1]{%
 \ifx #1\expandafter \@firstoftwo
 \else \expandafter \@secondoftwo
 \fi
}%
\providecommand \natexlab [1]{#1}%
\providecommand \enquote  [1]{``#1''}%
\providecommand \bibnamefont  [1]{#1}%
\providecommand \bibfnamefont [1]{#1}%
\providecommand \citenamefont [1]{#1}%
\providecommand \href@noop [0]{\@secondoftwo}%
\providecommand \href [0]{\begingroup \@sanitize@url \@href}%
\providecommand \@href[1]{\@@startlink{#1}\@@href}%
\providecommand \@@href[1]{\endgroup#1\@@endlink}%
\providecommand \@sanitize@url [0]{\catcode `\\12\catcode `\$12\catcode
  `\&12\catcode `\#12\catcode `\^12\catcode `\_12\catcode `\%12\relax}%
\providecommand \@@startlink[1]{}%
\providecommand \@@endlink[0]{}%
\providecommand \url  [0]{\begingroup\@sanitize@url \@url }%
\providecommand \@url [1]{\endgroup\@href {#1}{\urlprefix }}%
\providecommand \urlprefix  [0]{URL }%
\providecommand \Eprint [0]{\href }%
\providecommand \doibase [0]{https://doi.org/}%
\providecommand \selectlanguage [0]{\@gobble}%
\providecommand \bibinfo  [0]{\@secondoftwo}%
\providecommand \bibfield  [0]{\@secondoftwo}%
\providecommand \translation [1]{[#1]}%
\providecommand \BibitemOpen [0]{}%
\providecommand \bibitemStop [0]{}%
\providecommand \bibitemNoStop [0]{.\EOS\space}%
\providecommand \EOS [0]{\spacefactor3000\relax}%
\providecommand \BibitemShut  [1]{\csname bibitem#1\endcsname}%
\let\auto@bib@innerbib\@empty
\bibitem [{\citenamefont {Srednicki}(2007)}]{srednicki_2007}%
  \BibitemOpen
  \bibfield  {author} {\bibinfo {author} {\bibfnamefont {M.}~\bibnamefont
  {Srednicki}},\ }\href {https://doi.org/10.1017/CBO9780511813917} {\emph
  {\bibinfo {title} {Quantum Field Theory}}}\ (\bibinfo  {publisher} {Cambridge
  University Press},\ \bibinfo {year} {2007})\BibitemShut {NoStop}%
\bibitem [{\citenamefont {Griffiths}(2020)}]{griffiths2020introduction}%
  \BibitemOpen
  \bibfield  {author} {\bibinfo {author} {\bibfnamefont {D.}~\bibnamefont
  {Griffiths}},\ }\href@noop {} {\emph {\bibinfo {title} {Introduction to
  elementary particles}}}\ (\bibinfo  {publisher} {John Wiley \& Sons},\
  \bibinfo {year} {2020})\BibitemShut {NoStop}%
\bibitem [{\citenamefont {Rohlfing}\ and\ \citenamefont
  {Louie}(2000)}]{PhysRevB.62.4927}%
  \BibitemOpen
  \bibfield  {author} {\bibinfo {author} {\bibfnamefont {M.}~\bibnamefont
  {Rohlfing}}\ and\ \bibinfo {author} {\bibfnamefont {S.~G.}\ \bibnamefont
  {Louie}},\ }\bibfield  {title} {\bibinfo {title} {Electron-hole excitations
  and optical spectra from first principles},\ }\href
  {https://doi.org/10.1103/PhysRevB.62.4927} {\bibfield  {journal} {\bibinfo
  {journal} {Phys. Rev. B}\ }\textbf {\bibinfo {volume} {62}},\ \bibinfo
  {pages} {4927} (\bibinfo {year} {2000})}\BibitemShut {NoStop}%
\bibitem [{\citenamefont {Martin}\ \emph {et~al.}(2016)\citenamefont {Martin},
  \citenamefont {Reining},\ and\ \citenamefont
  {Ceperley}}]{martin_reining_ceperley_2016}%
  \BibitemOpen
  \bibfield  {author} {\bibinfo {author} {\bibfnamefont {R.~M.}\ \bibnamefont
  {Martin}}, \bibinfo {author} {\bibfnamefont {L.}~\bibnamefont {Reining}},\
  and\ \bibinfo {author} {\bibfnamefont {D.~M.}\ \bibnamefont {Ceperley}},\
  }\href {https://doi.org/10.1017/CBO9781139050807} {\emph {\bibinfo {title}
  {Interacting Electrons: Theory and Computational Approaches}}}\ (\bibinfo
  {publisher} {Cambridge University Press},\ \bibinfo {year}
  {2016})\BibitemShut {NoStop}%
\bibitem [{\citenamefont {Strinati}(1988)}]{strinati1988application}%
  \BibitemOpen
  \bibfield  {author} {\bibinfo {author} {\bibfnamefont {G.}~\bibnamefont
  {Strinati}},\ }\bibfield  {title} {\bibinfo {title} {Application of the
  {G}reen’s functions method to the study of the optical properties of
  semiconductors},\ }\href {https://doi.org/10.1007/BF02725962} {\bibfield
  {journal} {\bibinfo  {journal} {La Rivista del Nuovo Cimento (1978-1999)}\
  }\textbf {\bibinfo {volume} {11}},\ \bibinfo {pages} {1} (\bibinfo {year}
  {1988})}\BibitemShut {NoStop}%
\bibitem [{\citenamefont {Koch}\ \emph {et~al.}(2006)\citenamefont {Koch},
  \citenamefont {Kira}, \citenamefont {Khitrova},\ and\ \citenamefont
  {Gibbs}}]{koch2006semiconductor}%
  \BibitemOpen
  \bibfield  {author} {\bibinfo {author} {\bibfnamefont {S.}~\bibnamefont
  {Koch}}, \bibinfo {author} {\bibfnamefont {M.}~\bibnamefont {Kira}}, \bibinfo
  {author} {\bibfnamefont {G.}~\bibnamefont {Khitrova}},\ and\ \bibinfo
  {author} {\bibfnamefont {H.}~\bibnamefont {Gibbs}},\ }\bibfield  {title}
  {\bibinfo {title} {Semiconductor excitons in new light},\ }\href
  {https://doi.org/10.1038/nmat1658} {\bibfield  {journal} {\bibinfo  {journal}
  {Nature materials}\ }\textbf {\bibinfo {volume} {5}},\ \bibinfo {pages} {523}
  (\bibinfo {year} {2006})}\BibitemShut {NoStop}%
\bibitem [{\citenamefont {Kaindl}\ \emph {et~al.}(2003)\citenamefont {Kaindl},
  \citenamefont {Carnahan}, \citenamefont {H{\"a}gele}, \citenamefont
  {L{\"o}venich},\ and\ \citenamefont {Chemla}}]{kaindl2003ultrafast}%
  \BibitemOpen
  \bibfield  {author} {\bibinfo {author} {\bibfnamefont {R.~A.}\ \bibnamefont
  {Kaindl}}, \bibinfo {author} {\bibfnamefont {M.~A.}\ \bibnamefont
  {Carnahan}}, \bibinfo {author} {\bibfnamefont {D.}~\bibnamefont
  {H{\"a}gele}}, \bibinfo {author} {\bibfnamefont {R.}~\bibnamefont
  {L{\"o}venich}},\ and\ \bibinfo {author} {\bibfnamefont {D.~S.}\ \bibnamefont
  {Chemla}},\ }\bibfield  {title} {\bibinfo {title} {Ultrafast terahertz probes
  of transient conducting and insulating phases in an electron--hole gas},\
  }\href {https://www.nature.com/articles/nature01676} {\bibfield  {journal}
  {\bibinfo  {journal} {Nature}\ }\textbf {\bibinfo {volume} {423}},\ \bibinfo
  {pages} {734} (\bibinfo {year} {2003})}\BibitemShut {NoStop}%
\bibitem [{\citenamefont {Kira}\ \emph {et~al.}(1998)\citenamefont {Kira},
  \citenamefont {Jahnke},\ and\ \citenamefont {Koch}}]{PhysRevLett.81.3263}%
  \BibitemOpen
  \bibfield  {author} {\bibinfo {author} {\bibfnamefont {M.}~\bibnamefont
  {Kira}}, \bibinfo {author} {\bibfnamefont {F.}~\bibnamefont {Jahnke}},\ and\
  \bibinfo {author} {\bibfnamefont {S.~W.}\ \bibnamefont {Koch}},\ }\bibfield
  {title} {\bibinfo {title} {Microscopic theory of excitonic signatures in
  semiconductor photoluminescence},\ }\href
  {https://doi.org/10.1103/PhysRevLett.81.3263} {\bibfield  {journal} {\bibinfo
   {journal} {Phys. Rev. Lett.}\ }\textbf {\bibinfo {volume} {81}},\ \bibinfo
  {pages} {3263} (\bibinfo {year} {1998})}\BibitemShut {NoStop}%
\bibitem [{\citenamefont {Halperin}\ and\ \citenamefont
  {Rice}(1968{\natexlab{a}})}]{Halperin1968TheES}%
  \BibitemOpen
  \bibfield  {author} {\bibinfo {author} {\bibfnamefont {B.}~\bibnamefont
  {Halperin}}\ and\ \bibinfo {author} {\bibfnamefont {T.}~\bibnamefont
  {Rice}},\ }\bibfield  {title} {\bibinfo {title} {The excitonic state at the
  semiconductor-semimetal transition},\ }in\ \href
  {https://www.sciencedirect.com/science/article/abs/pii/S0081194708607407}
  {\emph {\bibinfo {booktitle} {Solid State Physics}}},\ Vol.~\bibinfo {volume}
  {21}\ (\bibinfo  {publisher} {Elsevier},\ \bibinfo {year} {1968})\ pp.\
  \bibinfo {pages} {115--192}\BibitemShut {NoStop}%
\bibitem [{\citenamefont {Halperin}\ and\ \citenamefont
  {Rice}(1968{\natexlab{b}})}]{RevModPhys.40.755}%
  \BibitemOpen
  \bibfield  {author} {\bibinfo {author} {\bibfnamefont {B.~I.}\ \bibnamefont
  {Halperin}}\ and\ \bibinfo {author} {\bibfnamefont {T.~M.}\ \bibnamefont
  {Rice}},\ }\bibfield  {title} {\bibinfo {title} {Possible anomalies at a
  semimetal-semiconductor transistion},\ }\href
  {https://doi.org/10.1103/RevModPhys.40.755} {\bibfield  {journal} {\bibinfo
  {journal} {Rev. Mod. Phys.}\ }\textbf {\bibinfo {volume} {40}},\ \bibinfo
  {pages} {755} (\bibinfo {year} {1968}{\natexlab{b}})}\BibitemShut {NoStop}%
\bibitem [{\citenamefont {Kozlov}\ and\ \citenamefont
  {Maksimov}(1965)}]{kozlov1965metal}%
  \BibitemOpen
  \bibfield  {author} {\bibinfo {author} {\bibfnamefont {A.}~\bibnamefont
  {Kozlov}}\ and\ \bibinfo {author} {\bibfnamefont {L.}~\bibnamefont
  {Maksimov}},\ }\bibfield  {title} {\bibinfo {title} {The metal-dielectric
  divalent crystal phase transition},\ }\href
  {http://jetp.ras.ru/cgi-bin/dn/e_021_04_0790.pdf} {\bibfield  {journal}
  {\bibinfo  {journal} {Sov. Phys. JETP}\ }\textbf {\bibinfo {volume} {21}},\
  \bibinfo {pages} {790} (\bibinfo {year} {1965})}\BibitemShut {NoStop}%
\bibitem [{\citenamefont {Hanamura}\ and\ \citenamefont
  {Haug}(1977)}]{HANAMURA1977209}%
  \BibitemOpen
  \bibfield  {author} {\bibinfo {author} {\bibfnamefont {E.}~\bibnamefont
  {Hanamura}}\ and\ \bibinfo {author} {\bibfnamefont {H.}~\bibnamefont
  {Haug}},\ }\bibfield  {title} {\bibinfo {title} {Condensation effects of
  excitons},\ }\href
  {https://doi.org/https://doi.org/10.1016/0370-1573(77)90012-6} {\bibfield
  {journal} {\bibinfo  {journal} {Physics Reports}\ }\textbf {\bibinfo {volume}
  {33}},\ \bibinfo {pages} {209} (\bibinfo {year} {1977})}\BibitemShut
  {NoStop}%
\bibitem [{\citenamefont {Kogar}\ \emph {et~al.}(2017)\citenamefont {Kogar},
  \citenamefont {Rak}, \citenamefont {Vig}, \citenamefont {Husain},
  \citenamefont {Flicker}, \citenamefont {Joe}, \citenamefont {Venema},
  \citenamefont {MacDougall}, \citenamefont {Chiang}, \citenamefont {Fradkin},
  \citenamefont {van Wezel},\ and\ \citenamefont
  {Abbamonte}}]{doi:10.1126/science.aam6432}%
  \BibitemOpen
  \bibfield  {author} {\bibinfo {author} {\bibfnamefont {A.}~\bibnamefont
  {Kogar}}, \bibinfo {author} {\bibfnamefont {M.~S.}\ \bibnamefont {Rak}},
  \bibinfo {author} {\bibfnamefont {S.}~\bibnamefont {Vig}}, \bibinfo {author}
  {\bibfnamefont {A.~A.}\ \bibnamefont {Husain}}, \bibinfo {author}
  {\bibfnamefont {F.}~\bibnamefont {Flicker}}, \bibinfo {author} {\bibfnamefont
  {Y.~I.}\ \bibnamefont {Joe}}, \bibinfo {author} {\bibfnamefont
  {L.}~\bibnamefont {Venema}}, \bibinfo {author} {\bibfnamefont {G.~J.}\
  \bibnamefont {MacDougall}}, \bibinfo {author} {\bibfnamefont {T.~C.}\
  \bibnamefont {Chiang}}, \bibinfo {author} {\bibfnamefont {E.}~\bibnamefont
  {Fradkin}}, \bibinfo {author} {\bibfnamefont {J.}~\bibnamefont {van Wezel}},\
  and\ \bibinfo {author} {\bibfnamefont {P.}~\bibnamefont {Abbamonte}},\
  }\bibfield  {title} {\bibinfo {title} {Signatures of exciton condensation in
  a transition metal dichalcogenide},\ }\href
  {https://doi.org/10.1126/science.aam6432} {\bibfield  {journal} {\bibinfo
  {journal} {Science}\ }\textbf {\bibinfo {volume} {358}},\ \bibinfo {pages}
  {1314} (\bibinfo {year} {2017})}\BibitemShut {NoStop}%
\bibitem [{\citenamefont {J\'erome}\ \emph {et~al.}(1967)\citenamefont
  {J\'erome}, \citenamefont {Rice},\ and\ \citenamefont
  {Kohn}}]{PhysRev.158.462}%
  \BibitemOpen
  \bibfield  {author} {\bibinfo {author} {\bibfnamefont {D.}~\bibnamefont
  {J\'erome}}, \bibinfo {author} {\bibfnamefont {T.~M.}\ \bibnamefont {Rice}},\
  and\ \bibinfo {author} {\bibfnamefont {W.}~\bibnamefont {Kohn}},\ }\bibfield
  {title} {\bibinfo {title} {Excitonic insulator},\ }\href
  {https://doi.org/10.1103/PhysRev.158.462} {\bibfield  {journal} {\bibinfo
  {journal} {Phys. Rev.}\ }\textbf {\bibinfo {volume} {158}},\ \bibinfo {pages}
  {462} (\bibinfo {year} {1967})}\BibitemShut {NoStop}%
\bibitem [{\citenamefont {Blatt}\ \emph {et~al.}(1962)\citenamefont {Blatt},
  \citenamefont {B\"oer},\ and\ \citenamefont {Brandt}}]{PhysRev.126.1691}%
  \BibitemOpen
  \bibfield  {author} {\bibinfo {author} {\bibfnamefont {J.~M.}\ \bibnamefont
  {Blatt}}, \bibinfo {author} {\bibfnamefont {K.~W.}\ \bibnamefont {B\"oer}},\
  and\ \bibinfo {author} {\bibfnamefont {W.}~\bibnamefont {Brandt}},\
  }\bibfield  {title} {\bibinfo {title} {Bose-{E}instein condensation of
  excitons},\ }\href {https://doi.org/10.1103/PhysRev.126.1691} {\bibfield
  {journal} {\bibinfo  {journal} {Phys. Rev.}\ }\textbf {\bibinfo {volume}
  {126}},\ \bibinfo {pages} {1691} (\bibinfo {year} {1962})}\BibitemShut
  {NoStop}%
\bibitem [{\citenamefont {Littlewood}\ \emph {et~al.}(2004)\citenamefont
  {Littlewood}, \citenamefont {Eastham}, \citenamefont {Keeling}, \citenamefont
  {Marchetti}, \citenamefont {Simons},\ and\ \citenamefont
  {Szymanska}}]{Littlewood_2004}%
  \BibitemOpen
  \bibfield  {author} {\bibinfo {author} {\bibfnamefont {P.~B.}\ \bibnamefont
  {Littlewood}}, \bibinfo {author} {\bibfnamefont {P.~R.}\ \bibnamefont
  {Eastham}}, \bibinfo {author} {\bibfnamefont {J.~M.~J.}\ \bibnamefont
  {Keeling}}, \bibinfo {author} {\bibfnamefont {F.~M.}\ \bibnamefont
  {Marchetti}}, \bibinfo {author} {\bibfnamefont {B.~D.}\ \bibnamefont
  {Simons}},\ and\ \bibinfo {author} {\bibfnamefont {M.~H.}\ \bibnamefont
  {Szymanska}},\ }\bibfield  {title} {\bibinfo {title} {Models of coherent
  exciton condensation},\ }\href {https://doi.org/10.1088/0953-8984/16/35/003}
  {\bibfield  {journal} {\bibinfo  {journal} {Journal of Physics: Condensed
  Matter}\ }\textbf {\bibinfo {volume} {16}},\ \bibinfo {pages} {S3597}
  (\bibinfo {year} {2004})}\BibitemShut {NoStop}%
\bibitem [{\citenamefont {Glutsch}\ and\ \citenamefont
  {Bechstedt}(1991)}]{PhysRevB.44.1368}%
  \BibitemOpen
  \bibfield  {author} {\bibinfo {author} {\bibfnamefont {S.}~\bibnamefont
  {Glutsch}}\ and\ \bibinfo {author} {\bibfnamefont {F.}~\bibnamefont
  {Bechstedt}},\ }\bibfield  {title} {\bibinfo {title} {Exciton redshift for
  coherent pumping near the absorption edge},\ }\href
  {https://doi.org/10.1103/PhysRevB.44.1368} {\bibfield  {journal} {\bibinfo
  {journal} {Phys. Rev. B}\ }\textbf {\bibinfo {volume} {44}},\ \bibinfo
  {pages} {1368} (\bibinfo {year} {1991})}\BibitemShut {NoStop}%
\bibitem [{\citenamefont {Bechstedt}\ and\ \citenamefont
  {Glutsch}(1991)}]{PhysRevB.44.3638}%
  \BibitemOpen
  \bibfield  {author} {\bibinfo {author} {\bibfnamefont {F.}~\bibnamefont
  {Bechstedt}}\ and\ \bibinfo {author} {\bibfnamefont {S.}~\bibnamefont
  {Glutsch}},\ }\bibfield  {title} {\bibinfo {title} {Nonperturbative treatment
  of excitons in semiconductors coherently pumped near the absorption edge},\
  }\href {https://doi.org/10.1103/PhysRevB.44.3638} {\bibfield  {journal}
  {\bibinfo  {journal} {Phys. Rev. B}\ }\textbf {\bibinfo {volume} {44}},\
  \bibinfo {pages} {3638} (\bibinfo {year} {1991})}\BibitemShut {NoStop}%
\bibitem [{\citenamefont {Arseev}\ and\ \citenamefont
  {Dzyubenko}(1998)}]{arseev1998}%
  \BibitemOpen
  \bibfield  {author} {\bibinfo {author} {\bibfnamefont {P.}~\bibnamefont
  {Arseev}}\ and\ \bibinfo {author} {\bibfnamefont {A.}~\bibnamefont
  {Dzyubenko}},\ }\bibfield  {title} {\bibinfo {title} {Exciton
  magnetotransport in two-dimensional systems: weak-localization effects},\
  }\href {https://link.springer.com/article/10.1134/1.558641} {\bibfield
  {journal} {\bibinfo  {journal} {Journal of Experimental and Theoretical
  Physics}\ }\textbf {\bibinfo {volume} {87}},\ \bibinfo {pages} {200}
  (\bibinfo {year} {1998})}\BibitemShut {NoStop}%
\bibitem [{\citenamefont {Lerner}\ and\ \citenamefont
  {Lozovik}(1981)}]{lerner1981two}%
  \BibitemOpen
  \bibfield  {author} {\bibinfo {author} {\bibfnamefont {I.}~\bibnamefont
  {Lerner}}\ and\ \bibinfo {author} {\bibfnamefont {Y.~E.}\ \bibnamefont
  {Lozovik}},\ }\bibfield  {title} {\bibinfo {title} {Two-dimensional
  electron-hole system in a strong magnetic field as an almost ideal exciton
  gas},\ }\href {http://jetp.ras.ru/cgi-bin/dn/e_053_04_0763.pdf} {\bibfield
  {journal} {\bibinfo  {journal} {Zh. Eksp. Teor. Fiz}\ }\textbf {\bibinfo
  {volume} {80}},\ \bibinfo {pages} {1488} (\bibinfo {year}
  {1981})}\BibitemShut {NoStop}%
\bibitem [{\citenamefont {Lozovik}\ \emph {et~al.}(1999)\citenamefont
  {Lozovik}, \citenamefont {Berman},\ and\ \citenamefont
  {Tsvetus}}]{PhysRevB.59.5627}%
  \BibitemOpen
  \bibfield  {author} {\bibinfo {author} {\bibfnamefont {Y.~E.}\ \bibnamefont
  {Lozovik}}, \bibinfo {author} {\bibfnamefont {O.~L.}\ \bibnamefont
  {Berman}},\ and\ \bibinfo {author} {\bibfnamefont {V.~G.}\ \bibnamefont
  {Tsvetus}},\ }\bibfield  {title} {\bibinfo {title} {Phase transitions of
  electron-hole and unbalanced electron systems in coupled quantum wells in
  high magnetic fields},\ }\href {https://doi.org/10.1103/PhysRevB.59.5627}
  {\bibfield  {journal} {\bibinfo  {journal} {Phys. Rev. B}\ }\textbf {\bibinfo
  {volume} {59}},\ \bibinfo {pages} {5627} (\bibinfo {year}
  {1999})}\BibitemShut {NoStop}%
\bibitem [{\citenamefont {Cloizeaux}(1965)}]{CLOIZEAUX1965259}%
  \BibitemOpen
  \bibfield  {author} {\bibinfo {author} {\bibfnamefont {J.~D.}\ \bibnamefont
  {Cloizeaux}},\ }\bibfield  {title} {\bibinfo {title} {Exciton instability and
  crystallographic anomalies in semiconductors},\ }\href
  {https://doi.org/https://doi.org/10.1016/0022-3697(65)90153-8} {\bibfield
  {journal} {\bibinfo  {journal} {Journal of Physics and Chemistry of Solids}\
  }\textbf {\bibinfo {volume} {26}},\ \bibinfo {pages} {259} (\bibinfo {year}
  {1965})}\BibitemShut {NoStop}%
\bibitem [{\citenamefont {Fetter}\ and\ \citenamefont
  {Walecka}(2012)}]{fetter2012quantum}%
  \BibitemOpen
  \bibfield  {author} {\bibinfo {author} {\bibfnamefont {A.~L.}\ \bibnamefont
  {Fetter}}\ and\ \bibinfo {author} {\bibfnamefont {J.~D.}\ \bibnamefont
  {Walecka}},\ }\href@noop {} {\emph {\bibinfo {title} {Quantum theory of
  many-particle systems}}}\ (\bibinfo  {publisher} {Courier Corporation},\
  \bibinfo {year} {2012})\BibitemShut {NoStop}%
\bibitem [{\citenamefont {Pekker}\ and\ \citenamefont
  {Varma}(2015)}]{annurev-conmatphys-031214-014350}%
  \BibitemOpen
  \bibfield  {author} {\bibinfo {author} {\bibfnamefont {D.}~\bibnamefont
  {Pekker}}\ and\ \bibinfo {author} {\bibfnamefont {C.}~\bibnamefont {Varma}},\
  }\bibfield  {title} {\bibinfo {title} {Amplitude/{H}iggs modes in condensed
  matter physics},\ }\href
  {https://doi.org/10.1146/annurev-conmatphys-031214-014350} {\bibfield
  {journal} {\bibinfo  {journal} {Annual Review of Condensed Matter Physics}\
  }\textbf {\bibinfo {volume} {6}},\ \bibinfo {pages} {269} (\bibinfo {year}
  {2015})}\BibitemShut {NoStop}%
\bibitem [{\citenamefont {Littlewood}\ and\ \citenamefont
  {Varma}(1981)}]{PhysRevLett.47.811}%
  \BibitemOpen
  \bibfield  {author} {\bibinfo {author} {\bibfnamefont {P.~B.}\ \bibnamefont
  {Littlewood}}\ and\ \bibinfo {author} {\bibfnamefont {C.~M.}\ \bibnamefont
  {Varma}},\ }\bibfield  {title} {\bibinfo {title} {Gauge-invariant theory of
  the dynamical interaction of charge density waves and superconductivity},\
  }\href {https://doi.org/10.1103/PhysRevLett.47.811} {\bibfield  {journal}
  {\bibinfo  {journal} {Phys. Rev. Lett.}\ }\textbf {\bibinfo {volume} {47}},\
  \bibinfo {pages} {811} (\bibinfo {year} {1981})}\BibitemShut {NoStop}%
\bibitem [{\citenamefont {Littlewood}\ and\ \citenamefont
  {Varma}(1982)}]{PhysRevB.26.4883}%
  \BibitemOpen
  \bibfield  {author} {\bibinfo {author} {\bibfnamefont {P.~B.}\ \bibnamefont
  {Littlewood}}\ and\ \bibinfo {author} {\bibfnamefont {C.~M.}\ \bibnamefont
  {Varma}},\ }\bibfield  {title} {\bibinfo {title} {Amplitude collective modes
  in superconductors and their coupling to charge-density waves},\ }\href
  {https://doi.org/10.1103/PhysRevB.26.4883} {\bibfield  {journal} {\bibinfo
  {journal} {Phys. Rev. B}\ }\textbf {\bibinfo {volume} {26}},\ \bibinfo
  {pages} {4883} (\bibinfo {year} {1982})}\BibitemShut {NoStop}%
\bibitem [{\citenamefont {Varma}(2002)}]{varma2002higgs}%
  \BibitemOpen
  \bibfield  {author} {\bibinfo {author} {\bibfnamefont {C.}~\bibnamefont
  {Varma}},\ }\bibfield  {title} {\bibinfo {title} {Higgs boson in
  superconductors},\ }\href@noop {} {\bibfield  {journal} {\bibinfo  {journal}
  {Journal of low temperature physics}\ }\textbf {\bibinfo {volume} {126}},\
  \bibinfo {pages} {901} (\bibinfo {year} {2002})}\BibitemShut {NoStop}%
\bibitem [{\citenamefont {Altland}\ and\ \citenamefont
  {Simons}(2010)}]{altland_simons_2010}%
  \BibitemOpen
  \bibfield  {author} {\bibinfo {author} {\bibfnamefont {A.}~\bibnamefont
  {Altland}}\ and\ \bibinfo {author} {\bibfnamefont {B.~D.}\ \bibnamefont
  {Simons}},\ }\href {https://doi.org/10.1017/CBO9780511789984} {\emph
  {\bibinfo {title} {Condensed Matter Field Theory}}},\ \bibinfo {edition}
  {2nd}\ ed.\ (\bibinfo  {publisher} {Cambridge University Press},\ \bibinfo
  {year} {2010})\BibitemShut {NoStop}%
\bibitem [{\citenamefont {Maialle}\ and\ \citenamefont
  {Sham}(1994)}]{MAIALLE1994256}%
  \BibitemOpen
  \bibfield  {author} {\bibinfo {author} {\bibfnamefont {M.}~\bibnamefont
  {Maialle}}\ and\ \bibinfo {author} {\bibfnamefont {L.}~\bibnamefont {Sham}},\
  }\bibfield  {title} {\bibinfo {title} {Exciton spin dynamics and polarized
  luminescence in quantum wells},\ }\href
  {https://doi.org/https://doi.org/10.1016/0039-6028(94)90896-6} {\bibfield
  {journal} {\bibinfo  {journal} {Surface Science}\ }\textbf {\bibinfo {volume}
  {305}},\ \bibinfo {pages} {256} (\bibinfo {year} {1994})}\BibitemShut
  {NoStop}%
\bibitem [{\citenamefont {Zhang}\ \emph {et~al.}(2015)\citenamefont {Zhang},
  \citenamefont {You}, \citenamefont {Zhao},\ and\ \citenamefont
  {Heinz}}]{PhysRevLett.115.257403}%
  \BibitemOpen
  \bibfield  {author} {\bibinfo {author} {\bibfnamefont {X.-X.}\ \bibnamefont
  {Zhang}}, \bibinfo {author} {\bibfnamefont {Y.}~\bibnamefont {You}}, \bibinfo
  {author} {\bibfnamefont {S.~Y.~F.}\ \bibnamefont {Zhao}},\ and\ \bibinfo
  {author} {\bibfnamefont {T.~F.}\ \bibnamefont {Heinz}},\ }\bibfield  {title}
  {\bibinfo {title} {Experimental evidence for dark excitons in monolayer
  {${\mathrm{WSe}}_{2}$}},\ }\href
  {https://doi.org/10.1103/PhysRevLett.115.257403} {\bibfield  {journal}
  {\bibinfo  {journal} {Phys. Rev. Lett.}\ }\textbf {\bibinfo {volume} {115}},\
  \bibinfo {pages} {257403} (\bibinfo {year} {2015})}\BibitemShut {NoStop}%
\bibitem [{\citenamefont {Giuliani}\ and\ \citenamefont
  {Vignale}(2005)}]{giuliani_vignale_2005}%
  \BibitemOpen
  \bibfield  {author} {\bibinfo {author} {\bibfnamefont {G.}~\bibnamefont
  {Giuliani}}\ and\ \bibinfo {author} {\bibfnamefont {G.}~\bibnamefont
  {Vignale}},\ }\href {https://doi.org/10.1017/CBO9780511619915} {\emph
  {\bibinfo {title} {Quantum Theory of the Electron Liquid}}}\ (\bibinfo
  {publisher} {Cambridge University Press},\ \bibinfo {year}
  {2005})\BibitemShut {NoStop}%
\bibitem [{\citenamefont {Mihaila}()}]{mihaila2011lindhard}%
  \BibitemOpen
  \bibfield  {author} {\bibinfo {author} {\bibfnamefont {B.}~\bibnamefont
  {Mihaila}},\ }\bibfield  {title} {\bibinfo {title} {Lindhard function of a
  d-dimensional {Fermi} gas},\ }\href {https://arxiv.org/abs/1111.5337}
  {\bibinfo  {journal} {arXiv:1111.5337}\ }\BibitemShut {NoStop}%
\bibitem [{\citenamefont {Jain}\ \emph {et~al.}(2013)\citenamefont {Jain},
  \citenamefont {Ong}, \citenamefont {Hautier}, \citenamefont {Chen},
  \citenamefont {Richards}, \citenamefont {Dacek}, \citenamefont {Cholia},
  \citenamefont {Gunter}, \citenamefont {Skinner}, \citenamefont {Ceder},\ and\
  \citenamefont {Persson}}]{Jain2013}%
  \BibitemOpen
\bibfield  {journal} {  }\bibfield  {author} {\bibinfo {author} {\bibfnamefont
  {A.}~\bibnamefont {Jain}}, \bibinfo {author} {\bibfnamefont {S.~P.}\
  \bibnamefont {Ong}}, \bibinfo {author} {\bibfnamefont {G.}~\bibnamefont
  {Hautier}}, \bibinfo {author} {\bibfnamefont {W.}~\bibnamefont {Chen}},
  \bibinfo {author} {\bibfnamefont {W.~D.}\ \bibnamefont {Richards}}, \bibinfo
  {author} {\bibfnamefont {S.}~\bibnamefont {Dacek}}, \bibinfo {author}
  {\bibfnamefont {S.}~\bibnamefont {Cholia}}, \bibinfo {author} {\bibfnamefont
  {D.}~\bibnamefont {Gunter}}, \bibinfo {author} {\bibfnamefont
  {D.}~\bibnamefont {Skinner}}, \bibinfo {author} {\bibfnamefont
  {G.}~\bibnamefont {Ceder}},\ and\ \bibinfo {author} {\bibfnamefont {K.~a.}\
  \bibnamefont {Persson}},\ }\bibfield  {title} {\bibinfo {title} {{The
  Materials Project: A materials genome approach to accelerating materials
  innovation}},\ }\href {https://doi.org/10.1063/1.4812323} {\bibfield
  {journal} {\bibinfo  {journal} {APL Materials}\ }\textbf {\bibinfo {volume}
  {1}},\ \bibinfo {pages} {011002} (\bibinfo {year} {2013})}\BibitemShut
  {NoStop}%
\bibitem [{\citenamefont {Wu}\ \emph {et~al.}(2019)\citenamefont {Wu},
  \citenamefont {Lou}, \citenamefont {Chang}, \citenamefont {Sullivan},\ and\
  \citenamefont {Du}}]{PhysRevB.99.085307}%
  \BibitemOpen
  \bibfield  {author} {\bibinfo {author} {\bibfnamefont {X.}~\bibnamefont
  {Wu}}, \bibinfo {author} {\bibfnamefont {W.}~\bibnamefont {Lou}}, \bibinfo
  {author} {\bibfnamefont {K.}~\bibnamefont {Chang}}, \bibinfo {author}
  {\bibfnamefont {G.}~\bibnamefont {Sullivan}},\ and\ \bibinfo {author}
  {\bibfnamefont {R.-R.}\ \bibnamefont {Du}},\ }\bibfield  {title} {\bibinfo
  {title} {Resistive signature of excitonic coupling in an electron-hole double
  layer with a middle barrier},\ }\href
  {https://doi.org/10.1103/PhysRevB.99.085307} {\bibfield  {journal} {\bibinfo
  {journal} {Phys. Rev. B}\ }\textbf {\bibinfo {volume} {99}},\ \bibinfo
  {pages} {085307} (\bibinfo {year} {2019})}\BibitemShut {NoStop}%
\bibitem [{\citenamefont {Jiang}\ \emph {et~al.}(2017)\citenamefont {Jiang},
  \citenamefont {Thapa}, \citenamefont {Sanders}, \citenamefont {Stanton},
  \citenamefont {Zhang}, \citenamefont {Kono}, \citenamefont {Lou},
  \citenamefont {Chang}, \citenamefont {Hawkins}, \citenamefont {Klem},
  \citenamefont {Pan}, \citenamefont {Smirnov},\ and\ \citenamefont
  {Jiang}}]{PhysRevB.95.045116}%
  \BibitemOpen
  \bibfield  {author} {\bibinfo {author} {\bibfnamefont {Y.}~\bibnamefont
  {Jiang}}, \bibinfo {author} {\bibfnamefont {S.}~\bibnamefont {Thapa}},
  \bibinfo {author} {\bibfnamefont {G.~D.}\ \bibnamefont {Sanders}}, \bibinfo
  {author} {\bibfnamefont {C.~J.}\ \bibnamefont {Stanton}}, \bibinfo {author}
  {\bibfnamefont {Q.}~\bibnamefont {Zhang}}, \bibinfo {author} {\bibfnamefont
  {J.}~\bibnamefont {Kono}}, \bibinfo {author} {\bibfnamefont {W.~K.}\
  \bibnamefont {Lou}}, \bibinfo {author} {\bibfnamefont {K.}~\bibnamefont
  {Chang}}, \bibinfo {author} {\bibfnamefont {S.~D.}\ \bibnamefont {Hawkins}},
  \bibinfo {author} {\bibfnamefont {J.~F.}\ \bibnamefont {Klem}}, \bibinfo
  {author} {\bibfnamefont {W.}~\bibnamefont {Pan}}, \bibinfo {author}
  {\bibfnamefont {D.}~\bibnamefont {Smirnov}},\ and\ \bibinfo {author}
  {\bibfnamefont {Z.}~\bibnamefont {Jiang}},\ }\bibfield  {title} {\bibinfo
  {title} {Probing the semiconductor to semimetal transition in {InAs/GaSb}
  double quantum wells by magneto-infrared spectroscopy},\ }\href
  {https://doi.org/10.1103/PhysRevB.95.045116} {\bibfield  {journal} {\bibinfo
  {journal} {Phys. Rev. B}\ }\textbf {\bibinfo {volume} {95}},\ \bibinfo
  {pages} {045116} (\bibinfo {year} {2017})}\BibitemShut {NoStop}%
\bibitem [{\citenamefont {Du}\ \emph {et~al.}(2017)\citenamefont {Du},
  \citenamefont {Li}, \citenamefont {Lou}, \citenamefont {Sullivan},
  \citenamefont {Chang}, \citenamefont {Kono},\ and\ \citenamefont
  {Du}}]{du2017evidence}%
  \BibitemOpen
  \bibfield  {author} {\bibinfo {author} {\bibfnamefont {L.}~\bibnamefont
  {Du}}, \bibinfo {author} {\bibfnamefont {X.}~\bibnamefont {Li}}, \bibinfo
  {author} {\bibfnamefont {W.}~\bibnamefont {Lou}}, \bibinfo {author}
  {\bibfnamefont {G.}~\bibnamefont {Sullivan}}, \bibinfo {author}
  {\bibfnamefont {K.}~\bibnamefont {Chang}}, \bibinfo {author} {\bibfnamefont
  {J.}~\bibnamefont {Kono}},\ and\ \bibinfo {author} {\bibfnamefont {R.-R.}\
  \bibnamefont {Du}},\ }\bibfield  {title} {\bibinfo {title} {Evidence for a
  topological excitonic insulator in {InAs/GaSb} bilayers},\ }\href
  {https://www.nature.com/articles/s41467-017-01988-1} {\bibfield  {journal}
  {\bibinfo  {journal} {Nature communications}\ }\textbf {\bibinfo {volume}
  {8}},\ \bibinfo {pages} {1} (\bibinfo {year} {2017})}\BibitemShut {NoStop}%
\bibitem [{\citenamefont {Li}\ \emph {et~al.}(2017)\citenamefont {Li},
  \citenamefont {Taniguchi}, \citenamefont {Watanabe}, \citenamefont {Hone},\
  and\ \citenamefont {Dean}}]{li2017excitonic}%
  \BibitemOpen
  \bibfield  {author} {\bibinfo {author} {\bibfnamefont {J.}~\bibnamefont
  {Li}}, \bibinfo {author} {\bibfnamefont {T.}~\bibnamefont {Taniguchi}},
  \bibinfo {author} {\bibfnamefont {K.}~\bibnamefont {Watanabe}}, \bibinfo
  {author} {\bibfnamefont {J.}~\bibnamefont {Hone}},\ and\ \bibinfo {author}
  {\bibfnamefont {C.}~\bibnamefont {Dean}},\ }\bibfield  {title} {\bibinfo
  {title} {Excitonic superfluid phase in double bilayer graphene},\ }\href
  {https://www.nature.com/articles/nphys4140} {\bibfield  {journal} {\bibinfo
  {journal} {Nature Physics}\ }\textbf {\bibinfo {volume} {13}},\ \bibinfo
  {pages} {751} (\bibinfo {year} {2017})}\BibitemShut {NoStop}%
\bibitem [{\citenamefont {Chen}\ and\ \citenamefont
  {Shindou}(2019)}]{PhysRevB.100.035130}%
  \BibitemOpen
  \bibfield  {author} {\bibinfo {author} {\bibfnamefont {K.}~\bibnamefont
  {Chen}}\ and\ \bibinfo {author} {\bibfnamefont {R.}~\bibnamefont {Shindou}},\
  }\bibfield  {title} {\bibinfo {title} {Helicoidal excitonic phase in an
  electron-hole double-layer system},\ }\href
  {https://doi.org/10.1103/PhysRevB.100.035130} {\bibfield  {journal} {\bibinfo
   {journal} {Phys. Rev. B}\ }\textbf {\bibinfo {volume} {100}},\ \bibinfo
  {pages} {035130} (\bibinfo {year} {2019})}\BibitemShut {NoStop}%
\bibitem [{\citenamefont {Zhang}\ and\ \citenamefont
  {Shindou}(2022)}]{zhang2021dissipationless}%
  \BibitemOpen
  \bibfield  {author} {\bibinfo {author} {\bibfnamefont {Y.}~\bibnamefont
  {Zhang}}\ and\ \bibinfo {author} {\bibfnamefont {R.}~\bibnamefont
  {Shindou}},\ }\bibfield  {title} {\bibinfo {title} {Dissipationless
  spin-charge conversion in excitonic pseudospin superfluid},\ }\href
  {https://doi.org/10.1103/PhysRevLett.128.066601} {\bibfield  {journal}
  {\bibinfo  {journal} {Phys. Rev. Lett.}\ }\textbf {\bibinfo {volume} {128}},\
  \bibinfo {pages} {066601} (\bibinfo {year} {2022})}\BibitemShut {NoStop}%
\bibitem [{\citenamefont {Kroemer}(2004)}]{kroemer20046}%
  \BibitemOpen
  \bibfield  {author} {\bibinfo {author} {\bibfnamefont {H.}~\bibnamefont
  {Kroemer}},\ }\bibfield  {title} {\bibinfo {title} {The 6.1 {A} family
  ({InAs, GaSb, AlSb}) and its heterostructures: a selective review},\ }\href
  {https://www.sciencedirect.com/science/article/pii/S1386947703004168}
  {\bibfield  {journal} {\bibinfo  {journal} {Physica E: Low-dimensional
  Systems and Nanostructures}\ }\textbf {\bibinfo {volume} {20}},\ \bibinfo
  {pages} {196} (\bibinfo {year} {2004})}\BibitemShut {NoStop}%
\bibitem [{\citenamefont {Perali}\ \emph {et~al.}(2013)\citenamefont {Perali},
  \citenamefont {Neilson},\ and\ \citenamefont {Hamilton}}]{Perali2013}%
  \BibitemOpen
  \bibfield  {author} {\bibinfo {author} {\bibfnamefont {A.}~\bibnamefont
  {Perali}}, \bibinfo {author} {\bibfnamefont {D.}~\bibnamefont {Neilson}},\
  and\ \bibinfo {author} {\bibfnamefont {A.~R.}\ \bibnamefont {Hamilton}},\
  }\bibfield  {title} {\bibinfo {title} {High-temperature superfluidity in
  double-bilayer graphene},\ }\href
  {https://doi.org/10.1103/PhysRevLett.110.146803} {\bibfield  {journal}
  {\bibinfo  {journal} {Phys. Rev. Lett.}\ }\textbf {\bibinfo {volume} {110}},\
  \bibinfo {pages} {146803} (\bibinfo {year} {2013})}\BibitemShut {NoStop}%
\bibitem [{\citenamefont {Conti}\ \emph {et~al.}(2017)\citenamefont {Conti},
  \citenamefont {Perali}, \citenamefont {Peeters},\ and\ \citenamefont
  {Neilson}}]{Conti2017}%
  \BibitemOpen
  \bibfield  {author} {\bibinfo {author} {\bibfnamefont {S.}~\bibnamefont
  {Conti}}, \bibinfo {author} {\bibfnamefont {A.}~\bibnamefont {Perali}},
  \bibinfo {author} {\bibfnamefont {F.~M.}\ \bibnamefont {Peeters}},\ and\
  \bibinfo {author} {\bibfnamefont {D.}~\bibnamefont {Neilson}},\ }\bibfield
  {title} {\bibinfo {title} {Multicomponent electron-hole superfluidity and the
  bcs-bec crossover in double bilayer graphene},\ }\href
  {https://doi.org/10.1103/PhysRevLett.119.257002} {\bibfield  {journal}
  {\bibinfo  {journal} {Phys. Rev. Lett.}\ }\textbf {\bibinfo {volume} {119}},\
  \bibinfo {pages} {257002} (\bibinfo {year} {2017})}\BibitemShut {NoStop}%
\bibitem [{\citenamefont {L\'opez~R\'{\i}os}\ \emph {et~al.}(2018)\citenamefont
  {L\'opez~R\'{\i}os}, \citenamefont {Perali}, \citenamefont {Needs},\ and\
  \citenamefont {Neilson}}]{Rios2018}%
  \BibitemOpen
  \bibfield  {author} {\bibinfo {author} {\bibfnamefont {P.}~\bibnamefont
  {L\'opez~R\'{\i}os}}, \bibinfo {author} {\bibfnamefont {A.}~\bibnamefont
  {Perali}}, \bibinfo {author} {\bibfnamefont {R.~J.}\ \bibnamefont {Needs}},\
  and\ \bibinfo {author} {\bibfnamefont {D.}~\bibnamefont {Neilson}},\
  }\bibfield  {title} {\bibinfo {title} {Evidence from quantum monte carlo
  simulations of large-gap superfluidity and bcs-bec crossover in double
  electron-hole layers},\ }\href
  {https://doi.org/10.1103/PhysRevLett.120.177701} {\bibfield  {journal}
  {\bibinfo  {journal} {Phys. Rev. Lett.}\ }\textbf {\bibinfo {volume} {120}},\
  \bibinfo {pages} {177701} (\bibinfo {year} {2018})}\BibitemShut {NoStop}%
\bibitem [{\citenamefont {Liang}(1970)}]{Liang_1970}%
  \BibitemOpen
  \bibfield  {author} {\bibinfo {author} {\bibfnamefont {W.~Y.}\ \bibnamefont
  {Liang}},\ }\bibfield  {title} {\bibinfo {title} {Excitons},\ }\href
  {https://doi.org/10.1088/0031-9120/5/4/003} {\bibfield  {journal} {\bibinfo
  {journal} {Physics Education}\ }\textbf {\bibinfo {volume} {5}},\ \bibinfo
  {pages} {226} (\bibinfo {year} {1970})}\BibitemShut {NoStop}%
\bibitem [{\citenamefont {Kira}\ \emph {et~al.}(1999)\citenamefont {Kira},
  \citenamefont {Jahnke},\ and\ \citenamefont {Koch}}]{PhysRevLett.82.3544}%
  \BibitemOpen
  \bibfield  {author} {\bibinfo {author} {\bibfnamefont {M.}~\bibnamefont
  {Kira}}, \bibinfo {author} {\bibfnamefont {F.}~\bibnamefont {Jahnke}},\ and\
  \bibinfo {author} {\bibfnamefont {S.~W.}\ \bibnamefont {Koch}},\ }\bibfield
  {title} {\bibinfo {title} {Quantum theory of secondary emission in optically
  excited semiconductor quantum wells},\ }\href
  {https://doi.org/10.1103/PhysRevLett.82.3544} {\bibfield  {journal} {\bibinfo
   {journal} {Phys. Rev. Lett.}\ }\textbf {\bibinfo {volume} {82}},\ \bibinfo
  {pages} {3544} (\bibinfo {year} {1999})}\BibitemShut {NoStop}%
\bibitem [{\citenamefont {Brar}\ \emph {et~al.}(1993)\citenamefont {Brar},
  \citenamefont {Kroemer}, \citenamefont {Ibbetson},\ and\ \citenamefont
  {English}}]{doi:10.1063/1.109053}%
  \BibitemOpen
  \bibfield  {author} {\bibinfo {author} {\bibfnamefont {B.}~\bibnamefont
  {Brar}}, \bibinfo {author} {\bibfnamefont {H.}~\bibnamefont {Kroemer}},
  \bibinfo {author} {\bibfnamefont {J.}~\bibnamefont {Ibbetson}},\ and\
  \bibinfo {author} {\bibfnamefont {J.~H.}\ \bibnamefont {English}},\
  }\bibfield  {title} {\bibinfo {title} {Photoluminescence from narrow
  {InAs‐AlSb} quantum wells},\ }\href {https://doi.org/10.1063/1.109053}
  {\bibfield  {journal} {\bibinfo  {journal} {Applied Physics Letters}\
  }\textbf {\bibinfo {volume} {62}},\ \bibinfo {pages} {3303} (\bibinfo {year}
  {1993})}\BibitemShut {NoStop}%
\bibitem [{\citenamefont {Wakisaka}\ \emph {et~al.}(2009)\citenamefont
  {Wakisaka}, \citenamefont {Sudayama}, \citenamefont {Takubo}, \citenamefont
  {Mizokawa}, \citenamefont {Arita}, \citenamefont {Namatame}, \citenamefont
  {Taniguchi}, \citenamefont {Katayama}, \citenamefont {Nohara},\ and\
  \citenamefont {Takagi}}]{PhysRevLett.103.026402}%
  \BibitemOpen
  \bibfield  {author} {\bibinfo {author} {\bibfnamefont {Y.}~\bibnamefont
  {Wakisaka}}, \bibinfo {author} {\bibfnamefont {T.}~\bibnamefont {Sudayama}},
  \bibinfo {author} {\bibfnamefont {K.}~\bibnamefont {Takubo}}, \bibinfo
  {author} {\bibfnamefont {T.}~\bibnamefont {Mizokawa}}, \bibinfo {author}
  {\bibfnamefont {M.}~\bibnamefont {Arita}}, \bibinfo {author} {\bibfnamefont
  {H.}~\bibnamefont {Namatame}}, \bibinfo {author} {\bibfnamefont
  {M.}~\bibnamefont {Taniguchi}}, \bibinfo {author} {\bibfnamefont
  {N.}~\bibnamefont {Katayama}}, \bibinfo {author} {\bibfnamefont
  {M.}~\bibnamefont {Nohara}},\ and\ \bibinfo {author} {\bibfnamefont
  {H.}~\bibnamefont {Takagi}},\ }\bibfield  {title} {\bibinfo {title}
  {Excitonic insulator state in {$\mathrm{Ta}_2{\mathrm{NiSe}}_5$} probed by
  photoemission spectroscopy},\ }\href
  {https://doi.org/10.1103/PhysRevLett.103.026402} {\bibfield  {journal}
  {\bibinfo  {journal} {Phys. Rev. Lett.}\ }\textbf {\bibinfo {volume} {103}},\
  \bibinfo {pages} {026402} (\bibinfo {year} {2009})}\BibitemShut {NoStop}%
\end{thebibliography}%

\end{document}